\documentclass[twocolumn,showpacs,showkeys,preprintnumbers,amsmath,amssymb,floatfix,aps]{revtex4-1}
\usepackage[latin9]{inputenc}
\usepackage{color}
\usepackage{graphicx}

\usepackage[colorlinks,urlcolor=blue]{hyperref}
\usepackage{breakurl}

\makeatletter

\providecommand{\tabularnewline}{\\}

\@ifundefined{textcolor}{}
{%
 \definecolor{BLACK}{gray}{0}
 \definecolor{WHITE}{gray}{1}
 \definecolor{RED}{rgb}{1,0,0}
 \definecolor{GREEN}{rgb}{0,1,0}
 \definecolor{BLUE}{rgb}{0,0,1}
 \definecolor{CYAN}{cmyk}{1,0,0,0}
 \definecolor{MAGENTA}{cmyk}{0,1,0,0}
 \definecolor{YELLOW}{cmyk}{0,0,1,0}
}

\newcounter{univ_counter}
\addtocounter{univ_counter} {1}
\edef\JLAB{$^{\arabic{univ_counter}}$ }

\addtocounter{univ_counter} {1} 
\edef\ANL{$^{\arabic{univ_counter}}$ }

\addtocounter{univ_counter} {1} 
\edef\ASU{$^{\arabic{univ_counter}}$ }

\addtocounter{univ_counter} {1} 
\edef\CSUDH{$^{\arabic{univ_counter}}$ }

\addtocounter{univ_counter} {1} 
\edef\CANISIUS{$^{\arabic{univ_counter}}$ }

\addtocounter{univ_counter} {1} 
\edef\CMU{$^{\arabic{univ_counter}}$ }

\addtocounter{univ_counter} {1} 
\edef\CUA{$^{\arabic{univ_counter}}$ }

\addtocounter{univ_counter} {1} 
\edef\SACLAY{$^{\arabic{univ_counter}}$ }

\addtocounter{univ_counter} {1} 
\edef\CNU{$^{\arabic{univ_counter}}$ }

\addtocounter{univ_counter} {1} 
\edef\UCONN{$^{\arabic{univ_counter}}$ }

\addtocounter{univ_counter} {1} 
\edef\DAMMAM{$^{\arabic{univ_counter}}$ }

\addtocounter{univ_counter} {1} 
\edef\DUKE{$^{\arabic{univ_counter}}$ }

\addtocounter{univ_counter} {1} 
\edef\EDINBURGH{$^{\arabic{univ_counter}}$ }

\addtocounter{univ_counter} {1}
\edef\FERRARAU{$^{\arabic{univ_counter}}$ }

\addtocounter{univ_counter} {1} 
\edef\FU{$^{\arabic{univ_counter}}$ }

\addtocounter{univ_counter} {1} 
\edef\FIU{$^{\arabic{univ_counter}}$ }

\addtocounter{univ_counter} {1} 
\edef\FSU{$^{\arabic{univ_counter}}$ }

\addtocounter{univ_counter} {1} 
\edef\GENOVA{$^{\arabic{univ_counter}}$ }

\addtocounter{univ_counter} {1} 
\edef\GLASGOW{$^{\arabic{univ_counter}}$ }

\addtocounter{univ_counter} {1} 
\edef\GWUI{$^{\arabic{univ_counter}}$ }

\addtocounter{univ_counter} {1} 
\edef\HAMPTON{$^{\arabic{univ_counter}}$ }

\addtocounter{univ_counter} {1} 
\edef\ISU{$^{\arabic{univ_counter}}$ }

\addtocounter{univ_counter} {1} 
\edef\INFNFE{$^{\arabic{univ_counter}}$ }

\addtocounter{univ_counter} {1} 
\edef\INFNFR{$^{\arabic{univ_counter}}$ }

\addtocounter{univ_counter} {1} 
\edef\INFNGE{$^{\arabic{univ_counter}}$ }

\addtocounter{univ_counter} {1} 
\edef\INFNRO{$^{\arabic{univ_counter}}$ }

\addtocounter{univ_counter} {1} 
\edef\INFNTUR{$^{\arabic{univ_counter}}$ }

\addtocounter{univ_counter} {1} 
\edef\ORSAY{$^{\arabic{univ_counter}}$ }

\addtocounter{univ_counter} {1}
\edef\RUDJER{$^{\arabic{univ_counter}}$ }

\addtocounter{univ_counter} {1} 
\edef\ITEP{$^{\arabic{univ_counter}}$ }

\addtocounter{univ_counter} {1} 
\edef\JMU{$^{\arabic{univ_counter}}$ }

\addtocounter{univ_counter} {1} 
\edef\KNU{$^{\arabic{univ_counter}}$ }

\addtocounter{univ_counter} {1} 
\edef\MISS{$^{\arabic{univ_counter}}$ }

\addtocounter{univ_counter} {1} 
\edef\UNH{$^{\arabic{univ_counter}}$ }

\addtocounter{univ_counter} {1} 
\edef\NSU{$^{\arabic{univ_counter}}$ }

\addtocounter{univ_counter} {1} 
\edef\OHIOU{$^{\arabic{univ_counter}}$ }

\addtocounter{univ_counter} {1} 
\edef\ODU{$^{\arabic{univ_counter}}$ }

\addtocounter{univ_counter} {1} 
\edef\URICH{$^{\arabic{univ_counter}}$ }

\addtocounter{univ_counter} {1} 
\edef\ROMAII{$^{\arabic{univ_counter}}$ }

\addtocounter{univ_counter} {1} 
\edef\MSU{$^{\arabic{univ_counter}}$ }

\addtocounter{univ_counter} {1} 
\edef\SCAROLINA{$^{\arabic{univ_counter}}$ }

\addtocounter{univ_counter} {1} 
\edef\TEMPLE{$^{\arabic{univ_counter}}$ }

\addtocounter{univ_counter} {1} 
\edef\UTFSM{$^{\arabic{univ_counter}}$ }

\addtocounter{univ_counter} {1} 
\edef\VT{$^{\arabic{univ_counter}}$ }

\addtocounter{univ_counter} {1} 
\edef\VIRGINIA{$^{\arabic{univ_counter}}$ }

\addtocounter{univ_counter} {1} 
\edef\WM{$^{\arabic{univ_counter}}$ }

\addtocounter{univ_counter} {1} 
\edef\YEREVAN{$^{\arabic{univ_counter}}$ }

\begin{document}

\preprint{Phys. Rev. C}

\title{Differential Cross Section Measurements for $\gamma n\to\pi^-p$ Above
the First Nucleon Resonance Region}

\author{P.~T.~Mattione,\JLAB\
D.~S.~Carman,\JLAB\
I.~I.~Strakovsky,\GWUI\
R.~L.~Workman,\GWUI\
A.~E.~Kudryavtsev,\ITEP\
A.~Svarc,\RUDJER\
V.~E.~Tarasov,\ITEP\
K.P.~Adhikari,\MISS
S.~Adhikari,\FIU
D.~Adikaram,\JLAB$\!\!^,$\ODU
% ...
\DUKE$\!\!^,$\ODU\
\\
(CLAS Collaboration)}

\affiliation{\JLAB Thomas Jefferson National Accelerator Facility, Newport News, 
Virginia 23606}
\affiliation{\ANL Argonne National Laboratory, Argonne, Illinois 60439}
\affiliation{\ASU Arizona State University, Tempe, Arizona 85287}
\affiliation{\CSUDH California State University, Dominguez Hills, Carson, California 90747}
\affiliation{\CANISIUS Canisius College, Buffalo, New York 14208}
\affiliation{\CMU Carnegie Mellon University, Pittsburgh, Pennsylvania 15213}
\affiliation{\CUA The Catholic University of America, Washington, D.C. 20064}
\affiliation{\SACLAY CEA, Centre de Saclay, Irfu/Service de Physique Nucl\'eaire, 91191 Gif-sur-Yvette, France}
\affiliation{\CNU Christopher Newport University, Newport News, Virginia 23606}
\affiliation{\UCONN University of Connecticut, Storrs, Connecticut 06269}
\affiliation{\DAMMAM University of Dammam, Industrial Jubail 31961, Saudi Arabia}
\affiliation{\DUKE Duke University, Durham, North Carolina 27708-0305}
\affiliation{\EDINBURGH Edinburgh University, Edinburgh EH9 3JZ, United Kingdom}
\affiliation{\FU Fairfield University, Fairfield, Connecticut 06824}
\affiliation{\FERRARAU Universita' di Ferrara, 44121 Ferrara, Italy}
\affiliation{\FIU Florida International University, Miami, Florida 33199}
\affiliation{\FSU Florida State University, Tallahassee, Florida 32306}
\affiliation{\GENOVA Universit$\grave{a}$ di Genova, 16146 Genova, Italy}
\affiliation{\GLASGOW University of Glasgow, Glasgow G12 8QQ, United Kingdom}
\affiliation{\GWUI The George Washington University, Washington, D.C. 20052}
\affiliation{\HAMPTON Hampton University, Hampton, VA 23668}
\affiliation{\ISU Idaho State University, Pocatello, Idaho 83209}
\affiliation{\INFNFE INFN, Sezione di Ferrara, 44100 Ferrara, Italy}
\affiliation{\INFNFR INFN, Laboratori Nazionali di Frascati, 00044 Frascati, Italy}
\affiliation{\INFNGE INFN, Sezione di Genova, 16146 Genova, Italy}
\affiliation{\INFNRO INFN, Sezione di Roma Tor Vergata, 00133 Rome, Italy}
\affiliation{\INFNTUR INFN, Sezione di Torino, 10125 Torino, Italy}
\affiliation{\ORSAY Institut de Physique Nucl\'eaire ORSAY, Orsay, France}
\affiliation{\RUDJER Rudjer Bo\v{s}kovi\'{c} Institute, Bijeni\v{c}ka Cesta 10002 Zagreb, Croatia} 
\affiliation{\ITEP National Research Center Kurchatov Institute, Institute of Theoretical 
and Experimental Physics, Moscow, 117218, Russia}
\affiliation{\JMU James Madison University, Harrisonburg, Virginia 22807}
\affiliation{\KNU Kyungpook National University, Daegu 702-701, Republic of Korea}
\affiliation{\MISS Mississippi State University, Mississippi State, MS 39762-5167}
%\affiliation{\LPSC LPSC, Universite Joseph Fourier, CNRS/IN2P3, INPG, Grenoble, France}
\affiliation{\UNH University of New Hampshire, Durham, New Hampshire 03824}
\affiliation{\NSU Norfolk State University, Norfolk, Virginia 23504}
\affiliation{\OHIOU Ohio University, Athens, Ohio  45701}
\affiliation{\ODU Old Dominion University, Norfolk, Virginia 23529}
%\affiliation{\RPI Rensselaer Polytechnic Institute, Troy, New York 12180}
\affiliation{\URICH University of Richmond, Richmond, Virginia 23173}
\affiliation{\ROMAII Universita' di Roma Tor Vergata, 00133 Rome, Italy}
\affiliation{\MSU Skobeltsyn Nuclear Physics Institute, 119899 Moscow, Russia}
\affiliation{\SCAROLINA University of South Carolina, Columbia, South Carolina 29208}
\affiliation{\TEMPLE Temple University, Philadelphia, PA 19122}
\affiliation{\UTFSM Universidad T\'{e}cnica Federico Santa Mar\'{i}a, Casilla 110-V Valpara\'{i}so, Chile}
\affiliation{\VT Virginia Tech, Blacksburg, Virginia 24061-0435}
\affiliation{\VIRGINIA University of Virginia, Charlottesville, Virginia 22901}
\affiliation{\WM College of William and Mary, Williamsburg, Virginia 23187}
\affiliation{\YEREVAN Yerevan Physics Institute, 375036 Yerevan, Armenia}
 
\date{\today}

\begin{abstract}
The quasi-free $\gamma d\to\pi^{-}p(p)$ differential cross section has been 
measured with CLAS at photon beam energies $E_\gamma$ from 0.445~GeV to 2.510~GeV 
(corresponding to $W$ from 1.311~GeV to 2.366~GeV) for pion center-of-mass angles 
$\cos\theta_\pi^{c.m.}$ from $-0.72$ to 0.92. A correction for final state interactions 
has been applied to this data to extract the $\gamma n\to\pi^-p$ differential cross 
sections. These cross sections are quoted in 8428 $(E_\gamma,\cos\theta_\pi^{c.m.})$ 
bins, a factor of nearly three increase in the world statistics for this channel in 
this kinematic range. These new data help to constrain coupled-channel analysis fits 
used to disentangle the spectrum of $N^*$ resonances and extract their properties. 
Selected photon decay amplitudes $N^* \to \gamma n$ at the resonance poles are 
determined for the first time and are reported here.
\end{abstract}

%\pacs{13.60.Le, 14.20.Gk, 13.30.Eg, 11.80.Et}
%PACS codes for: meson production, baryon resonances, hadronic decays, partial-wave analysis

%\keywords{CLAS, kaon electroproduction, structure functions, hyperons, 

\maketitle

%------------------------------------------
\section{Introduction}
\label{sec:intro}

The determination of the resonance properties for all accessible baryon states is a 
central objective in nuclear physics. The extracted resonance parameters provide 
a crucial body of information for understanding the nucleon excitation spectrum and 
for testing models of the nucleon inspired by Quantum Chromodynamics (QCD) and, more 
recently, lattice QCD calculations. The spectrum of $N^*$ and $\Delta^*$ baryon 
resonances has been extensively studied through meson-nucleon scattering and meson 
photoproduction experiments. Properties of the known resonances continue to become 
better determined as experiments involving polarized beams, targets, and recoil 
measurements are expanded and refined~\cite{joint}. Extracted quantities include 
resonance masses, widths, branching fractions, pole positions, and associated 
residues, as well as photo-decay amplitudes~\cite{PDG16}. New states have also been 
found, mainly through multi-channel analyses that are sensitive to states having a 
relatively weak coupling to the $\pi N$ decay channel~\cite{Julich,BnGa14,EBAC}.

Knowledge of the $N^*$ and $\Delta^*$ resonance photo-decay amplitudes has 
largely been restricted to the charged states. 
Apart from lower-energy inverse reaction $\pi^-p\to \gamma n$ measurements, 
the extraction of the two-body $\gamma n\to\pi^-p$ and $\gamma n\to\pi^0n$ 
observables requires the use of a model-dependent nuclear correction, 
which mainly comes from final state interaction (FSI) effects within the
target deuteron. Most $\gamma n$ data are unpolarized and cover fairly narrow 
energy ranges. Of these, only about 400 $\pi^{0}n$ measurement data points exist, 
spanning the full nucleon resonance region~\cite{SAID}.

The importance of improving the $\gamma n$ database relative to the $\gamma p$
database is directly related to the fact that the electromagnetic interaction 
does not conserve isospin symmetry. The amplitude for the reactions 
$\gamma N \to \pi X$ factors into distinct $I=1/2$ and $I=3/2$ isospin components, 
$A_{\gamma,\pi^\pm} = \sqrt{2} ( A_{p/n}^{I=1/2} \mp A^{I=3/2} )$. This expression
indicates that the excitation of the $I=3/2$ $\Delta^*$ states can be entirely 
determined from proton target data. However, measurements from datasets with both 
neutron and proton targets are required to determine the isospin $I=1/2$ amplitudes 
and to separate the $\gamma p N^*$ and $\gamma n N^*$ photocouplings.

This work focuses on negative pion photoproduction off the neutron using a
deuteron target. A large body of new precision $\gamma n\to\pi^-p$ differential 
cross sections for $E_\gamma$ = 0.445~GeV to 2.510~GeV in laboratory photon energy, 
corresponding to an invariant energy range from $W$ = 1.311~GeV to 2.366~GeV, are 
reported. Pion center-of-mass (c.m.) production angles, ranging from 
$\theta_\pi^{c.m.}$ = 26$^{\circ}$ to 135$^{\circ}$, have been measured during the 
CLAS Collaboration g13 run period~\cite{g13_Report}. These new cross section data have 
nearly tripled the world $\gamma n\to\pi^-p$ database below $E_\gamma$ = 2.700~GeV
\cite{SAID}.

The $\gamma n \to \pi^- p$ differential cross section was previously measured by
the CLAS g10~\cite{gb12} experiment. Those measurements contained 855 data points 
in 50- and 100-MeV-wide bins of beam energy $E_\gamma$ from 1.050~GeV to 3.500~GeV, 
corresponding to a $W$ range from 1.690~GeV to 2.731~GeV. However, the 8428 data 
points from g13 are a precision measurement of this cross section, with a factor 
of $\sim$10 increase in data points. These data are reported in 10- and 20-MeV-wide 
bins of beam energy $E_\gamma$, with overall normalization uncertainties of $\sim$3.4\%, 
compared to the $\sim$6\% to $\sim$10\% overall normalization uncertainties achieved 
by g10. Also, unlike the g10 measurements, the g13 data cover the $W$ range of the 
low-mass $N^*$ resonances, and can be used to investigate their helicity amplitudes 
and resonance parameters.

The present dataset, together with completed polarized measurements for both 
$\pi^-p$ and $\pi^0n$ from Jefferson Lab~\cite{FR-HD} and MAMI~\cite{A2}, are 
expected to lead to the determination of well-constrained $\gamma n$ decay 
amplitudes in the near future. However, these new CLAS $\gamma n \to \pi^- p$ 
data allow for the first determination of selected photon decay amplitudes 
$N^* \to \gamma n$ at their pole on the complex plane.

The organization for this paper is as follows. In Section~\ref{sec:Experiment},
details of the g13 experiment and the CLAS detector are provided. 
Section~\ref{sec:eventsel} outlines the event selection and 
Section~\ref{sec:eff} provides the tracking and triggering efficiency 
corrections. Section~\ref{sec:yields} describes the extraction of the 
event yields and the acceptance corrections, and Section~\ref{sec:lum} 
describes how the beam-target luminosity was determined. 
Section~\ref{sec:Quasi-Free-Cross} presents and discusses the measured 
differential cross sections for the reaction $\gamma n\to\pi^{-}p$, while 
Section~\ref{sec:FSI} reviews the approach for determining the final 
state interaction corrections. Sections~\ref{sec:Legendre} and~\ref{sec:PWA} 
describe the Legendre fits and multipole fit results, respectively. Finally, 
Section~\ref{sec:conc} provides a summary of this work and the conclusions.

%------------------------------------------
\section{Experiment}
\label{sec:Experiment}

The CLAS g13 experiment~\cite{g13_Report} ran from October 2006 to
June 2007 in Hall~B at the Continuous Electron Beam Accelerator 
Facility (CEBAF) at Jefferson Lab in Newport News, Virginia. Circularly 
and linearly polarized tagged bremsstrahlung photon beams were incident on 
a liquid-deuterium ($LD_2$) target located near the center of the CLAS 
detector~\cite{NIM_CLAS}. The circularly polarized photon beam portion 
of this experiment, called g13a, was used for this analysis.

For g13a, the CEBAF electron beam was supplied at two different energies,
1.990~GeV and 2.655~GeV. These electrons were delivered at currents between 
33~nA and 45~nA in beam bunches separated by about 2~ns. The electron beam 
was incident on a $10^{-4}$ radiation-length-thick gold foil radiator to 
produce the bremsstrahlung photon beam.

The dipole magnet of the Hall~B photon tagger deflected the electron beam 
and post-bremsstrahlung electrons in order to tag photons produced with 
energies between $\sim$20\% and $\sim$95\% of the incident electron beam 
energy~\cite{NIM_Tagger}. The tagging system provided a photon beam energy 
resolution of $\sim$0.1\% of the electron beam energy with a 150~ps timing 
resolution. A 6.4~mm diameter nickel collimator downstream of the radiator 
provided $\sim$90\% beam transmission to the 40-cm-long $LD_2$ target, 
which was centered 20~cm upstream from the center of the CLAS detector.
This resulted in a tagged photon flux on the order of $10^{7}$~Hz on the
target.

\begin{figure}
\begin{centering}
\includegraphics[width=1\columnwidth]{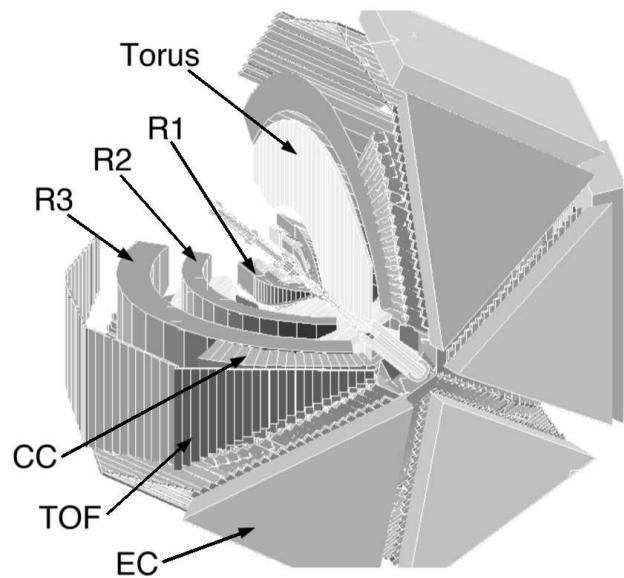} 
\par\end{centering}

\protect\caption{\label{fig:The-CLAS-detector}Cut-away view of the CLAS 
	detector~\protect\cite{NIM_CLAS} illustrating the torus magnet, 
        three regions of drift chambers (R1 - R3), Cherenkov counters (CC), 
        time-of-flight scintillators (TOF), and electromagnetic calorimeters 
        (EC). The CLAS detector is roughly 10~m in diameter.}
\end{figure}

The CLAS detector, shown in Fig.~\ref{fig:The-CLAS-detector}, was designed 
around six superconducting coils arranged in a hexagonal configuration that 
produced an approximately toroidal magnetic field surrounding the beamline. 
The magnetic field bent charged particles through the three regions of 
multi-layer drift chambers for momentum measurements. The drift chambers were 
positioned between the superconducting coils within six sectors in azimuthal 
$\phi$, each spanning roughly $60^\circ$. Charged particles produced at a momentum
of 1~GeV/c were measured with a momentum resolution of $\sigma(p)/p$ $\leq$ 0.5\%, 
and with average angular resolutions in the fiducial volume of $\sigma(\theta)$, 
$\sigma(\phi)$ $\sim$2~mrad~\cite{NIM_DC}. For the g13 experiment, the torus 
magnet operated at $\sim$40\% of its maximum current with reversed field polarity 
(such that negatively charged particles were bent away from the beamline), 
producing an integrated magnetic field of 0.972~T~m along the track path length 
at forward angles and 0.233~T~m at $90^\circ$.
%------------------------------------------
%------------------------------------------

The start counter (ST) surrounding the target had a timing resolution of 260~ps, 
and was used to determine which of the 2~ns electron beam bunches was associated 
with the recorded physics event~\cite{NIM_StartCounter}. The time-of-flight (TOF) 
scintillator paddles had a timing resolution between 150~ps - 250~ps, depending on 
the length of the paddle, and were used for particle identification~\cite{NIM_TOF}. 
At forward angles, Cherenkov counters (not used for this experiment) could be used 
to identify electrons~\cite{NIM_CC}, and the electromagnetic calorimeters could be 
used to detect electrons and neutral particles~\cite{NIM_EC}.

A coincidence between the start counter and TOF scintillators in at least two of the 
six CLAS sectors was required for triggering of the data acquisition. With slightly 
more than two months of running, 20~billion physics events were recorded in the g13a 
dataset.

%------------------------------------------
\section{Event Selection\label{sec:eventsel}}

The $\gamma d\to\pi^{-}p(p)$ differential cross section was measured separately 
for the 1.990~GeV and 2.655~GeV beam data, and these cross section results were 
combined as discussed in Section~\ref{sec:Quasi-Free-Cross}. The yields of the 
$\gamma d\to\pi^{-}p(p)$ reaction were determined by reconstructing the $\pi^-$ 
and scattered (higher momentum) proton, with the lower momentum proton missing. 
The proton in the deuteron typically has a momentum from Fermi-motion of less than 
200~MeV/c~\cite{ParisPotential} (and peaks at $\sim$50~MeV/c), and was often 
stopped before it could escape the $LD_2$ target.

The reconstructed beam energy and track momenta were slightly distorted by
effects not taken into account during the event reconstruction. These effects 
included uncertainties in the incident electron beam energy, unaccounted for 
energy losses as the tracks traversed the detector, and drift chamber 
misalignments and inaccuracies in the magnetic field map that affected the 
reconstructed track momenta. Each of these effects was studied and resulted in 
beam energy and track momentum corrections on the order of a few percent
\cite{CLAS_g13_Studies}.

%------------------------------------------
\subsection{Particle Identification}

\begin{figure}
\begin{centering}
\includegraphics[width=1\columnwidth]{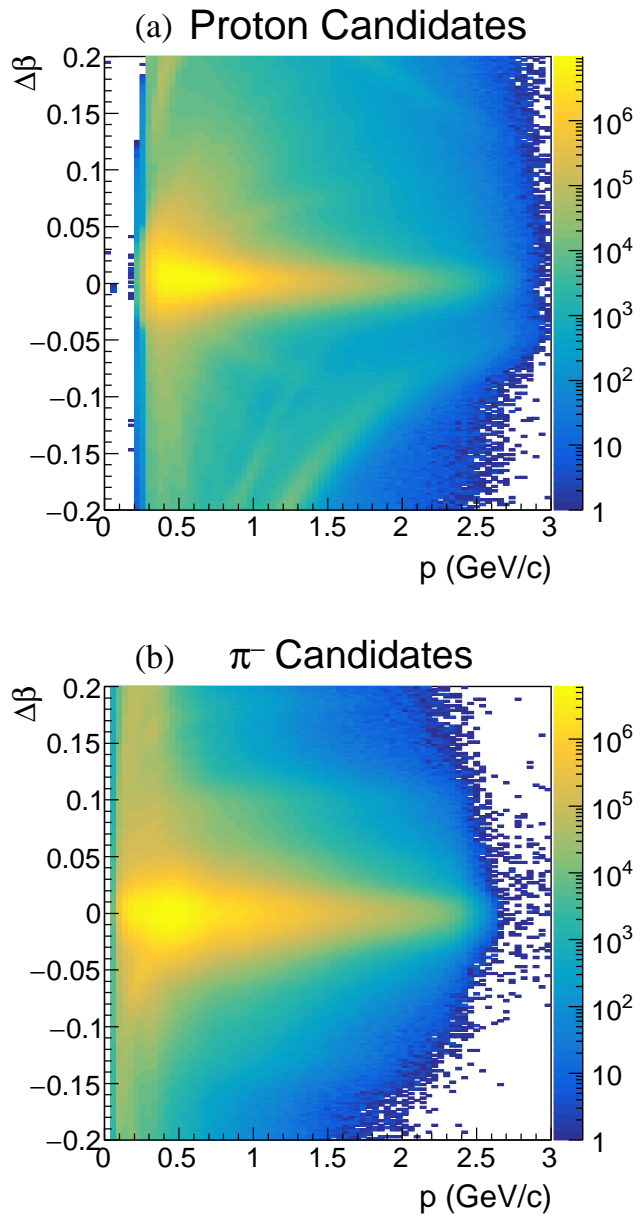} 
\par\end{centering}

\protect\caption{\label{fig:PID}(Color online) $\Delta\beta$ vs. momentum 
        for proton (a) and $\pi^-$ (b) candidates. $\Delta\beta$ is centered 
        at zero for the protons and $\pi^-$'s, and the neighboring bands are 
	from other particle types, such as $\pi^+$'s, or from choosing the 
        wrong beam bunch.}
\end{figure}

Initially, all reconstructed positively and negatively charged tracks were 
treated as candidates for the proton and $\pi^-$, respectively. Then, for 
each combination of proton and $\pi^-$ candidates, their start counter hits 
were used to select the beam bunch corresponding to the event. The arrival 
time of the beam bunches was known to a resolution of 50~ps, and was used 
as a reference time for the particle identification. Fig.~\ref{fig:PID} 
shows the $\Delta\beta$ vs. momentum distributions of the proton and $\pi^-$ 
candidates, where $\Delta\beta$ is the difference between $\beta=p/E$ using
the candidate mass, and $\beta$ determined from the track path length (from 
the event vertex to the TOF system) and the track hit time from the TOF paddle. 
$\Delta\beta$ is centered at zero for the protons and $\pi^-$'s, and the 
neighboring bands are from other particle types, such as $\pi^+$'s, or from 
choosing the wrong beam bunch. 
%------------------------------------------
%------------------------------------------

For proton identification, a momentum-dependent $\pm5\sigma$ cut was 
applied on $\Delta\beta$. No particle identification cut was used to 
identify $\pi^-$'s since the background from electron, muon, and kaon 
events was negligible, as seen in Fig.~\ref{fig:PID}(b). Poorly 
performing or miscalibrated TOF counters were excluded from the analysis.

%------------------------------------------
\subsection{Vertex Cuts and Missing Momentum}

\begin{figure}
\begin{centering}
\includegraphics[width=1\columnwidth]{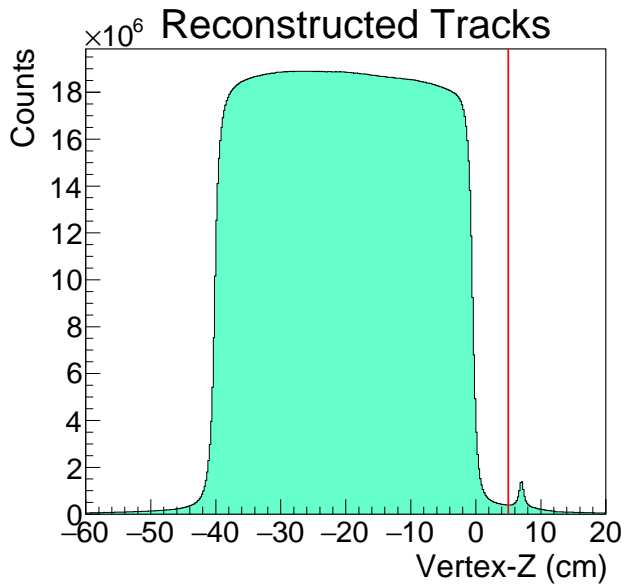} 
\par\end{centering}

\protect\caption{\label{fig:VertexZ}(Color online) The vertex-$z$ of 
	the reconstructed tracks in $\gamma d\to \pi^{-}p(p)$. Tracks 
	with $z>5$~cm were cut, removing backgrounds from 
	the aluminum endcap of the target assembly at $z=7$~cm.}
\end{figure}

The vertex-$z$ distribution of the reconstructed tracks, defined as their 
distance-of-closest-approach to the nominal beamline (defined as the $z$-axis), 
is shown in Fig.~\ref{fig:VertexZ}. A cut was applied requiring that the 
reconstructed vertex-$z$ of both the proton and the $\pi^-$ be less than 5~cm. 
The target extended from -40~cm to 0~cm in vertex-$z$, so this cut was used to 
remove backgrounds from beam photons striking the aluminum endcap of the target 
assembly.
%------------------------------------------
%------------------------------------------

\begin{figure}
\begin{centering}
\includegraphics[width=1\columnwidth]{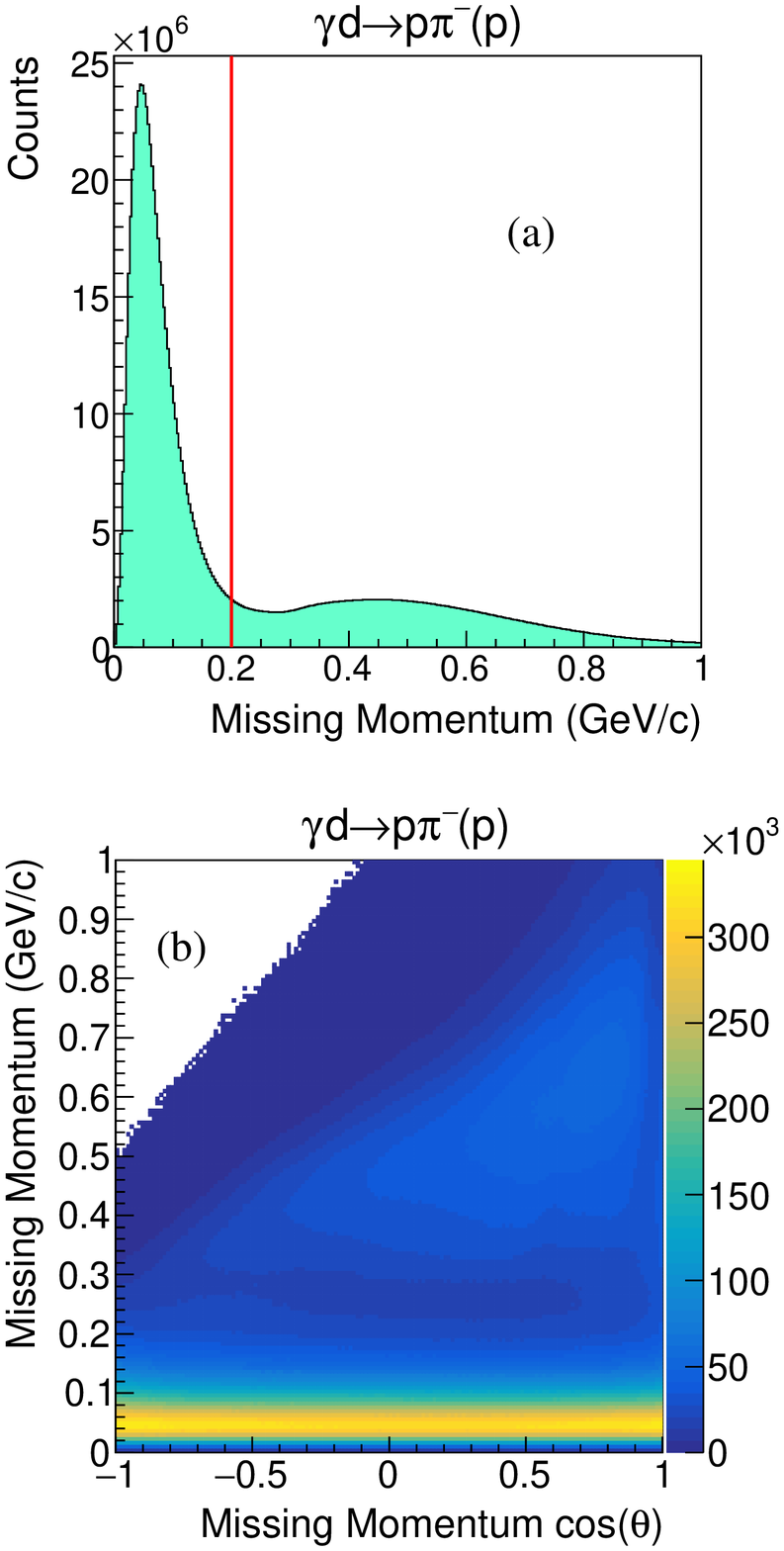} 
\par\end{centering}

\protect\caption{\label{fig:MissingP}(Color online) (a) The 
        $\gamma d\to\pi^-p(p)$ missing momentum and (b) the same missing 
        momentum vs. $\cos\theta$ of the missing momentum in the laboratory 
        frame. The low-momentum, quasi-free protons followed the Fermi 
	motion distribution and were relatively uniformly distributed 
	in $\cos\theta$. The high-momentum tail of rescattered protons 
	was removed by the 200~MeV/c cut on the missing momentum 
	(indicated by the red vertical line in (a)).}
\end{figure}

To illustrate the missing momentum distribution of $\gamma d\to\pi^-p(p)$ events, 
a $\pm3\sigma$ cut was applied around the missing mass peak of the proton. 
Fig.~\ref{fig:MissingP}(a) shows the missing momentum distribution after this 
cut. The missing momentum is primarily peaked at low momenta due to Fermi motion, 
and the high-momentum tail is primarily from rescattering events. 
Fig.~\ref{fig:MissingP}(b) shows that the slow-proton momentum is uniformly 
distributed in $\cos\theta$, where $\theta$ is the angle between the missing 
momentum and the beam in the laboratory frame. A cut was applied at 200~MeV/c to 
reject the majority of the rescattering events. Since there are still rescattering 
effects present after this cut, the $\gamma d\to\pi^-p(p)$ cross section is quoted 
as ``quasi-free."
%------------------------------------------
%------------------------------------------

%------------------------------------------
\section{CLAS Efficiency Studies}
\label{sec:eff}

%------------------------------------------
\subsection{Tracking Efficiency}

To determine the charged particle tracking efficiency, the CLAS drift chamber wire hit 
efficiencies were studied by determining how often a given sense wire recorded 
a hit when a reconstructed track passed nearby. To make sure that the study was 
unbiased, the efficiencies were only evaluated when there were significantly 
more hits on the track than the minimum needed for reconstruction. 

%------------------------------------------
%\begin{figure}
%\begin{centering}
%\includegraphics[width=1\columnwidth]{DCWireHitEff} 
%\par\end{centering}
%
%\protect\caption{\label{fig:DCHitEff}(Color online) The drift chamber wire 
%        hit efficiencies for Sector~1 of CLAS in the g13 experiment. Wire 
%        number increases with polar angle $\theta$ and wire layers 5 and 
%	6 do not exist. Horizontal groups of weak or dead wires correspond 
%        to bad high-voltage connections. The small, inefficient vertical 
%        groups of 10 to 20 wires correspond to bad amplifier/discriminator 
%        boards, cable disconnects, or cable swaps.}
%\end{figure}
%------------------------------------------

These studies allowed issues associated with missing wires due to bad 
high-voltage connections and amplifier low-voltage shorts to be taken 
into account. Furthermore, these studies were able to determine tracking 
inefficiencies due to readout electronics problems, cable disconnects, and 
cable swaps. Groups of wires that were correlated with a common problem were 
grouped together in the simulation so that they were either kept or rejected 
as a whole. The efficiency calculated and applied for the regions with cable 
swaps or disconnects does not properly model the experimental data, so these 
regions were eliminated from the analysis. In this manner a very good match of 
the tracking efficiency in the simulation code to the CLAS hardware was 
possible. A comparison of the tracking efficiencies for each CLAS drift 
chamber sector are available in Ref.~\cite{CLAS_g13_Studies}.

\begin{figure}
\begin{centering}
\includegraphics[width=1\columnwidth]{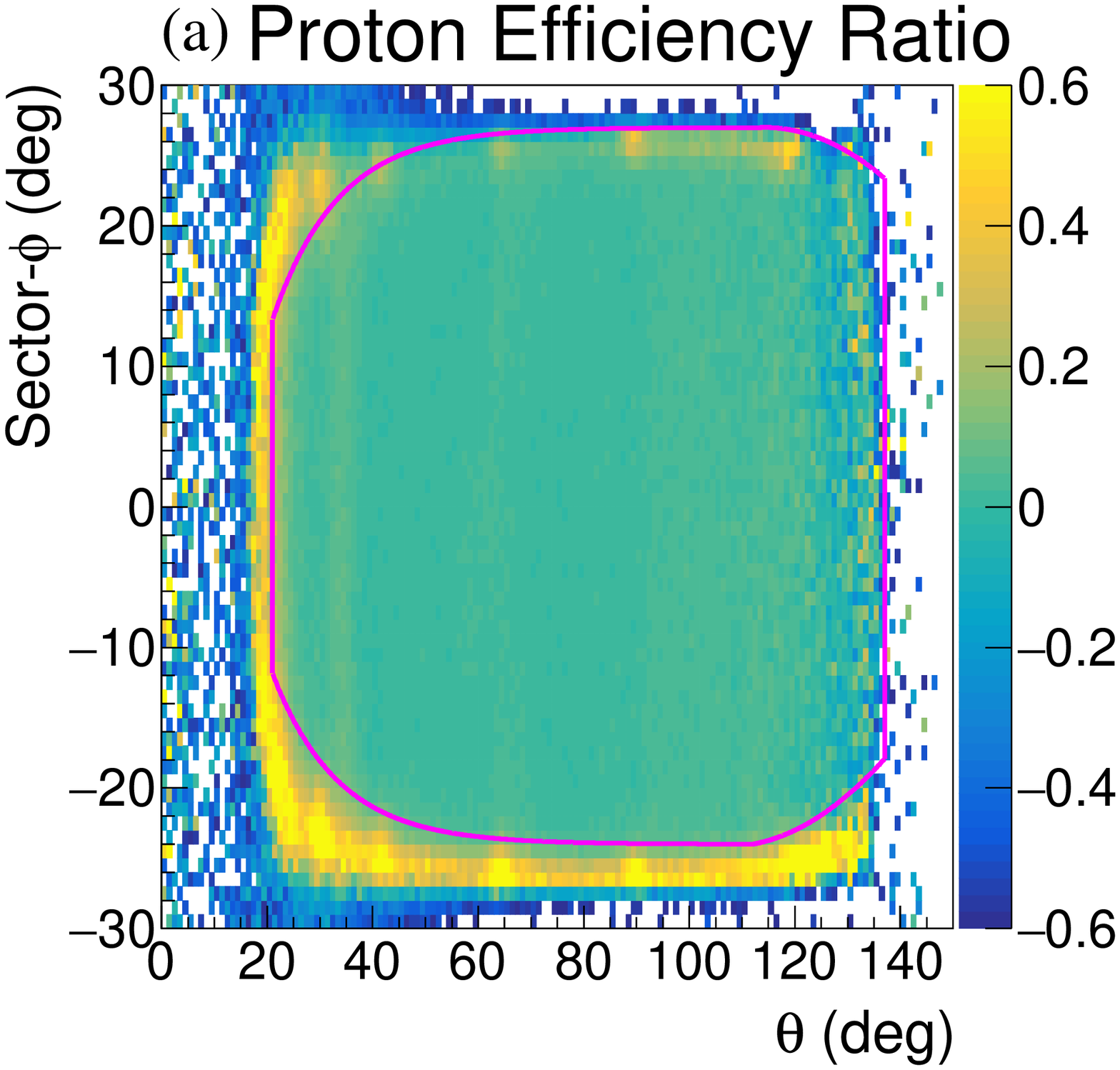} 
\par\end{centering}
\begin{centering}
\includegraphics[width=1\columnwidth]{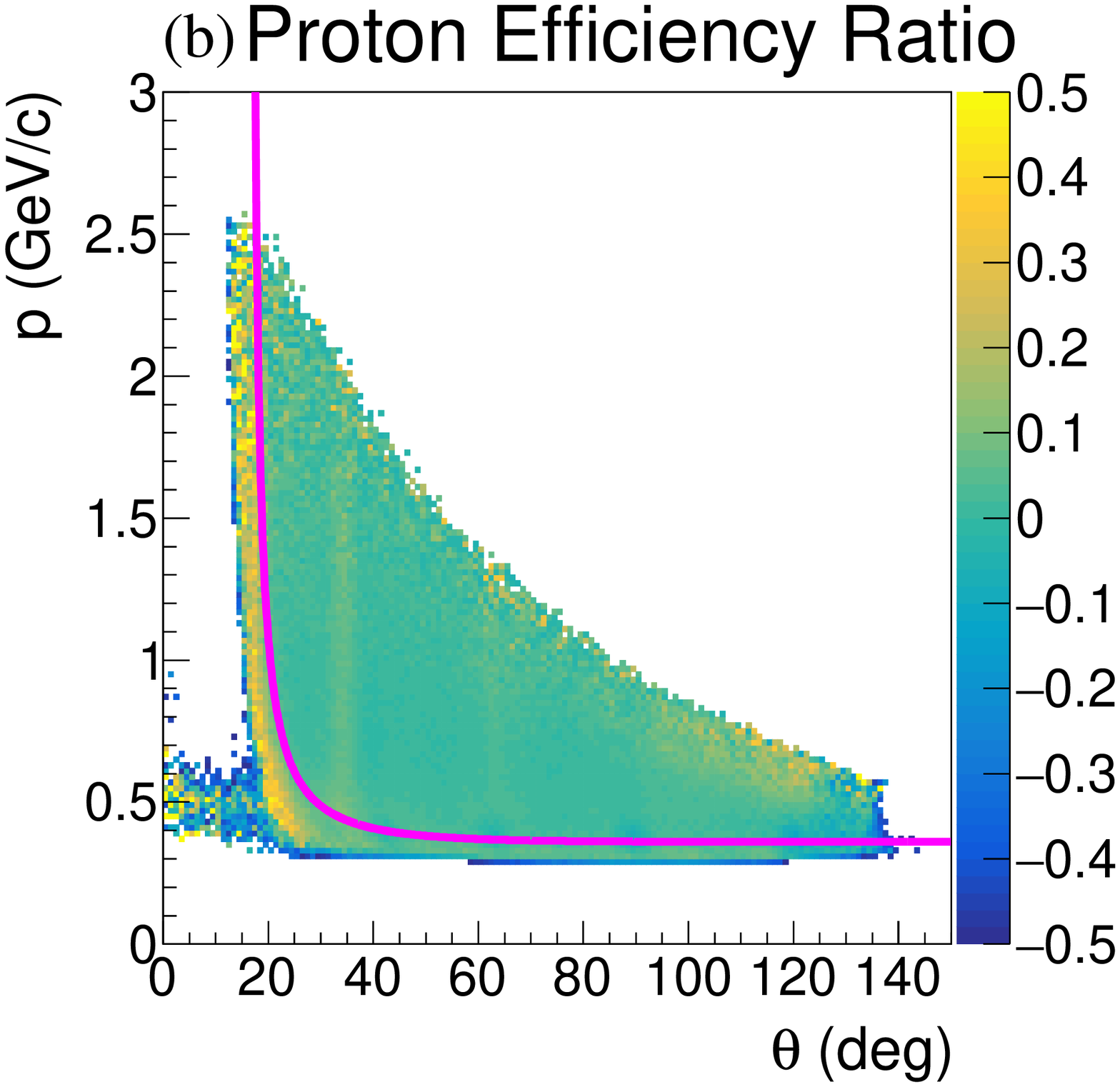} 
\par\end{centering}

\protect\caption{\label{fig:Fiducial}(Color online) The proton reconstruction 
        efficiency ratio defined by Eq.~(\protect\ref{eq:Relative_Eff}) as a 
        function of lab polar angle. In (a), sector-$\phi$ is the azimuthal 
        angle relative to the center of the CLAS sector. Both plots are summed 
        over all CLAS sectors. The magenta lines indicate the cuts used to 
	reject data that were not accurately modeled by the simulation.}
\end{figure}

The track reconstruction efficiencies for protons and $\pi^-$'s were studied 
by analyzing the $\gamma d\to\pi^-p(p)$ and $\gamma d\to pp(\pi^-)$ topologies, 
respectively, and determining how often the missing particle was reconstructed 
when it was in the fiducial region of the detector. These studies were performed 
with both experimental and phase-space Monte Carlo (MC) simulated data as a 
function of track momentum and direction. For the g13 experiment, these 
reconstruction efficiencies were 95\% or higher in the nominal fiducial regions 
of the detector. Fig.~\ref{fig:Fiducial} shows the ratio of these reconstruction 
efficiencies $\varepsilon$ for the proton, which was computed as:
\begin{equation}
	\mathrm{\varepsilon_{Ratio}}=\frac{\mathrm{\varepsilon_{Simulation}}
	-\mathrm{\varepsilon_{Experiment}}}{\mathrm{\varepsilon_{Experiment}}}.
	\label{eq:Relative_Eff}
\end{equation}
%------------------------------------------
%------------------------------------------

Thus, regions with an efficiency ratio significantly greater (less) than
zero are regions where the reconstruction efficiency was much lower (higher)
in the experiment than in the MC simulation. The discrepancies seen in the 
figure at the edges of the acceptance are due to a mismatch between the 
simulated and experimental geometry. These regions were cut from the analysis 
so that only regions that were accurately modeled in the simulation were 
included in the cross section measurement. The efficiency ratio distributions 
for the $\pi^-$'s, which were bent differently in the CLAS magnetic field, are 
similar but required separate cuts. The absolute minimum accepted proton and 
$\pi^-$ momenta were 360~MeV/c and 100~MeV/c, respectively.

%------------------------------------------
\subsection{Triggering Efficiency}

As discussed in Section~\ref{sec:Experiment}, the g13a trigger was designed 
to record events with a ST and TOF coincidence in at least two sectors of 
CLAS. To determine the triggering efficiency, the $\gamma d\to \pi^- pp$ 
topology was studied, with the requirement that the three final state 
particles be in different sectors. For every pair of particles that 
registered as contributing to the trigger, the triggering rate of the third 
particle was studied.

\begin{figure}
\begin{centering}
\includegraphics[width=1\columnwidth]{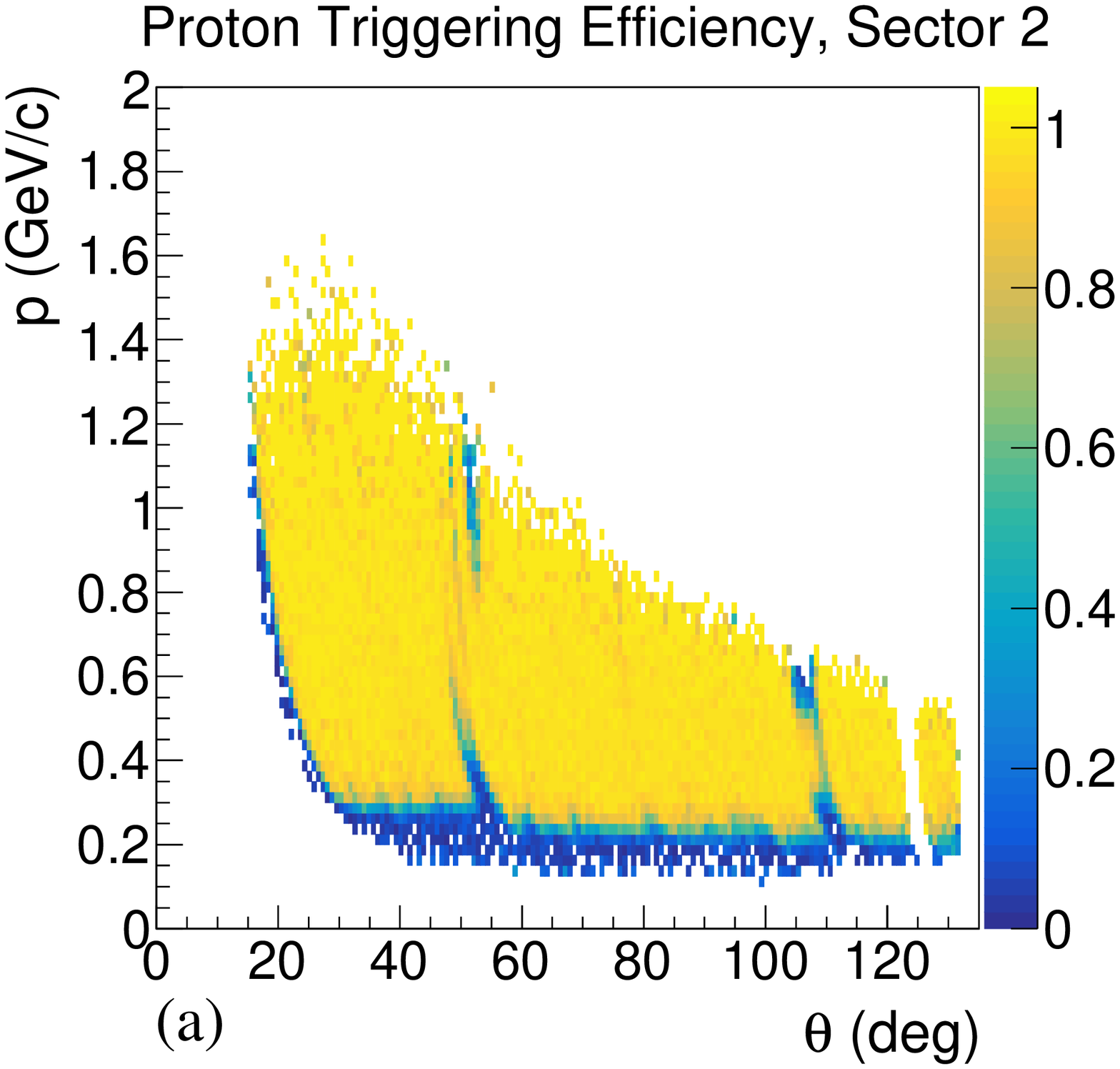}
\par\end{centering}
\begin{centering}
\includegraphics[width=1\columnwidth]{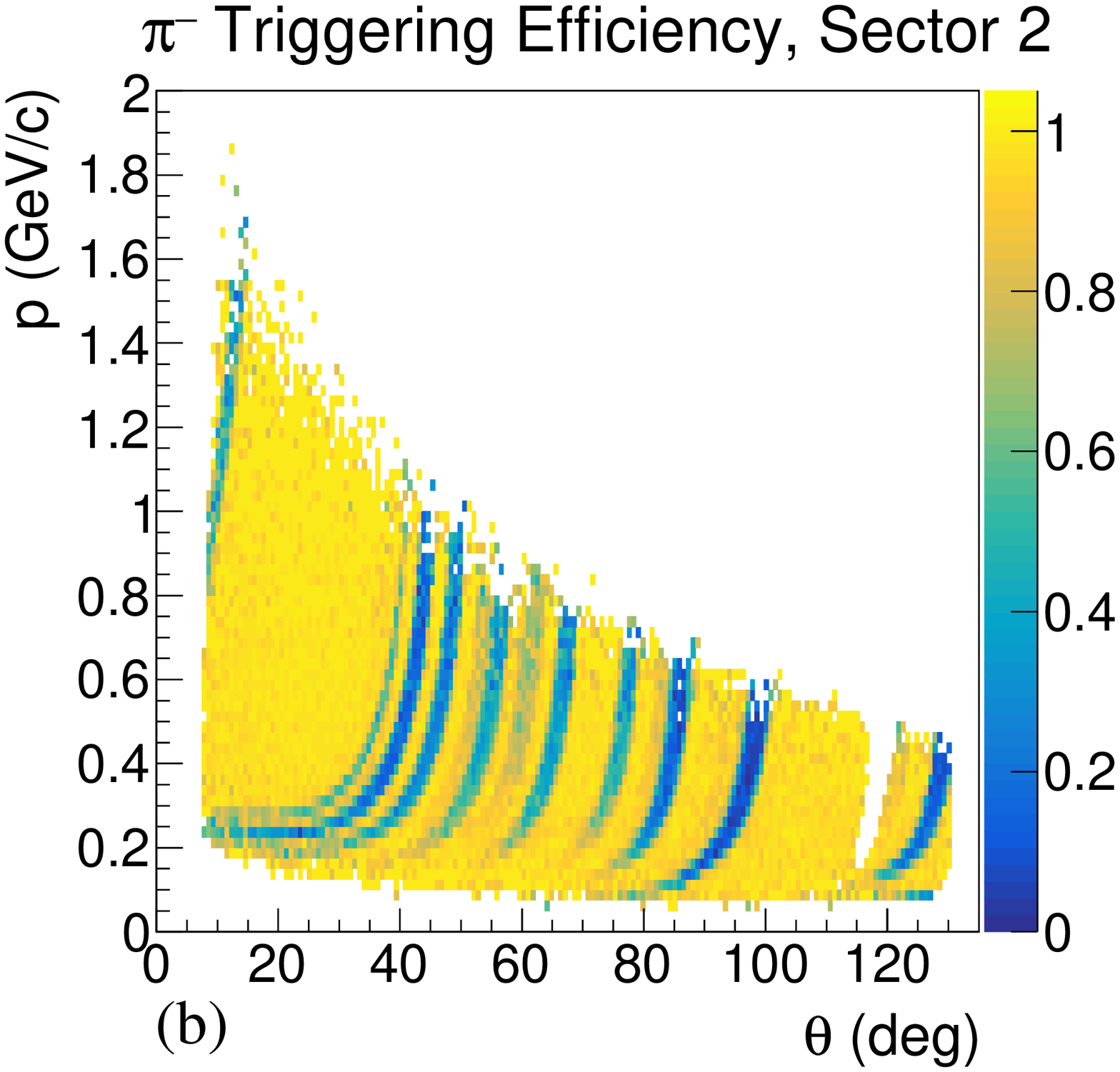}
\par\end{centering}

\protect\caption{\label{fig:Trigger_Eff}(Color online) The proton (a) 
	and $\pi^-$ (b) triggering efficiencies in CLAS Sector~2 in
        terms of momentum $p$ vs. lab polar angle $\theta$. The proton 
	triggering was efficient in general, but the $\pi^-$ efficiency 
        was affected over portions of the acceptance due to low-gain 
        TOF PMTs.}
\end{figure}

Figure~\ref{fig:Trigger_Eff} shows the proton and $\pi^-$ triggering
efficiencies in a representative sector, as a function of the track angle 
and momentum. Because these efficiencies were studied as a function of all 
kinematics, they include both TOF and ST efficiency effects. The proton 
triggering was efficient in general, but was low in a few of the TOF 
paddles, due to one or both of the photomultiplier tubes (PMTs) at the end 
of the scintillators having low gain. 
%------------------------------------------
%------------------------------------------

However, the $\pi^-$ efficiencies were significantly worse than those of 
the proton. This was because $\pi^-$'s deposited much less energy than 
protons of the same momentum in the scintillators, due to their higher 
velocity. A number of inefficient channels were present due to low gain 
TOF PMTs (even though they were set at their maximum voltage). These PMTs 
were still efficient for hit readout, as the 100~mV triggering threshold 
was much higher than the 20~mV detection threshold. The efficiencies for 
the other sectors are available in Ref.~\cite{CLAS_g13_Studies}. These 
triggering efficiencies were applied to the MC simulation to model these 
event losses.

%------------------------------------------
\section{Yields and Acceptance\label{sec:yields}}

\begin{figure}
\begin{centering}
\includegraphics[width=1\linewidth]{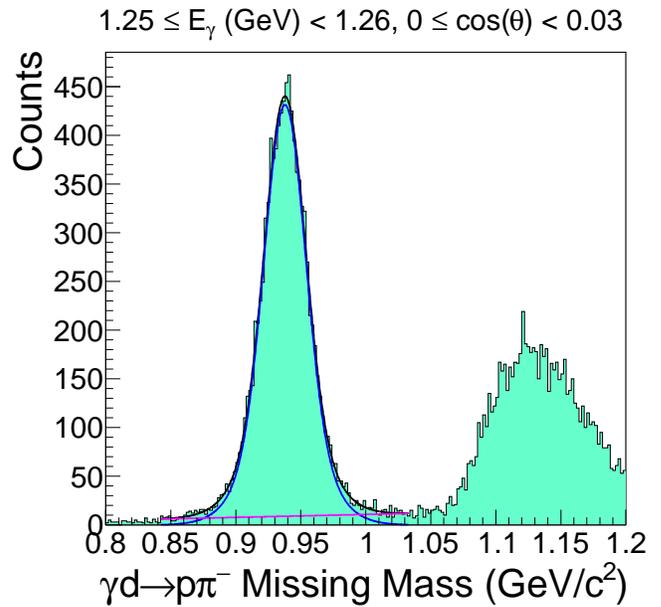} 
\par\end{centering}

\protect\caption{\label{fig:PPM_YieldFit}(Color online) Sample fit of the 
        missing-proton peak in the missing mass off of $\gamma d\to\pi^-p$. 
        The signals were fit to a double-Gaussian function over a linear 
        background. The blue lineshape represents the double-Gaussian fit 
	function, the magenta lineshape represents the linear fit function, 
        and the black lineshape represents the total fit function.}
\end{figure}

The $\gamma d\to\pi^-p(p)$ data were separated into 10- and 20-MeV-wide 
$E_\gamma$ bins and 0.02- to 0.04-wide bins in $\cos\theta_\pi^{c.m.}$, 
where $\theta_\pi^{c.m.}$ is the angle between the $\pi^-$ and the beam in 
the $\pi^- p$ c.m. frame. These data spanned the range from 0.440~GeV to 
2.520~GeV in beam energy and $-0.72$ to $0.92$ in $\cos\theta_\pi^{c.m.}$ 
for a total of 8428 bins. In each bin, the missing-proton peaks were fit 
to double-Gaussian functions over a linear background, an example of which 
is shown in Fig.~\ref{fig:PPM_YieldFit}. A double-Gaussian function is 
defined as the sum of two Gaussians with identical means, but different 
heights and widths. The larger, primary Gaussian was used to model the 
Gaussian-scattering portion of the signal distribution, and the smaller, 
secondary Gaussian was used to fit the tails of the signal distribution.
%------------------------------------------
%------------------------------------------

The backgrounds are primarily due to misidentified protons and $\pi^-$'s, 
or selection of the wrong beam photon. The $\gamma d\to\pi^-p(p)$ yield 
was defined as the number of events above the background within the 
$\pm4.5\sigma$ fit range about the missing-proton peak. There were over 
$400$ million $\gamma d\to\pi^-p(p)$ events in the g13a experimental data 
sample used for this analysis.

\begin{figure}
\begin{centering}
\includegraphics[width=1\linewidth]{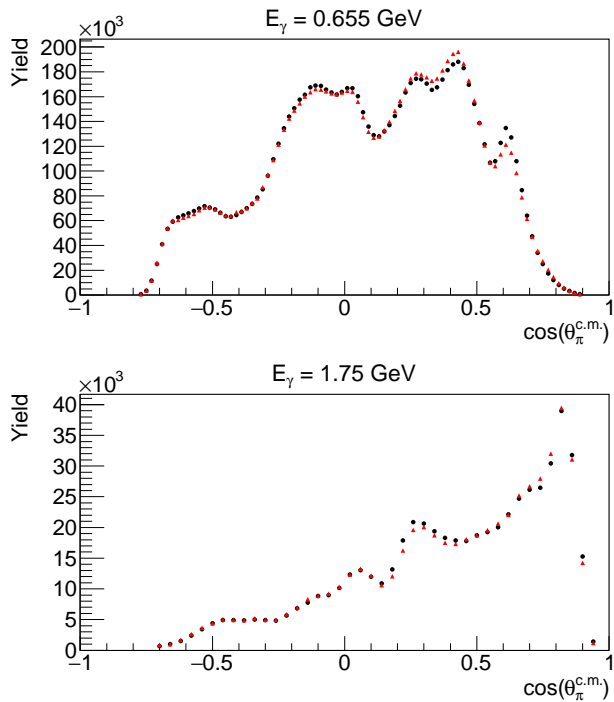} 
\par\end{centering}

\protect\caption{\label{fig:YieldCompare}(Color online) A comparison of the reconstructed
        $\gamma d\to\pi^-p(p)$ yields between the experimental data (black) and the 
        simulation (red) as a function of the pion c.m. angle $\cos\theta_\pi^{c.m.}$ 
        in two selected beam energy bins for the 2.655 GeV data. The simulated yields have 
        been scaled such that their integral matches that of the experimental data.}
\end{figure}

A total of 1.8~billion MC $\gamma d\to \pi^- pp$ events were simulated for
each electron beam energy to calculate the acceptance corrections. These
data were evaluated separately in order to individually compute the cross 
sections for the different run ranges. The George Washington University 
(GWU) SAID GB12 cross section predictions~\cite{gb12}, based on the world 
data of the $\gamma n\to\pi^-p$ reaction, were used to generate the event 
distributions. After a preliminary quasi-free $\gamma d\to\pi^-p(p)$ cross 
section measurement was obtained from the g13 data, this measurement was 
used to generate the final simulated data. A comparison of the reconstructed 
$\gamma d\to\pi^-p(p)$ yield between the experimental data and the MC is shown in Fig.~\ref{fig:YieldCompare}.
This shows that the inefficient regions of the detector are well modeled by the simulation. 
Thus, any variations of the detector acceptance across the widths of the narrow yield-extraction 
bins did not cause an incorrect modeling of the CLAS acceptance.

%------------------------------------------
%------------------------------------------

The same analysis procedure and cuts used to select the $\gamma d\to\pi^-p(p)$ 
final state in the experimental data were used for the simulated data. However, 
a $\sim$5\% yield correction factor $Y_{\mathrm{CF}}$ was applied to the 
experimental yields to correct for event losses from choosing the incorrect 
beam bunch, which was not modeled in the simulation. This correction factor 
was determined by studying the $\gamma d \to \pi^- p(p)$ yield from all other 
beam bunches recorded in the event. The uncertainties on this correction factor 
were $\sim$0.003\% from statistics and $\sim$0.88\% from systematics, 
determined by studying the variation in the correction factor with beam energy.

\begin{figure}
\begin{centering}
\includegraphics[width=1\linewidth]{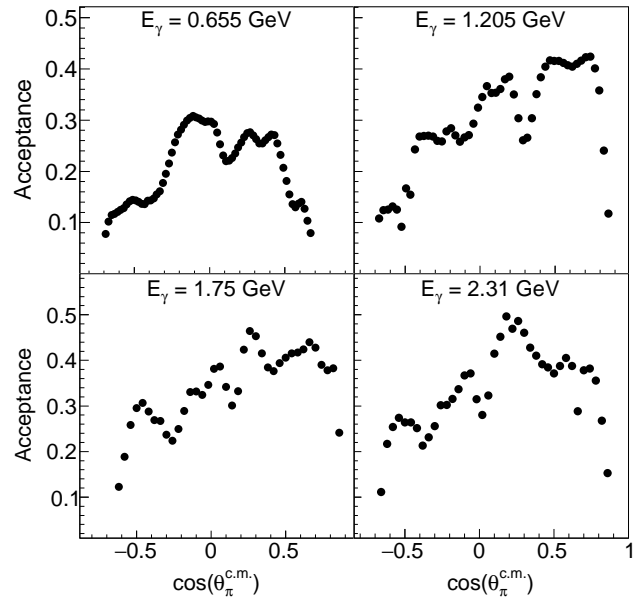} 
\par\end{centering}

\protect\caption{\label{fig:PPM_Acceptance}The CLAS acceptance in selected 
	energy bins as a function of the pion c.m. angle $\cos\theta_\pi^{c.m.}$.
	The dips in the acceptance are primarily due to triggering inefficiencies
	and problematic areas of the drift chambers.}
\end{figure}

Since the CLAS acceptance rapidly falls off near the edges of the detector, 
cross section measurements in these regions had systematic uncertainties that
were difficult to quantify. In addition, small mismatches between the generated MC 
distribution and the experimental data could cause large uncertainties in 
regions of low acceptance. To remove these regions, bins with an acceptance 
less than 20\% of the maximum acceptance within each $E_\gamma$ bin were rejected 
from the analysis. The CLAS acceptance of the $\gamma d\to\pi^-p(p)$ reaction 
for the CLAS g13 experiment after this cut is shown in 
Fig.~\ref{fig:PPM_Acceptance} for selected beam energy bins. Overall, the 
acceptance varied between 5\% and 50\%, and the large dips were primarily due 
to triggering inefficiencies and drift chamber problem areas. 
%------------------------------------------
%------------------------------------------

The systematic uncertainty due to event selection was typically less
than 2\%, although it increased to 10\% near the edges of the detector.
The uncertainty due to the yield extraction was less than 5\%. These
uncertainties were determined by varying the widths of the cuts used,
and the range and starting parameters of the missing mass fits.

The systematic uncertainties due to the acceptance corrections were
typically less than 5\%, but increased to 10\% in problematic regions
with low triggering or drift chamber acceptance. These uncertainties
were determined by studying how the acceptance-corrected yields 
changed when individual CLAS sectors and target vertex-$z$ bins were 
removed from the analysis. A small number of bins had large ($\geq5\%$) 
systematic uncertainties for half of the CLAS sectors, and were 
removed from the results. Overall, the angular-dependent systematic 
uncertainties varied between 1\% and 15\%.

%------------------------------------------
\section{Luminosity Determination}
\label{sec:lum}

The number of tagged photons incident on the target while the data
acquisition (DAQ) system was ready to record events, $N_\gamma$, was 
calculated separately for each tagger counter as~\cite{gflux}:
\begin{equation}
	N_{\gamma}=\epsilon N_e,
\end{equation}
\noindent 
where $\epsilon$ is the tagging ratio of the given tagger counter and $N_e$ is 
the number of detected electron hits in that counter while the DAQ was 
ready. $N_e$ was calculated from the rate of ``out-of-time'' electron hits and 
the livetime of the DAQ. ``Out-of-time'' hits are from electrons that did not 
coincide in time with the trigger, and were used so that the rate calculation 
was not biased by the trigger.

The tagging ratios $\epsilon$ were determined by taking several normalization 
runs throughout the g13 experiment. During these runs, a total absorption 
counter (TAC) was inserted into the beamline to determine the number of photons 
incident on the target. The TAC was positioned about 25~m downstream of CLAS and 
consisted of a single lead-glass block. A PMT attached to the block was used to 
count the number of photons incident on the TAC, which was 100\% efficient
\cite{NIM_CLAS}. A low beam current of 0.1~nA was necessary to prevent radiation 
damage to the TAC during these normalization runs.

For each normalization run the tagging ratios were calculated for each 
tagger counter as~\cite{gflux}:
\begin{equation}
	\epsilon=\frac{N_{\mathrm{TAC}}}{N(1-\alpha)},
\end{equation}\noindent 
where $N$ is the total number of electron hits in a given tagger counter, 
$N_{\mathrm{TAC}}$ is the total number of these hits that represent 
coincident matched photon hits in the TAC for that tagger counter, and 
$\alpha$ is the photon attenuation factor. This factor takes into account 
the fraction of the photons incident on the target that did not reach the 
TAC. This photon attenuation factor was $\sim$4\% and was energy-independent
\cite{PhotonLoss_ToTAC}. These losses were primarily due to electron-positron 
pair production and Compton scattering as the photons interacted with the 
target and the beamline components. The tagging ratios were typically 60\% -
72\% for the 1.990~GeV data and between 73\% and 82\% for the 2.655~GeV
data. Because the beam was collimated through a 6.4~mm opening, the
tagging ratios were lower for the 1.990~GeV data due to the larger
beam dispersion of the lower-energy beam.

In total, approximately 46.8~trillion tagged photons were incident
on the CLAS target in this analysis. The statistical uncertainty on
the flux measurement ranged between 0.0024\% and 0.14\%, and are 
reported as energy-dependent normalization uncertainties. The systematic 
uncertainty of the photon flux was determined by examining the stability 
of the flux-normalized yields of $\gamma d\to\pi^-p(p)$ throughout the 
experimental run. These systematic uncertainties were 0.4\% and 0.7\% for 
the 1.990~GeV and 2.655~GeV data, respectively, and are reported as 
energy-independent normalization uncertainties.

In addition, the systematic uncertainties on the target length and
density determinations were each 0.4\%, and were dominated by thermal
contraction and temperature variation, respectively. These uncertainties
are reported as energy-independent normalization uncertainties.

%------------------------------------------
\section{Differential Cross Sections
\label{sec:Quasi-Free-Cross}}

The data from the 1.990~GeV and 2.655~GeV electron beam energies were merged
together to produce the final set of measured $\gamma d\to\pi^-p(p)$
differential cross sections. This merging was performed by calculating
an uncertainty-weighted average of the two cross section measurements
in bins where both were available. In bins where data was only available
from one beam energy, only that result was used.

The differential cross section of the $\gamma d\to\pi^-p(p)$ reaction
was calculated for each bin of photon beam energy $E_\gamma$ and 
$\cos\theta_\pi^{c.m.}$ as:
\begin{multline}
	\frac{d\sigma}{d\Omega}(E_\gamma,\:\cos\theta_\pi^{c.m.})=\\
\frac{1}{2\pi(\Delta\cos\theta_\pi^{c.m.})}\frac{A_r}{\rho LN_A}
	\frac{Y(E_\gamma,\:\cos\theta_\pi^{c.m.})Y_{\mathrm{CF}}}{\Phi(E_\gamma)
	A(E_\gamma,\:\cos\theta_\pi^{c.m.})},
	\label{eq:Cross-Section}
\end{multline}
\noindent 
where $\Delta\cos\theta_\pi^{c.m.}$ is 
the bin width in $\cos\theta_\pi^{c.m.}$, $A_r$ is the effective atomic weight 
of the neutrons in the deuterium target, $\rho$ is the target density, $L$ 
is the target length, $N_A$ is Avogadro's number, $\Phi$ is the photon flux 
in the given photon energy bin, $Y$ is the experimental yield in the given 
bin, $A$ is the simulated acceptance in the given bin, and $Y_{\mathrm{CF}}$ is 
the yield correction factor discussed in Section~\ref{sec:yields}. The factor 
of $2\pi$ is due to the integration over the azimuthal angle $\phi$ in the 
binning used for the cross section calculation. The statistical uncertainty 
of the cross section was calculated for each bin by combining the statistical 
uncertainties of the experimental yield and simulated acceptance in quadrature, 
and ranged between 0.3\% and 5\%. These uncertainties were dominated by the 
yield uncertainties. All data from this measurement are included in the CLAS 
physics database~\cite{database}. 

To study the stability of the overall normalization of the $\gamma d\to\pi^-p(p)$ 
cross section measurements, it was calculated separately for several different 
run ranges throughout both beam energy settings of
the experiment. Overall, the total spread between the 
measurements was 2.4\%, and is reported as an energy-independent normalization 
uncertainty. This uncertainty takes into account any systematic 
differences between the 1.990 GeV and 2.655 GeV data that were merged together.
The total normalization uncertainties were about 3.4\%, and were 
primarily due to this run range-dependent variation in the cross section 
measurements and the FSI corrections, which are discussed in 
Section~\ref{sec:FSI}. The total uncertainty on the $\gamma d\to\pi^-p(p)$ cross 
sections is typically between 4.2\% and 15\%.

\begin{figure*}
\begin{centering}
\includegraphics[width=1\linewidth]{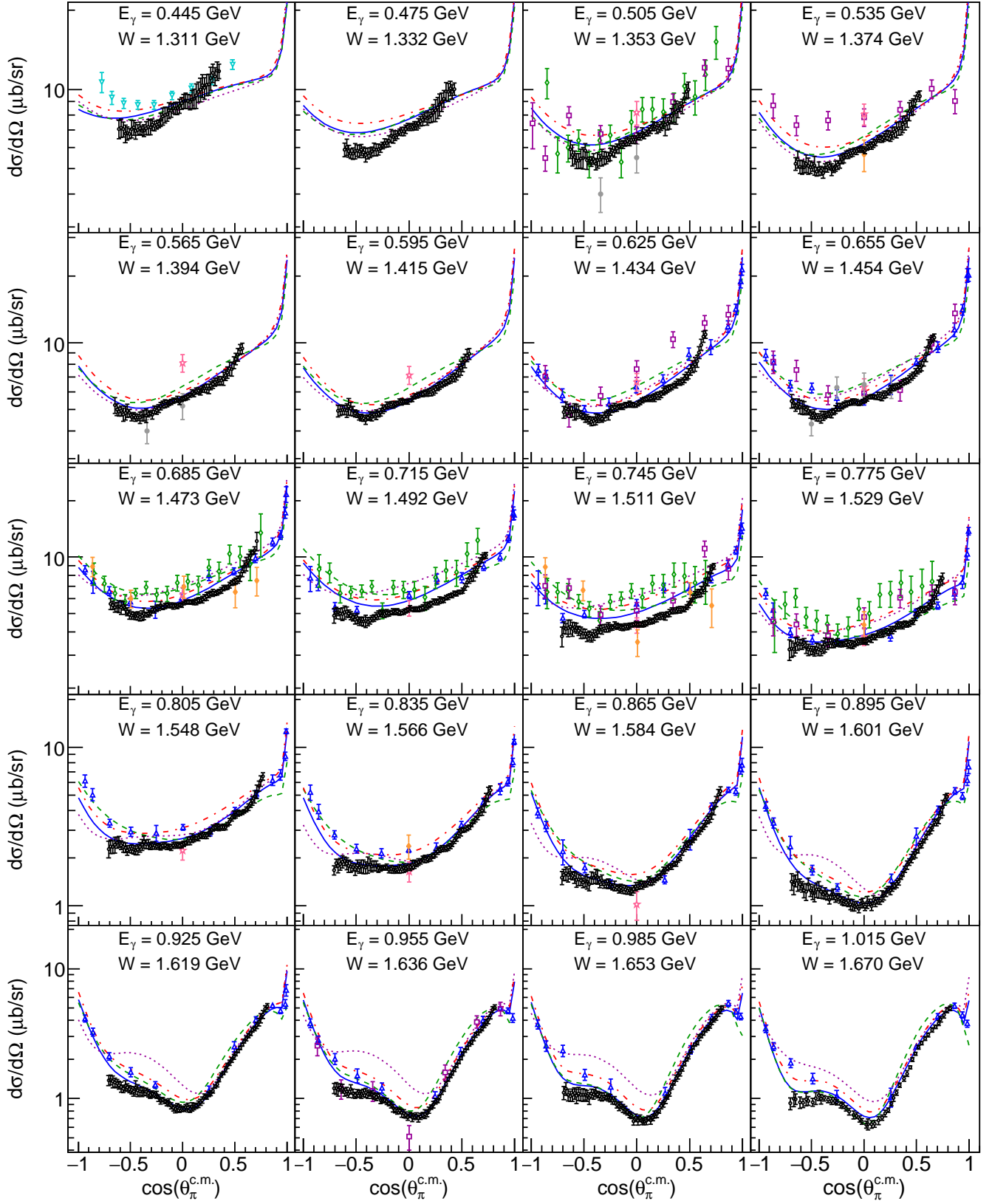} 
\par\end{centering}

\protect\caption{\label{fig:CrossSections1}(Color online) Selected cross 
        section data for $\gamma n\to \pi^-p$ vs. $\cos\theta_\pi^{c.m.}$ 
        below $E_\gamma=1.05$~GeV: CLAS g13 (black open circles), 
        SLAC~\protect\cite{Scheffler} (blue open triangles), 
        DESY~\protect\cite{Benz} (violet open squares), 
	MAMI-B~\protect\cite{Briscoe} (cyan open down-triangles), and 
	Frascati~\protect\cite{Beneventano} (pink open stars); 
	$\pi^-p\to\gamma n$ data: BNL~\protect\cite{Shafi} (green
	open diamonds), LBL~\protect\cite{Weiss} (orange closed diamonds), 
	and LAMPF~\protect\cite{Kim} (gray closed circles); fits: SAID 
        MA27 (blue solid lines), SAID PR15~\protect\cite{pr15} (red 
        dot-dashed lines), BG2014-02~\protect\cite{BnGa14} (green dashed 
        lines), and MAID2007~\protect\cite{MAID07} (violet dotted lines). 
        The $y$-axes are log scale. Only angle-dependent uncertainties 
        are shown for all data. The total normalization uncertainties for 
        the CLAS g13 data are about 3.4\%.}
\end{figure*}

\begin{figure*}
\begin{centering}
\includegraphics[width=1\linewidth]{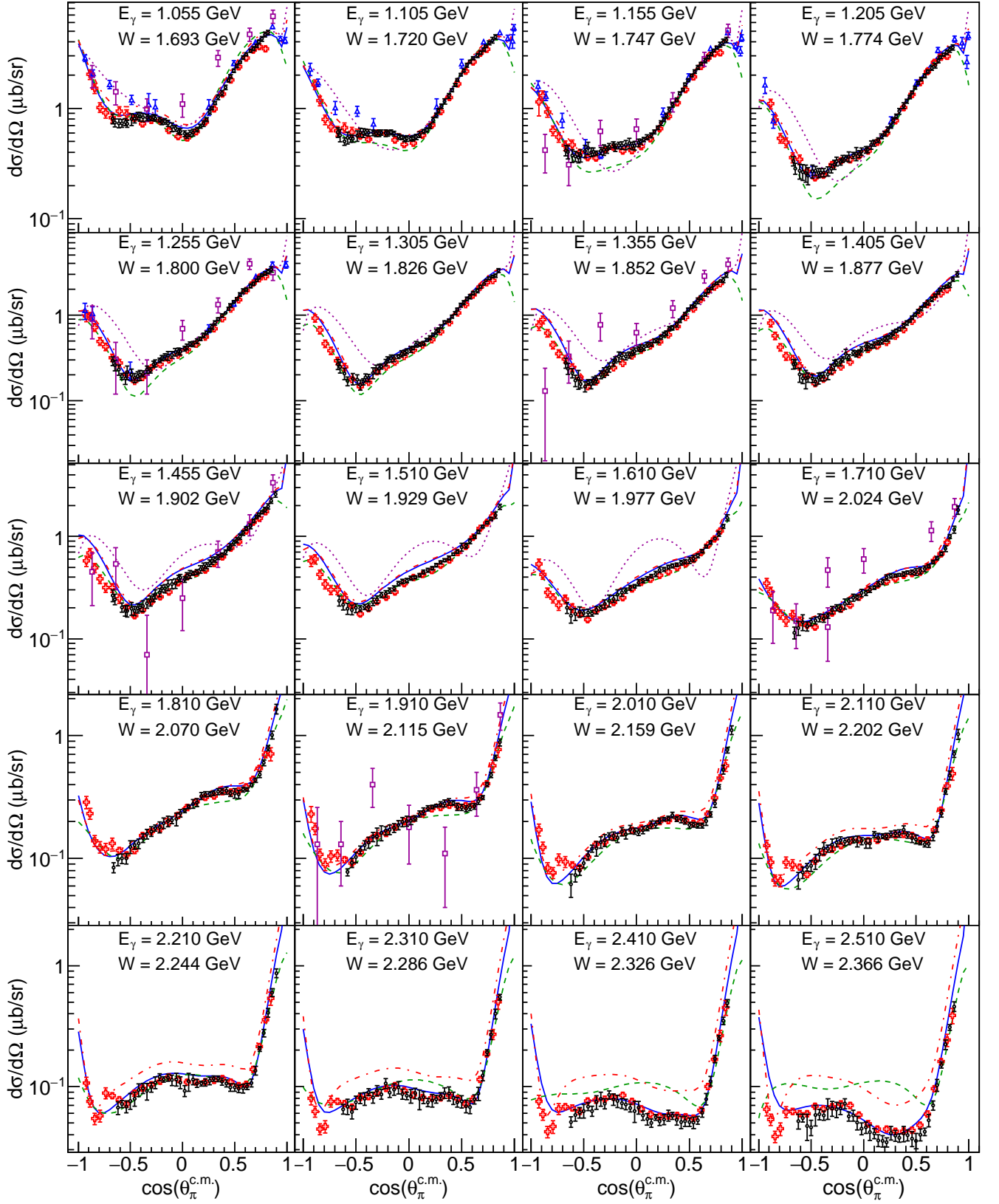} 
\par\end{centering}

\protect\caption{\label{fig:CrossSections2}(Color online) Selected cross 
        section data for $\gamma n\to \pi^- p$ vs. $\cos\theta_\pi^{c.m.}$ 
        above $E_\gamma=1.05$~GeV: CLAS g13 (black open circles), CLAS 
        g10~\protect\cite{gb12} (red open pluses), 
        SLAC~\protect\cite{Scheffler} (blue open triangles), and 
        DESY~\protect\cite{Benz} (violet open squares); fits: SAID MA27 
        (blue solid lines), SAID PR15~\protect\cite{pr15} (red dot-dashed 
        lines), BG2014-02~\protect\cite{BnGa14} (green dashed lines), and
        MAID2007~\protect\cite{MAID07} (violet dashed lines). The $y$-axes 
        are log scale. Only angle-dependent uncertainties are shown for all 
        data. The total normalization uncertainties for the CLAS g13 data 
        are about 3.4\%.}
\end{figure*}

\begin{figure*}
\begin{centering}
\includegraphics[width=0.7\linewidth]{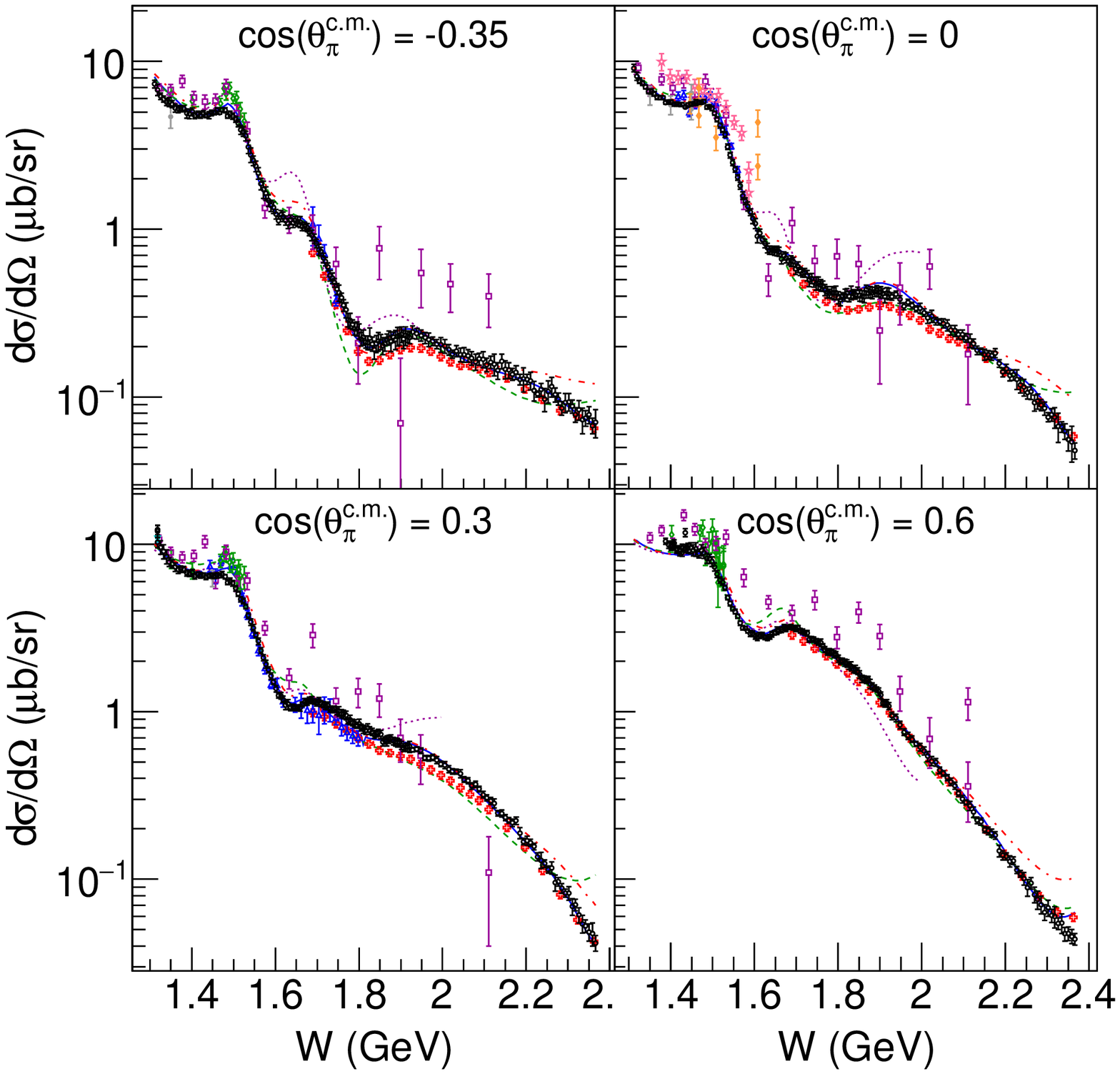} 
\par\end{centering}

\protect\caption{\label{fig:CrossSectionsVsW}(Color online) Selected cross 
        section data for $\gamma n\to \pi^- p$ vs. $W$: CLAS g13 (black open 
        circles), CLAS g10~\protect\cite{gb12} (red open pluses), 
	SLAC~\protect\cite{Scheffler} (blue open triangles), 
	DESY~\protect\cite{Benz} (violet open squares), 
	MAMI-B~\protect\cite{Briscoe} (cyan open down-triangles), 
	and Frascati~\protect\cite{Beneventano} (pink open stars);
	$\pi^{-}p\to\gamma n$ data: BNL~\protect\cite{Shafi} (green 
	open diamonds), LBL~\protect\cite{Weiss} (orange closed diamonds), 
	and LAMPF~\protect\cite{Kim} (gray closed circles); fits: SAID MA27 
        (blue solid lines), SAID PR15~\protect\cite{pr15} (red dot-dashed 
        lines), BG2014-02~\protect\cite{BnGa14} (green dashed lines), and
        MAID2007~\protect\cite{MAID07}(which terminates at $W$ = 2~GeV or 
        $E_\gamma=1.65$~GeV) (violet dotted lines). The $y$-axes are log scale. 
        Only angle-dependent uncertainties are shown for all data. The 
        total normalization uncertainties for the CLAS g13 data are about 3.4\%.}
\end{figure*}

To extract the $\gamma n\to\pi^-p$ differential cross sections, model-dependent 
final state interaction corrections were applied to the $\gamma d\to\pi^-p(p)$ 
data, as discussed in Section~\ref{sec:FSI}. These data were split up into 157 
photon energy bins from 0.440~GeV to 2.520~GeV, 10-MeV-wide below 1.5~GeV and 
20-MeV-wide above. The $\gamma n\to\pi^-p$ differential cross section measurements 
are shown for 40 of these $E_\gamma$ bins in Figs.~\ref{fig:CrossSections1} and
\ref{fig:CrossSections2}, compared against previous measurements and available 
partial wave analysis solutions. They are also shown in 
Fig.~\ref{fig:CrossSectionsVsW} vs. $W$ in four bins of $\cos\theta_\pi^{c.m.}$. 
These figures include $\gamma n\to\pi^-p$ measurements from CLAS g10~\cite{gb12}, 
SLAC~\cite{Scheffler}, DESY~\cite{Benz}, MAMI-B~\cite{Briscoe}, and Frascati
\cite{Beneventano}, and $\pi^-p\to\gamma n$ measurements from BNL~\cite{Shafi}, 
LBL~\cite{Weiss}, and LAMPF~\cite{Kim}. Only the angle-dependent uncertainties 
are shown for all measurements. All non-CLAS g13 data shown in 
Figs.~\ref{fig:CrossSections1} and \ref{fig:CrossSections2} are within 
$\pm$10~MeV of the selected g13 energy bin, and all non-CLAS g13 data shown in 
Fig.~\ref{fig:CrossSectionsVsW} are within $\cos \theta_\pi^{c.m.}$ of $\pm$0.05 
of the g13 angle bin.

%------------------------------------------
%------------------------------------------
%------------------------------------------
%------------------------------------------
%------------------------------------------
%------------------------------------------

The differential cross section peaks at low energy due to $\Delta(1232)$
and $N^*$ resonance production, and at forward angles due to 
$t$-channel pion exchange. Below $E_\gamma$ = 1~GeV, the new CLAS g13 
data dominate the previous world measurements, with $\cos\theta_\pi^{c.m.}$ 
bins 0.02-wide and total uncertainties typically less than 10\% in this 
range. The CLAS g13 data are systematically lower than the DESY~\cite{Benz}, 
BNL~\cite{Shafi}, and SLAC~\cite{Scheffler} measurements in several energy bins, 
and each of these measurements quote normalization uncertainties of about 5\%. 
There is also a discrepancy in the trend of the data at forward angles between 
the CLAS g13 and SLAC measurements below $E_\gamma$ = 0.800~GeV, with the g13 data 
rising more sharply at forward angles. 

Above $E_\gamma=1$~GeV, the g13 data are reported in bins that are 0.03-wide 
in $\cos\theta_\pi^{c.m.}$ up to 1.5~GeV, and 0.04-wide in 
$\cos\theta_\pi^{c.m.}$ above 1.5~GeV. Here, the CLAS g10 data~\cite{gb12}
were the previous highest-statistics measurement, reported in 50- and 
100-MeV-wide beam energy bins. The g13 data are in excellent agreement with 
these measurements, as the g10 data have normalization uncertainties of 
$\sim$6\% to $\sim$10\% that are not shown in the figures.

The SAID PR15~\cite{pr15}, Bonn-Gatchina BG2014-02~\cite{BnGa14}, and MAID2007
\cite{MAID07} curves shown in these figures did not include the new CLAS g13 
data in their fits, and the MAID2007 fit does not include the CLAS g10 
measurements either. The data in these previous fits, and in the new SAID MA27 
fit that includes the g13 data, are discussed in Section~\ref{sec:PWA}.

%------------------------------------------
\section{Final State Interactions
\label{sec:FSI}}

\begin{figure}
\begin{centering}
\includegraphics[height=2.4cm]{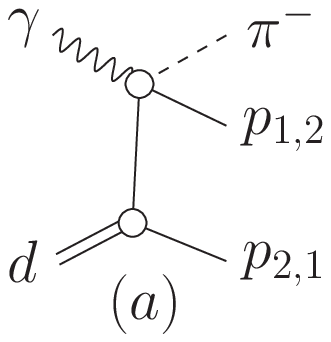}~~
\includegraphics[height=2.4cm]{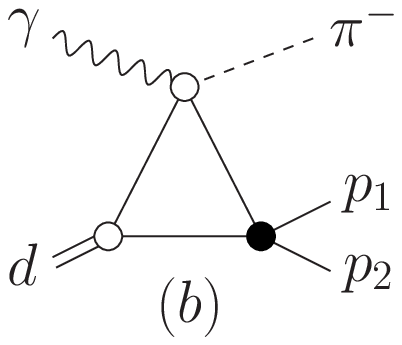}~~
\includegraphics[height=2.4cm]{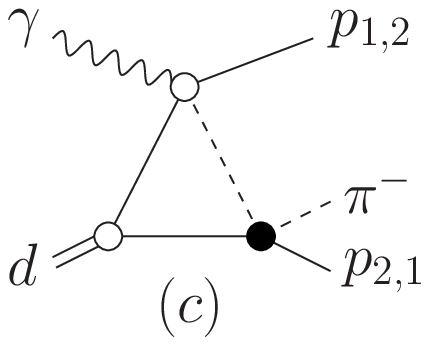} 
\par\end{centering}

\protect\caption{Feynman diagrams for the leading terms of the 
	$\gamma d\to\pi^-pp$ amplitude: (a) IA, (b) $pp$-FSI, and 
	(c) $\pi$N-FSI. The filled black circles represent the FSI 
	vertices. The wavy, dashed, solid, and double lines correspond 
	to the photons, pions, nucleons, and deuterons, respectively. 
	\label{fig:g3}}
\end{figure}

The $\gamma n\to\pi^-p$ cross sections were extracted on a free neutron
from the deuteron data in the quasi-free kinematic region of
the $\gamma d\to\pi^-pp$ reaction, which has a fast knocked-out proton
$p_1$ and a slow proton spectator $p_2$, assumed not to be involved in 
the pion production process. In this quasi-free region, the reaction mechanism 
corresponds to the ``dominant'' Impulse Approximation (IA) diagram in 
Fig.~\ref{fig:g3}(a) with the slow proton $p_2$ emerging from the deuteron 
vertex. Here, the differential cross section on the deuteron can be related to 
that on the neutron target in a well understood way (see, e.g., Eq.~(22) of 
Ref.~\cite{FSI} and references therein). Fig.~\ref{fig:g3}(a) illustrates 
this ``dominant'' IA diagram, as well as the ``suppressed'' IA diagram with 
the protons interchanged. This approximation, with the additional assumption
that the neutron is at rest in the deuteron, allows for the identification
of the quasi-free cross section $\frac{d\sigma}{d\Omega}$ on the deuteron with
that on the neutron, where $d\Omega$ is the solid angle of the outgoing
pion in the $\gamma n$ rest frame. The $\gamma n$ cross section can be
calculated as
\begin{equation}
	\frac{d\sigma}{d\Omega}(\gamma n)=R(E_\gamma,\theta_\pi^{c.m.})^{-1}
	\frac{d\sigma}{d\Omega}(\gamma d),\label{eq:1}
\end{equation}
\noindent
where $\frac{d\sigma}{d\Omega}(\gamma d)$ is the quasi-free CLAS g13 measurement
on the deuteron and $R(E_\gamma,\theta_\pi^{c.m.})$ is the FSI correction factor 
that takes into account the FSI effects discussed below, as well as the identity 
of the two protons in the $\gamma d$ reaction. This factor is defined as the 
ratio between the full contribution of the three diagrams in Fig.~\ref{fig:g3} and 
that of the ``dominant'' IA diagram in Fig.~\ref{fig:g3}(a). There are two critical 
factors to consider when using this approach: 
\begin{enumerate}
\item[1)] the neutron is bound in the deuteron and not at rest, and \vspace{-2mm}
\item[2)] there are $NN$- and $\pi N$-FSI effects. 
\end{enumerate}
Factor 1) means that the effective mass of the neutron
\begin{equation}
	m_{eff}=\sqrt{(p_d - p_s)^2}\approx m_n - \epsilon_d -\vec{p}_s^{\;2}/m_N
\end{equation}\noindent 
is not equal to the mass of the free neutron $m_n$. Here, $p_d$, $p_s$, 
$\vec{p}_s$, $\epsilon_d$, and $m_N$ are the deuteron 4-momentum, 4- and 
3-momenta of the spectator proton, the deuteron binding energy, and the 
nucleon mass, respectively. Also, the invariant mass $\sqrt{s_{\pi N}}$ of 
the final $\pi N$-system,
\begin{equation}
\sqrt{s_{\pi N}}=\sqrt{s_{\gamma N}}=\sqrt{[(E_\gamma+m_d
-E_s)^2-(\vec{p}_{\gamma}-\vec{p}_s)^2]},
\end{equation}
\noindent 
depends on the proton-spectator momentum $\vec{p}_s$ ($s_{\gamma N}$ 
is the invariant mass squared of the initial $\gamma N$ state). Here, 
$E_\gamma$ ($E_s$), $m_d$, and $\vec{p}_\gamma$ are the total 
energy of the initial photon (proton-spectator), the deuteron mass, 
and the photon 3-momentum, respectively, and $E_\gamma =|\vec{p}_\gamma|$.

%------------------------------------------
%------------------------------------------

Since $\sqrt{s_{\pi N}}$ depends on $\vec{p}_s$, the $\gamma N\to\pi N$
cross section extracted from the deuteron data, with an undetected 
nucleon-spectator, is averaged over an energy range that depends on the 
kinematic cuts employed for $\vec{p}_s$. Thus, the effective photon laboratory 
energy $E_{\gamma n}$ (defined through the relation $s_{\gamma N}=m_n^2
+2m_nE_{\gamma n}$ for the $\gamma n\to\pi^-p$ reaction) and the pion c.m.
angle $\theta_\pi^{c.m.}$ are smeared due to the deuteron wave function (DWF). 
This smearing has been estimated from a simplified calculation, where the 
$\gamma d\to\pi^-pp$ amplitude is proportional to the DWF and depends only on 
the laboratory momentum of one of the final protons, say $p_2$. Here, 
$E_{\gamma n}$ is determined through the above-mentioned relation with the 
effective mass of the pion-proton pair with the other proton $p_1$. The 
distortion of the extracted $\gamma n\to\pi^-p$ cross sections due to the 
smearing effect is negligible, as was shown in Ref.~\cite{Briscoe}.

Factor 2) corresponds to the inclusion of the FSI corrections. Their
leading terms correspond to the Feynman diagrams shown in 
Figs.~\ref{fig:g3}(b,c). The GWU SAID database contains phenomenological 
amplitudes for the reactions $\pi N\to\pi N$~\cite{piN}, $NN\to NN$~\cite{NN}, 
and $\gamma N\to\pi N$~\cite{du07}, which were used as inputs to calculate the 
dominant diagrams of the GWU-ITEP FSI approach. The full Bonn potential
\cite{Bonn} was used for the deuteron description. 

Calculations of the $\gamma d\to\pi^-pp$ differential cross sections with the 
FSI taken into account (including all diagrams in Fig.~\ref{fig:g3}) were 
done for the present g13 data as they were done previously for the CLAS g10 
data ($E_\gamma=1.050$~GeV to 2.700~GeV and $\theta_\pi^{c.m.}$ = 32$^\circ$ to 
157$^\circ$)~\cite{gb12} and MAMI-B data ($E_\gamma=0.301$~GeV to 0.455~GeV and 
$\theta_\pi^{c.m.}=58^{\circ}$ to 141$^\circ$)~\cite{Briscoe}. 

\begin{figure*}[th]
\centerline{ \includegraphics[height=0.8\textwidth,angle=90]{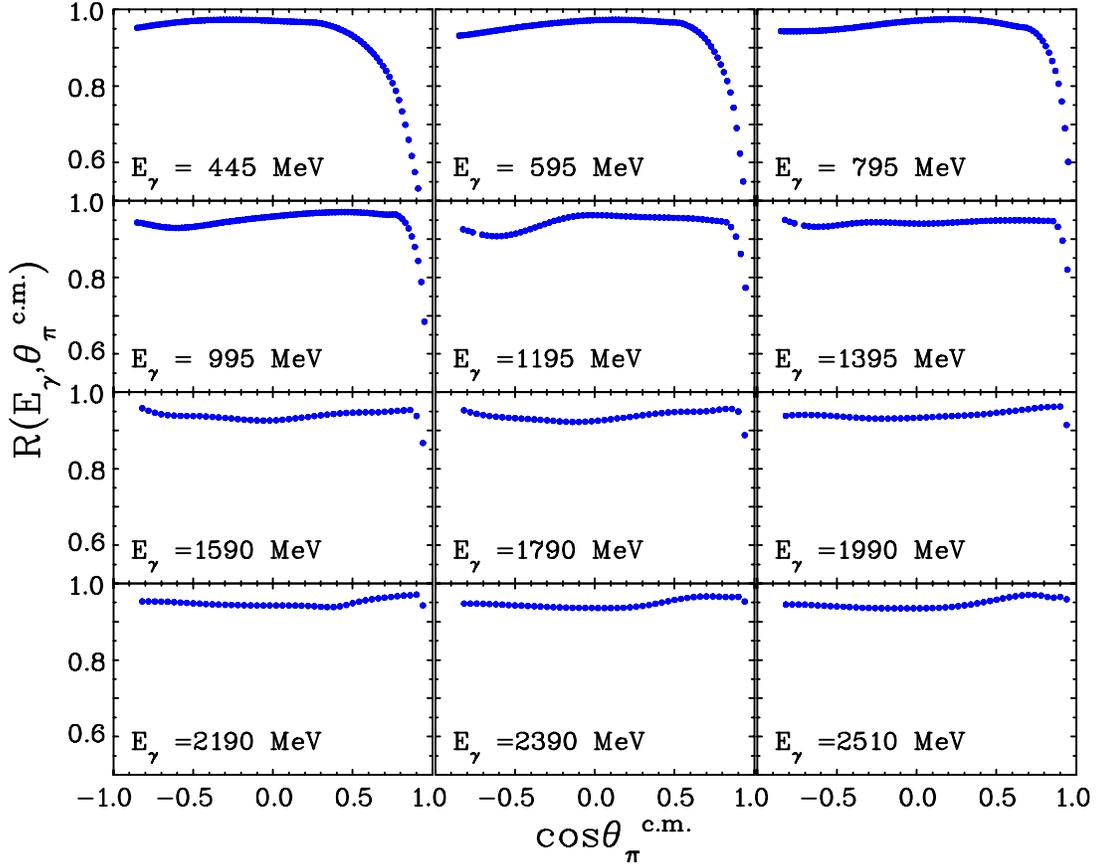}}

\protect\caption{(Color online) The FSI correction factor $R(E_\gamma,\theta_\pi^{c.m.})$
        for selected beam energies vs. $\cos \theta_\pi^{c.m.}$, where $\theta_\pi^{c.m.}$ 
        is the polar angle of the outgoing $\pi^-$ in the rest frame of the 
	$\pi^-$ and the fast proton. The fast knocked-out protons $p_1$ 
	with momentum $>200$~MeV/c were selected, while the slow proton spectators 
	$p_2$ have momentum $<200$~MeV/c. The 2\% normalization uncertainties are 
        not shown. 
        \label{fig:ratio}}
\end{figure*}

The GWU-ITEP FSI calculations~\cite{FSI} are available over a broad
energy range (threshold to $E_\gamma=2.700$~GeV) and for the full c.m.
angular range ($\theta_\pi^{c.m.}$ = 0$^{\circ}$ to 180$^{\circ}$).
Fig.~\ref{fig:ratio} shows the FSI correction factor
$R=R(E_\gamma,\theta_\pi^{c.m.})$ for the $\gamma n\to\pi^-p$ differential
cross section as a function of $\theta_\pi^{c.m.}$ for 
different energies over the range of the CLAS g13 experiment. Overall, the 
FSI correction factor $R<1$, while the value of $R$ varied from 70\% to 
90\% depending on the kinematics. The behavior of $R$ is very smooth vs. pion 
production angle. Note that $R(E_\gamma,\theta_\pi^{c.m.})$ is the FSI 
correction factor for the CLAS quasi-free $\gamma d \to\pi^-pp$ cross section 
averaged over the laboratory photon energy $E_\gamma$ bin width.
%------------------------------------------
%------------------------------------------

The contribution of FSI calculations~\cite{FSI} to the overall systematic 
normalization uncertainty is estimated to be about 2.2\% (the sensitivity to 
the DWF is 1\% and to the number of steps in the integration of the five-fold 
integrals is 2\%). No sensitivity was found to the kinematic cuts used for the 
detected protons in CLAS.

%------------------------------------------
\section{Legendre Analysis}
\label{sec:Legendre}

Legendre expansions provide a model-independent approach suitable for presentation 
of modern detailed (high-precision, high-statistics, and narrow energy and angular 
binning) data for pion photoproduction reactions~\cite{leg}. This approach is 
applicable both to cross sections and to polarization observables; it is much more 
compact and visual than traditional methods (see, for instance, 
Figs.~\ref{fig:CrossSections1} to \ref{fig:CrossSectionsVsW}), at least at energies 
within the nucleon resonance region. The Legendre coefficients reveal specific 
correlations and interferences between resonant states of definite parities.

The small statistical uncertainties of the g13 data obtained here allow a 
correspondingly robust determination of the Legendre polynomial coefficients 
$A_J(W)$. These coefficients were very difficult to determine unambiguously with 
previously published $\pi^-$ photoproduction data of lower statistical accuracy. 
Because of the limited angular range of the g13 data, several sets of quasi-data 
were generated using the MA27 SAID solution (see Section~\ref{sec:PWA} for details) 
in bins with width $\Delta \cos\theta_\pi^{c.m.} = 0.05$ for the forward and 
backward directions to cover the full angular range. 

It is important to note that the MA27 solution was constrained at the forward and
backward angular ranges beyond the extent of the g13 data by the existing world data
shown in Figs.~\ref{fig:CrossSections1} to \ref{fig:CrossSectionsVsW}. However, as 
the available data does not span the full $\cos \theta_\pi^{c.m.}$ range for the $W$ 
range of the g13 data, the MA27 quasi-data were conservatively assigned 10\% 
uncertainties, which matches the largest of the experimental uncertainties reported 
within the g13 data, excepting a few regions with a problematic acceptance determination. 
Conservative assignment of uncertainties in these regions is 
important as these regions are quite sensitive to the highest partial waves.

As expected for such a fit using orthogonal polynomials, the Legendre
coefficients $A_J(W)$ decrease markedly for large $J$. With the energy range
and precision of the g13 data, a maximum value of $J = 10$ was found to be 
sufficient to describe the data (similar to the analysis of the CLAS $\pi^0$ 
and $\pi^+$ $\Sigma$ beam asymmetry measurements~\cite{ASU}). Thus, 
the infinite series is truncated as
\begin{equation}
        \frac{d\sigma(W,\cos\theta_\pi^{c.m.})}{d\Omega}=\sum_{J=0}^{10}\,A_J(W)\,
        P_J(\cos\theta_\pi^{c.m.}), \label{decomp}
\end{equation}
where the total cross section $\sigma^{\rm tot} = 4\pi A_0(W)$.
%-----------------------------------------------------
\begin{figure*}[th]
\centerline{\includegraphics[height=0.8\textwidth,angle=90]{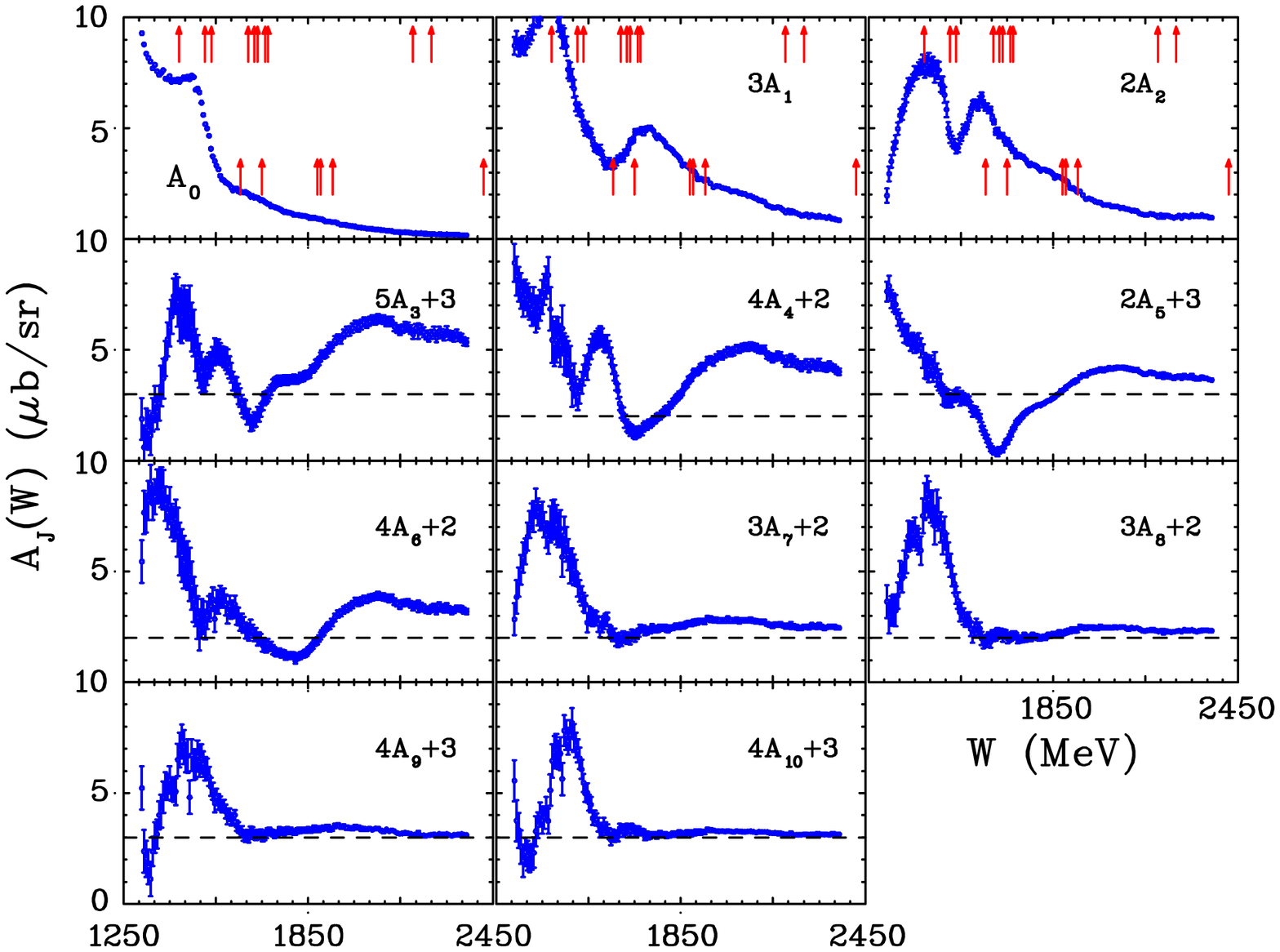}}

\protect\caption{(Color online) Coefficients of the Legendre polynomials
	$A_J(W)$ (blue filled circles) from the fits of the CLAS g13
        $\gamma n \to \pi^- p$ cross section data. The error bars 
	represent the $A_J(W)$ uncertainties from the fits in which only 
	the statistical uncertainties were used. The $A_J(W)$ coefficients
        have been scaled (dashed horizontal lines) by $n\times A_J+m$ to
        enable easy visualization. The red vertical arrows in the top 
        row of plots indicate the masses of the PDG four-star resonances 
        (Breit-Wigner masses) in this energy range~\protect\cite{PDG16}. 
        The upper row of arrows corresponds to $N^*$ states with isospin $I = 1/2$ 
        and the lower row corresponds to $\Delta^*$ states with $I = 3/2$.
        \label{fig:leg}}
\end{figure*}
%-----------------------------------------------------

In Fig.~\ref{fig:leg}, the Legendre coefficients $A_0(W)$ to $A_{10}(W)$ are 
shown as a function of $W$ from the fit of the CLAS g13 $d\sigma/d\Omega$ 
data and the $d\sigma/d\Omega$ data generated from the MA27 predictions. The 
individual Legendre coefficients have been scaled by $n \times A_J +m$ to allow
plotting on a common abscissa. The $n$ and $m$ scaling values are given on
the subplots of Fig.~\ref{fig:leg}.

The results of our fits yield unprecedented detail on the energy dependence 
of the Legendre coefficients $A_J(W)$, and should prove useful for
performing a phase shift analysis of pion photoproduction data for the present 
energy range. As expected from the form of Eq.~(28) of Ref.~\cite{leg},
resonance contributions from the second, third, and fourth resonance regions
combine to produce clear peaks in the coefficient $A_0(W)$. It is interesting 
that all $A_J(W)$ coefficients show structure for the $W$ = 1.3~GeV to 1.8~GeV 
range, which was also seen in the MAMI A2 $\pi^0$ data~\cite{pr15}.
However, wide structures are also visible in the range $W$ = 1.8~GeV to 2.0~GeV, 
most likely attributable to contributions from one or more nucleon 
resonances known in this energy range with spin up to 7/2, as was seen in the 
recent CLAS g8 $\pi^0$ and $\pi^+$ $\Sigma$ beam asymmetry measurement Legendre 
analysis~\cite{ASU}.

The Legendre fit results shown in Fig.~\ref{fig:leg} do not include any
assignment of model uncertainties associated with the extrapolations of
the MA27 model beyond the range of the available data. Such assignments
could be expected to be non-negligible for the higher Legendre moments
shown here. However, our purpose in displaying the Legendre fit results is
not to perform a quantitative amplitude analysis, but to showcase how the
precision g13 cross section measurements can provide significant constraints 
on the resonance contributions over a broad range in $W$.

%------------------------------------------
\section{Multipole Analysis}
\label{sec:PWA}

The SAID parameterization of the transition amplitude $T_{\alpha\beta}$ 
used in the hadronic fits to the $\pi N$ scattering data is given as
\begin{equation}
        T_{\alpha\beta} = \sum_\sigma [1-\overline{K} C]^{-1}_{\alpha\sigma}
        \overline{K}_{\sigma\beta},
\end{equation}
where $\alpha$, $\beta$, and $\sigma$ are channel indices for the 
$\pi N$, $\pi\Delta$, $\rho N$, and $\eta N$ channels. Here 
$\overline{K}_{\alpha\beta}$ are the Chew-Mandelstam $K$-matrices, which are 
parameterized as polynomials in the scattering energy. $C_\alpha$ is the 
Chew-Mandelstam function, an element of a diagonal matrix $C$ in channel-space, 
which is expressed as a dispersion integral with an imaginary part equal to the 
two-body phase space~\cite{ford}.

In Ref.~\cite{cm12}, it was shown that this form could be extended to 
$T_{\alpha\gamma}$ to include the electromagnetic channel as
\begin{equation}
	T_{\alpha\gamma} = \sum_\sigma [1-\overline{K} C]^{-1}_{\alpha\sigma}
	\overline{K}_{\sigma\gamma}.
\end{equation}
Here, the Chew-Mandelstam $K$-matrix elements associated with the hadronic channels 
are kept fixed from the previous SAID solution SP06~\cite{piN}, and only the 
electromagnetic elements are varied. The resonance pole and cut structures are also 
fixed from hadronic scattering. This provides a minimal description of the 
photoproduction process, where only the $N^*$ and $\Delta^*$ states present in the 
SAID $\pi N$ scattering amplitudes are included in this multipole analysis.

%------------------------------------------
\begin{table*}[htb!]

\centering \protect\caption{Comparison of $\chi^2$ per data point ($d.p.$) below 
        $E_\gamma$ = 2.7~GeV ($W$ = 2.5~GeV) for the $\gamma n \to \pi^- p$ channel 
        using predictions for the recent SAID PR15~\protect\cite{pr15} and the
	current MA27 solution. The first row of solutions compares the fit
        quality to the available data not including the CLAS g13 data. The 
        second row compares the solutions to the available data including the 
        g13 data. The last row compares the solutions only to the g13 data.}
\vspace{2mm}
{%
\begin{tabular}{|c|c|c|}
\hline
Data          & Solution & $\chi^2/(\pi^-p~d.p.)$ \tabularnewline \hline
Existing data & PR15     & 6541/3162 = 2.07       \tabularnewline
without g13   & MA27     & 7112/3162 = 2.25       \tabularnewline \hline
Existing data & PR15     & 24052/11590 = 2.08     \tabularnewline
with g13      & MA27     & 16442/11590 = 1.42     \tabularnewline \hline
Only          & PR15     & 17511/8452 = 2.07      \tabularnewline
g13           & MA27     &  9330/8452 = 1.10      \tabularnewline \hline
\end{tabular}} \label{tab:tbl1a}
\end{table*}
%------------------------------------------

%------------------------------------------
\begin{table*}[htb!]

\centering \protect\caption{Comparison of $\chi^2$ per data point ($d.p.$) below 
        $E_\gamma$ = 2.7~GeV ($W$ = 2.5~GeV) for all $\gamma N \to \pi N$ channels 
        using predictions for the recent SAID PR15~\protect\cite{pr15} and the
	current MA27 solution. The fit quality for the $\pi^0 p$, $\pi^+n$, 
        $\pi^- p$, and $\pi^0 n$ channels is compared to the available data 
        including the g13 data.}
\vspace{2mm}
{%
\begin{tabular}{|c|c|c|c|c|c|}
\hline
Data       & Solution & $\chi^2/(\pi^0p~d.p.)$  & $\chi^2/(\pi^+n~d.p.)$  & $\chi^2/(\pi^-p~d.p.)$   & $\chi^2/(\pi^0n~d.p.)$ \tabularnewline \hline
Existing data  & PR15     & 54985/25540 = 2.15& 23558/9859 = 2.39& 24052/11590 = 2.08& 1152/364 = 3.16 \tabularnewline
with g13       & MA27     & 55530/25540 = 2.17& 20736/9859 = 2.10& 16442/11590 = 1.42& 1540/364 = 4.23 \tabularnewline \hline 
\end{tabular}} \label{tab:tbl1b}
\end{table*}
%------------------------------------------

For each angular distribution, a normalization constant $(X)$ and its uncertainty 
$(\epsilon_X)$ were assigned. The quantity $\epsilon_X$ is generally associated 
with the normalization uncertainty (if known). The modified $\chi^2$ function to 
be minimized is given by
\begin{eqnarray}
        \chi^2=\sum_i\left(\frac{{X\theta_i-
        \theta_i^{exp}}}{{\epsilon_i}}\right)^2
        +\left(\frac{{X-1}}{{\epsilon_X}}\right)^2,\label{eq:norm}
\end{eqnarray}\noindent
where the subscript $i$ labels the data points within the distribution,
$\theta_i^{exp}$ is an individual measurement, $\theta_i$ is the corresponding 
calculated value, and $\epsilon_i$ represents the total angle-dependent 
uncertainty. The total $\chi^2$ is then found by summing over all measurements. 
This renormalization freedom is essential for obtaining the best SAID fit results. 
For other data analyzed in the fit, such as the total cross sections and excitation 
data, the statistical and systematic uncertainties were combined in quadrature and 
no renormalization was allowed.

\begin{figure}[htb!]
\includegraphics[width=2in,height=2.5in,angle=90]{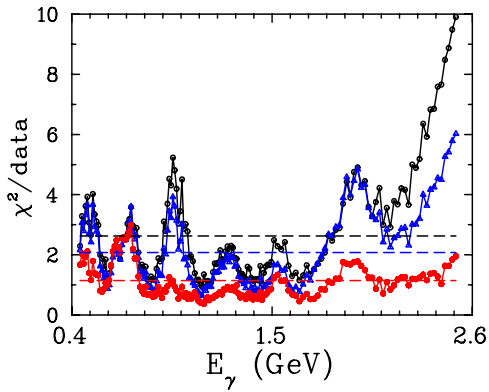}
 
\protect\caption{(Color online) Comparison of the previous SAID solution 
	PR15~\protect\cite{pr15} applied to the present g13 data with 
	(blue filled triangles) and without FSI corrections (black open circles), 
        and the new SAID MA27 (red full circles) solution obtained after adding 
        the present g13 data with FSI corrections into the fit (the solid lines 
        connecting the points are included only to guide the eye). Shown are the 
        fit $\chi^2$ per data point values averaged within each energy bin 
        $E_\gamma$, where the horizontal dashed lines (blue (black) for PR15 and 
        red for MA27) show the overall $\chi^2$ per data point values from 
        Table~\protect\ref{tab:tbl1b}.}
\label{fig:chi2} 
\end{figure}

In the previous fits to the $\gamma n\to\pi^-p$ differential cross sections of 
Ref.~\cite{gb12}, the unrestricted best fit gave renormalization constants $X$ 
significantly different from unity. As can be seen from Eq.~(\ref{eq:norm}), 
if an angular distribution contains many measurements with small statistical 
uncertainties, a change in the renormalization may improve the fit with only a 
modest $\chi^2$ penalty. Here, however, the weight of the second term in 
Eq.~(\ref{eq:norm}) has been adjusted by the fit for each dataset to keep the 
renormalization constants approximately within $\epsilon_X$ of unity. This was 
possible without degrading the overall fit $\chi^2$, as can be seen in 
Fig.~\ref{fig:chi2}. 

%------------------------------------------
%------------------------------------------

With the new high-precision $\gamma n \to \pi^- p$ cross sections from the CLAS 
g13 dataset, a new SAID multipole analysis has been completed. This new global 
energy-dependent solution has been labeled as MA27. The overall fit quality of 
the present MA27 and previous SAID PR15 solutions are compared in 
Tables~\ref{tab:tbl1a} and \ref{tab:tbl1b}. The inclusion of the g13 dataset 
shows significant improvement in the comparisons between the $\pi^- p$ fits and 
data ($\chi^2/d.p.$ for PR15 = 2.08 and $\chi^2/d.p.$ for MA27 = 1.10) as shown 
in Fig.~\ref{fig:chi2} and Table~\ref{tab:tbl1a}. This demonstrates the power 
of these cross section measurements with their small uncertainties. The overall 
comparison of the PR15 and MA27 solutions in Table~\ref{tab:tbl1b} shows the fit 
$\chi^2/d.p.$ values are essentially unchanged for the $\pi^0 p$ and $\pi^+ n$ 
channels but are notably worse for the $\pi^0 n$ channel, which has very low 
statistics. The overall $\chi^2$ per data point including all available data 
and the new g13 data for PR15 is $\chi^2/d.p.=2.19$ (103747/47353) and for MA27 
is $\chi^2/d.p.=1.99$ (94248/47353). 

\begin{figure*}[th]
\centerline{\includegraphics[height=0.4\textwidth,angle=90]{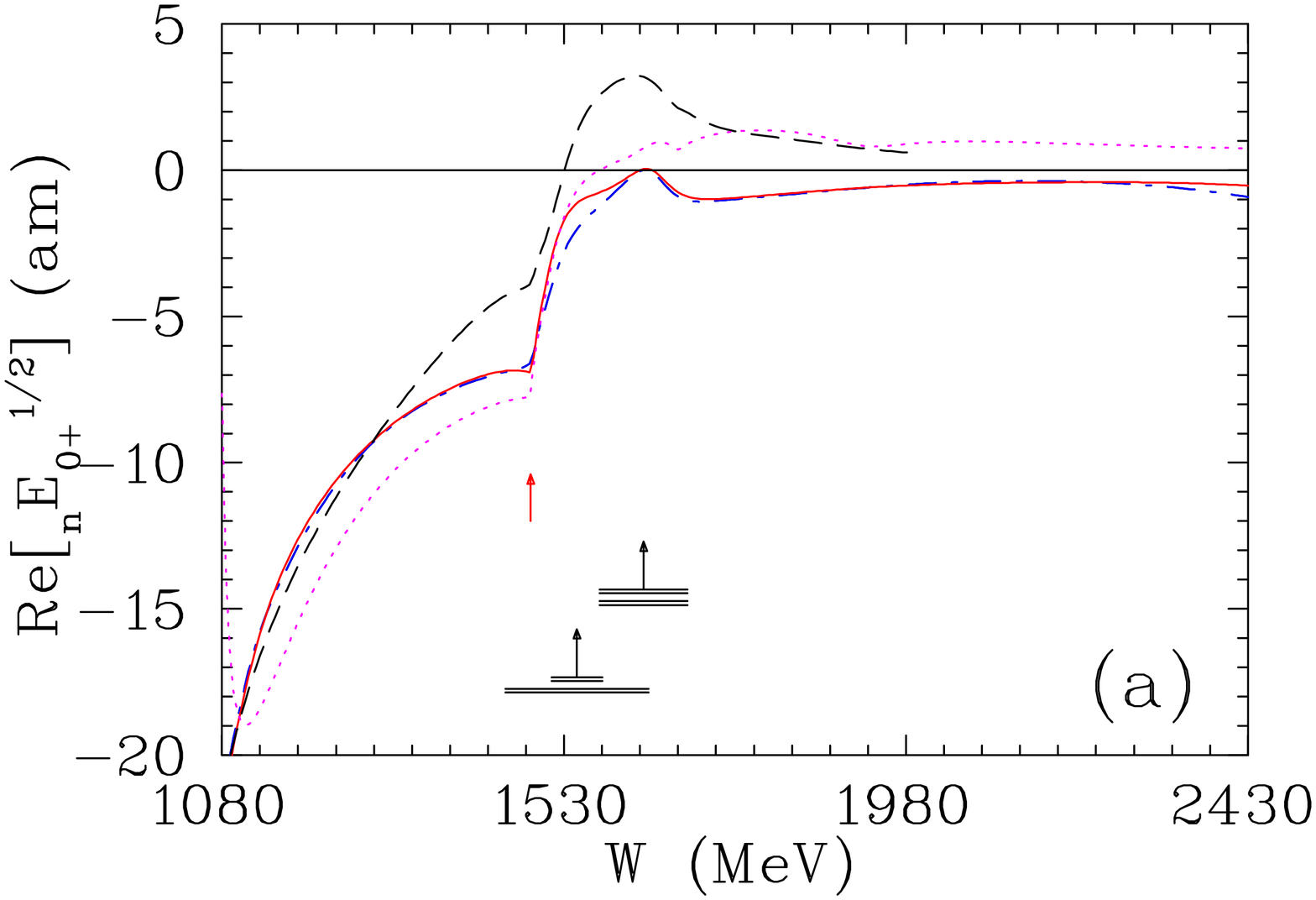}~~\includegraphics[height=0.4\textwidth,angle=90]{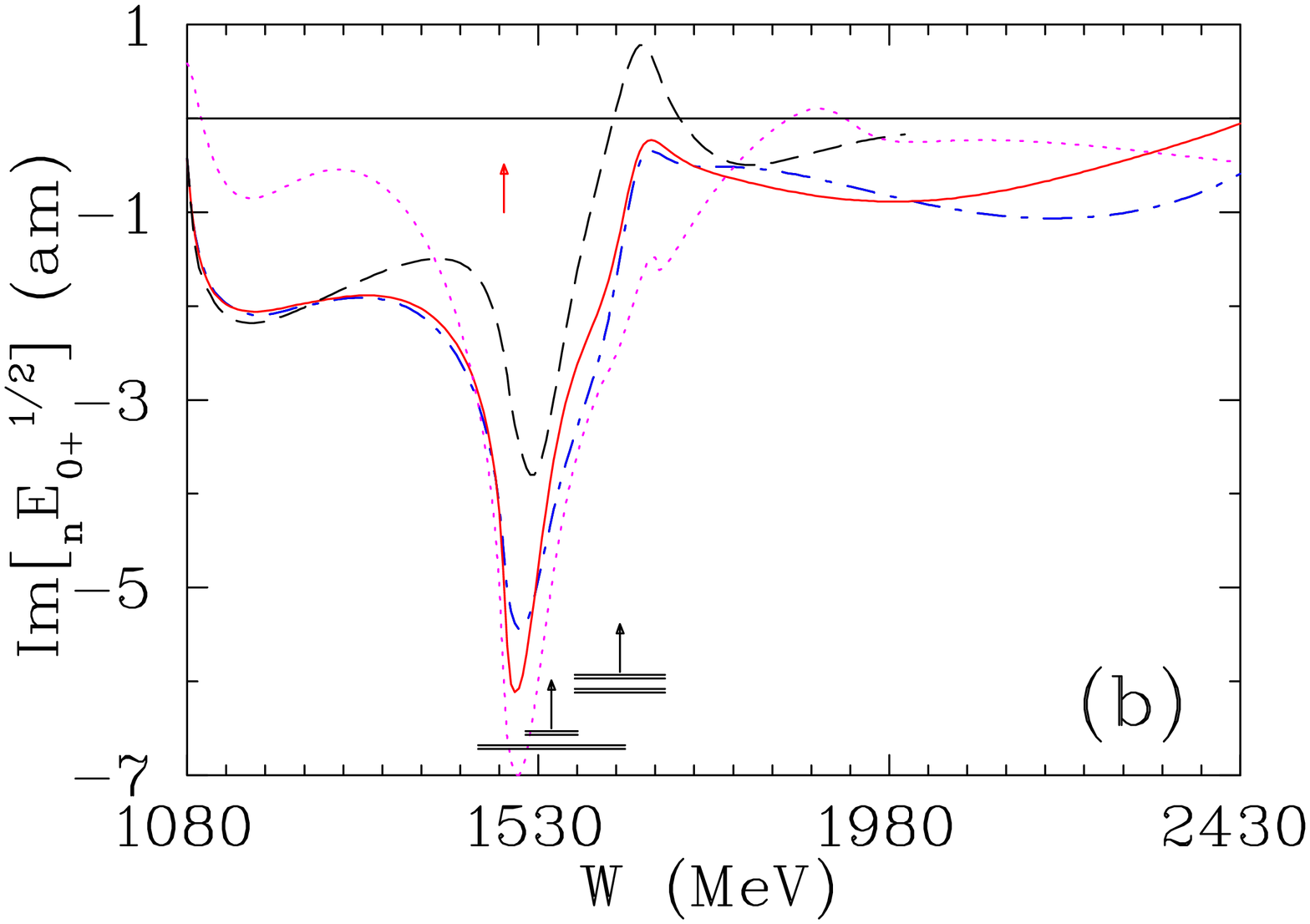}}
\centerline{\includegraphics[height=0.4\textwidth,angle=90]{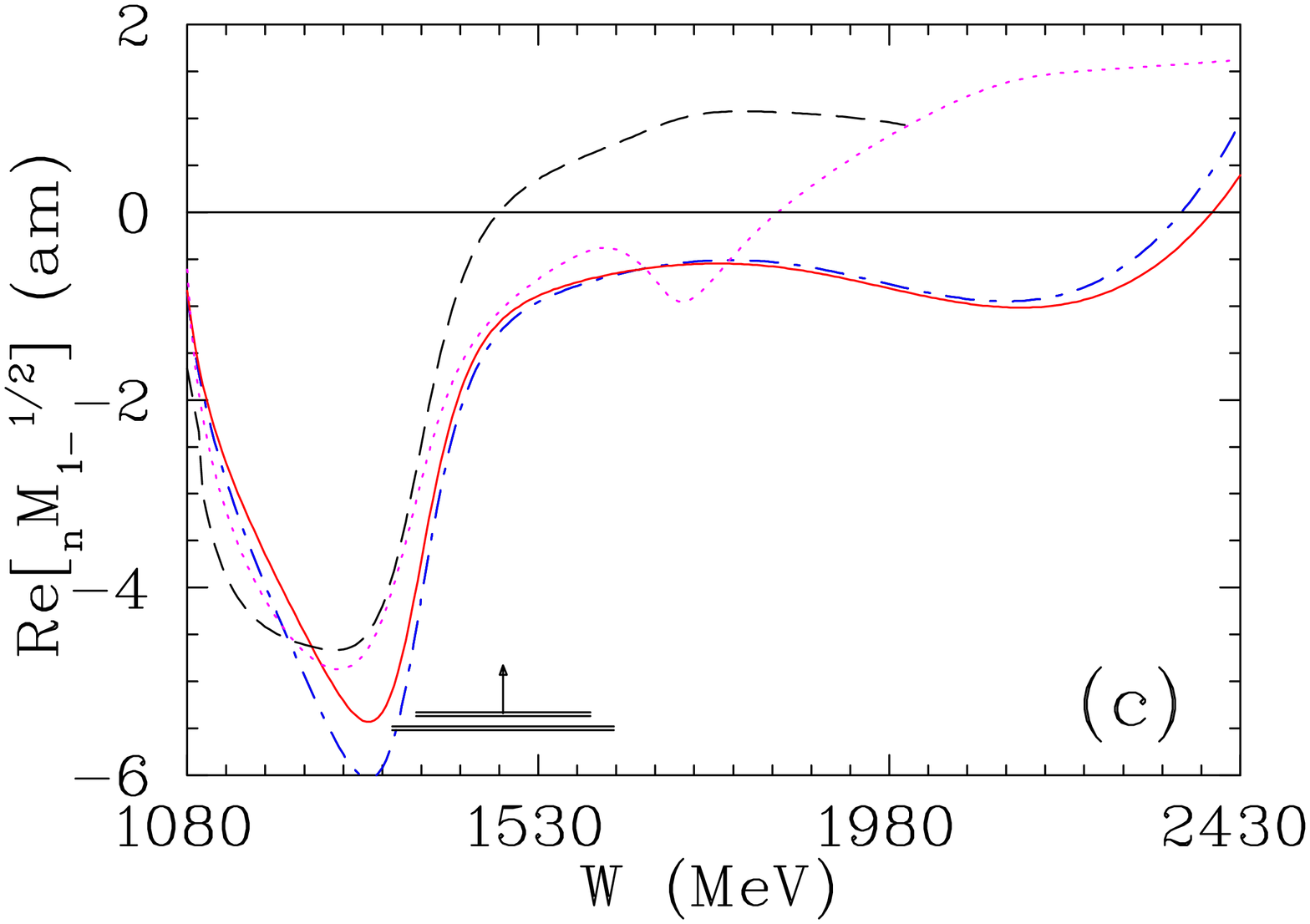}~~\includegraphics[height=0.4\textwidth,angle=90]{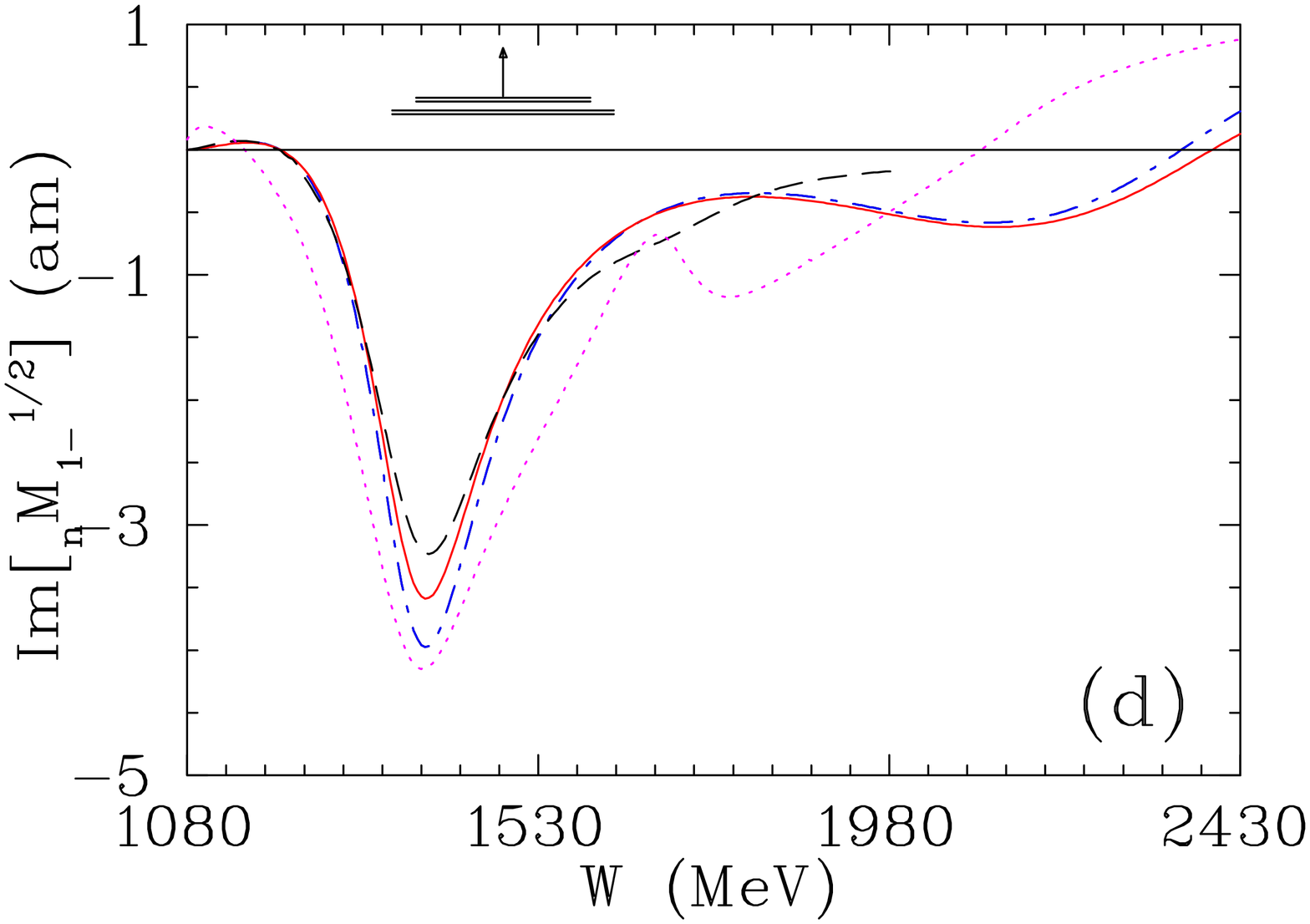}}
\centerline{\includegraphics[height=0.4\textwidth,angle=90]{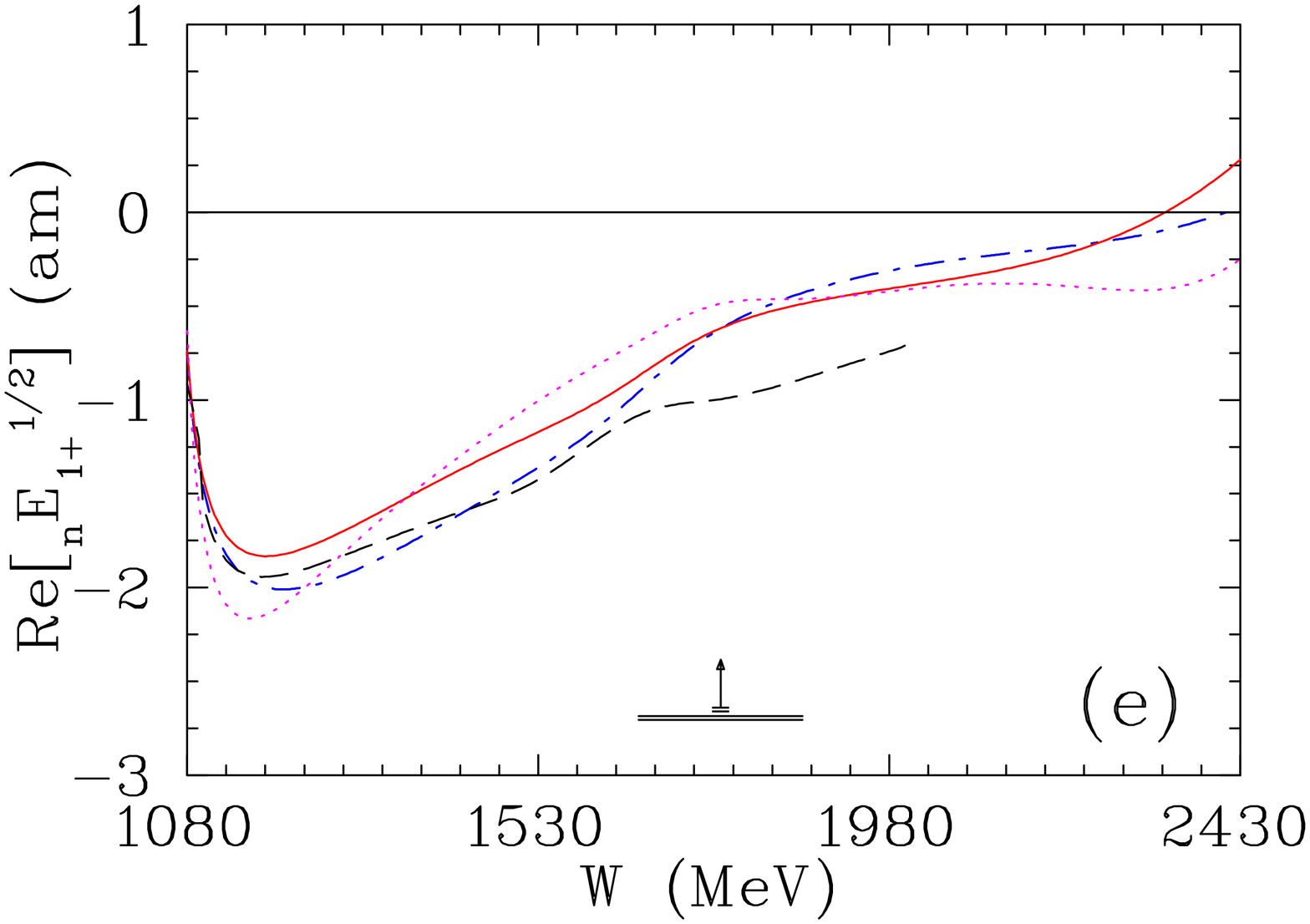}~~\includegraphics[height=0.4\textwidth,angle=90]{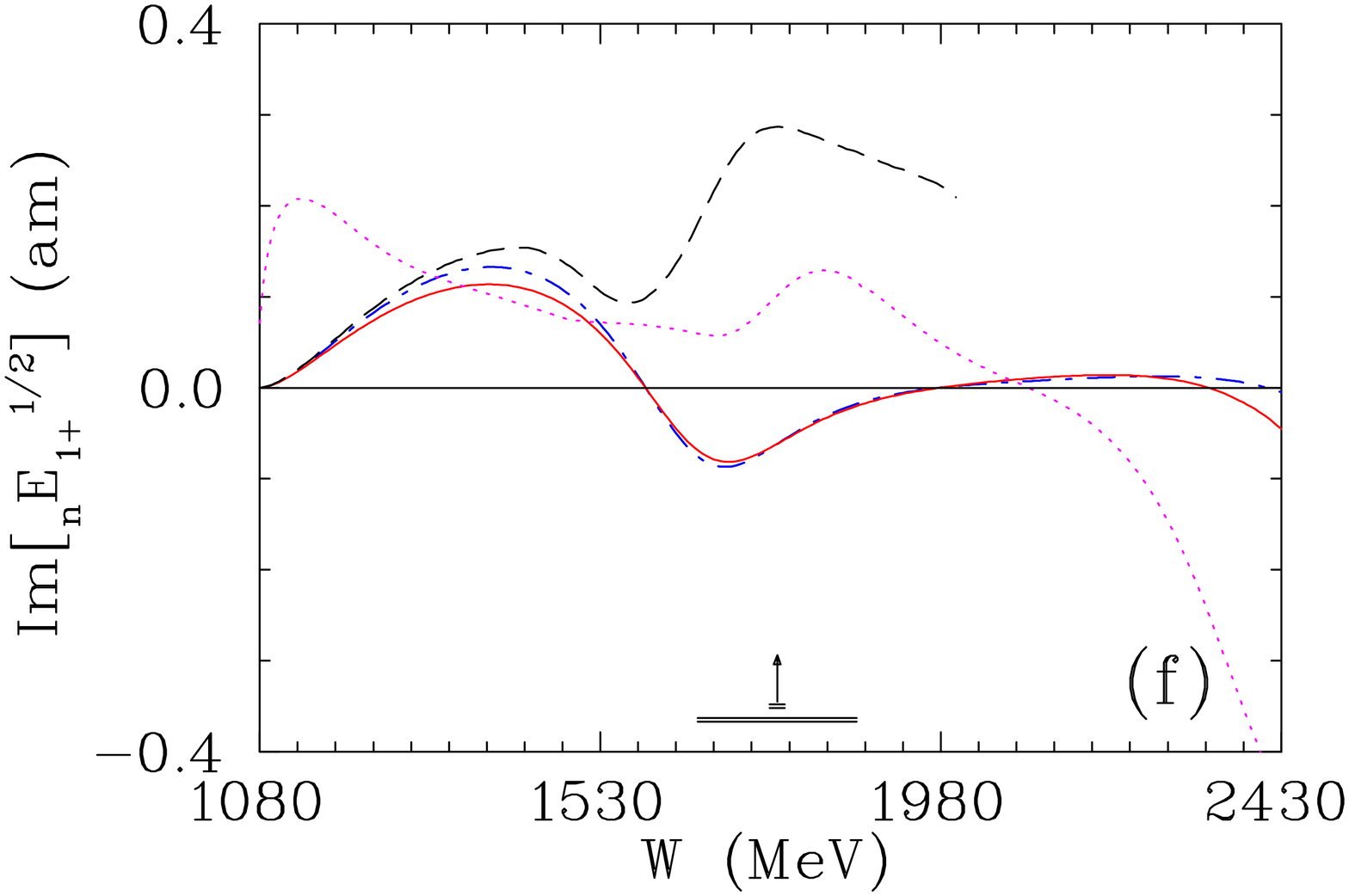}}
\centerline{\includegraphics[height=0.4\textwidth,angle=90]{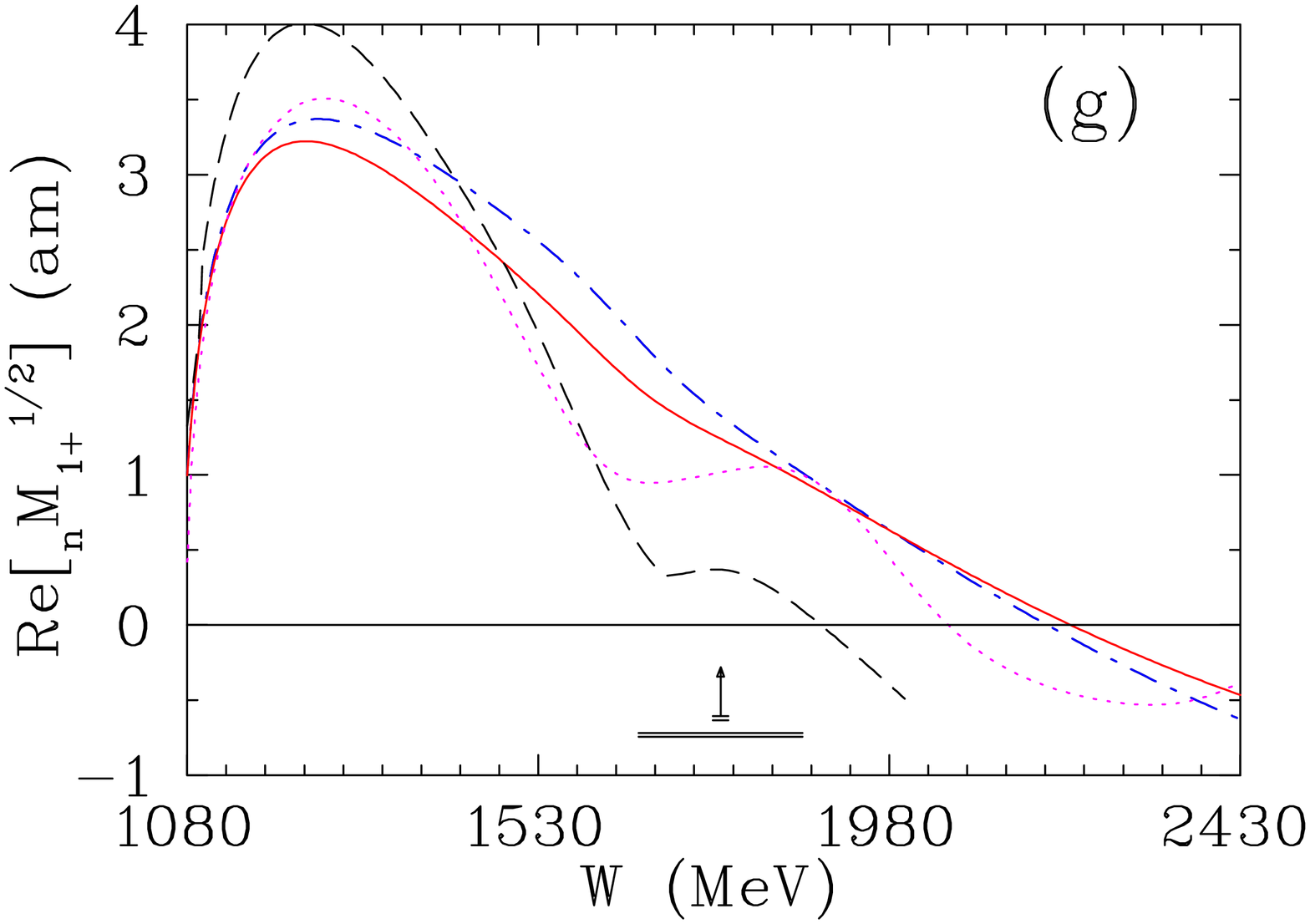}~~\includegraphics[height=0.4\textwidth,angle=90]{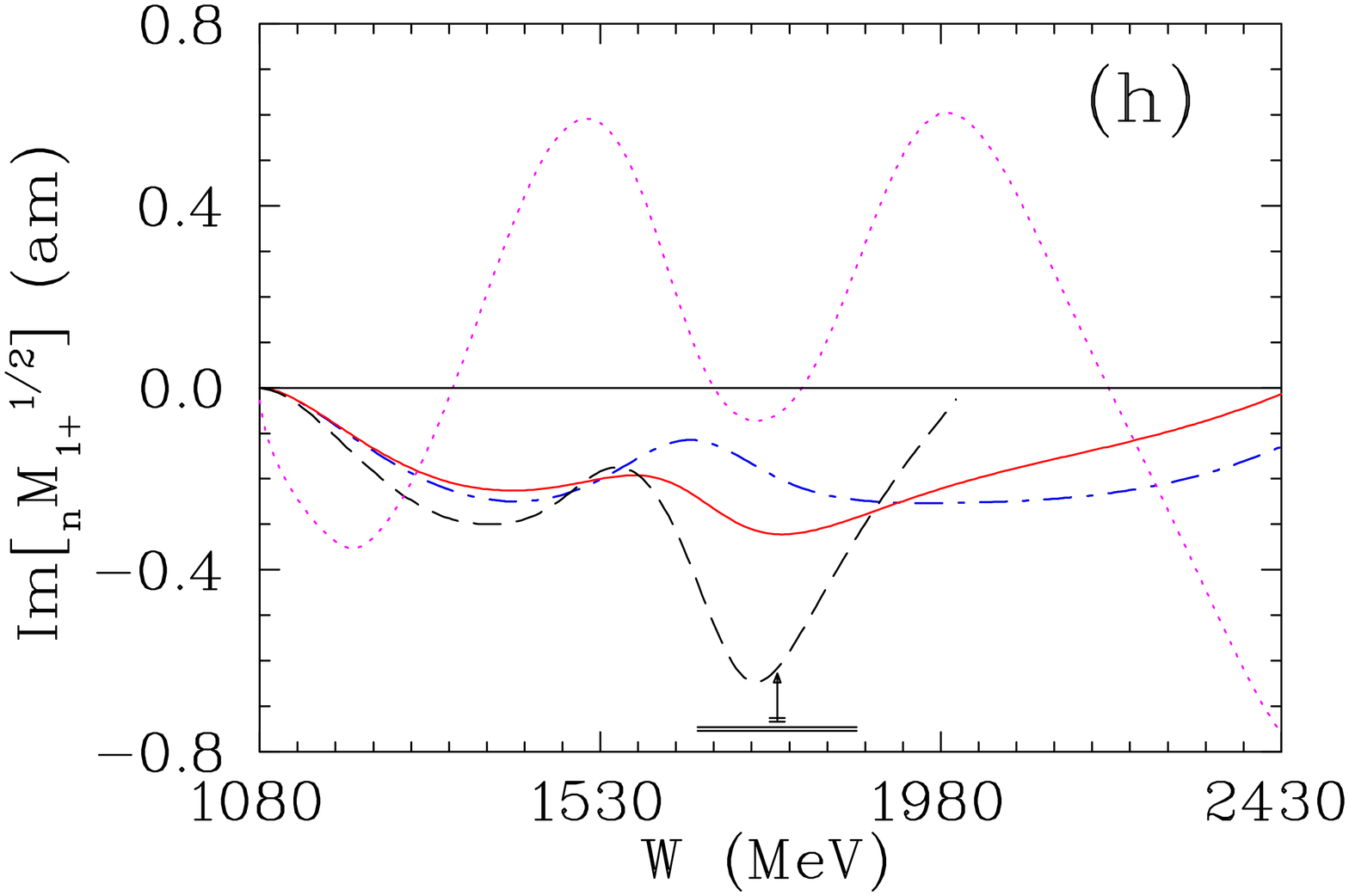}}

\protect\caption{(Color online) Neutron multipole $I$ = 1/2 amplitudes (in attometer - am units)
	from threshold to $W$ = 2.43~GeV ($E_\gamma$ = 2.7~GeV). 
	For the amplitudes, the subscript $n$ denotes a neutron target, the subscript $\ell \pm$ 
	gives the value of $j=\ell \pm 1/2$, and the superscript gives the isospin index.
	The red solid (blue dash-dotted) lines correspond to the new SAID MA27 
	(old PR15~\protect\cite{pr15}) solution. The magenta dotted (black 
	dashed) lines give the BG2014-02~\protect\cite{BnGa14} 
	(MAID2007~\protect\cite{MAID07}, which terminates at $W$ = 2~GeV) 
        solution. The vertical arrows indicate the Breit-Wigner mass 
	($W_{R}$), and the upper and lower horizontal bars show the partial 
        ($\Gamma_{\pi N}$) and the full ($\Gamma$) widths, respectively, of the
        resonances extracted by the Breit-Wigner fit of the $\pi N$ data associated with 
        the SAID solution SP06~\protect\cite{piN}. The red vertical arrows for 
        (a) and (b) indicate the $\eta$ production threshold. 
	\label{fig:g5}}
\end{figure*}

\begin{figure*}[th]
\centerline{\includegraphics[height=0.4\textwidth,angle=90]{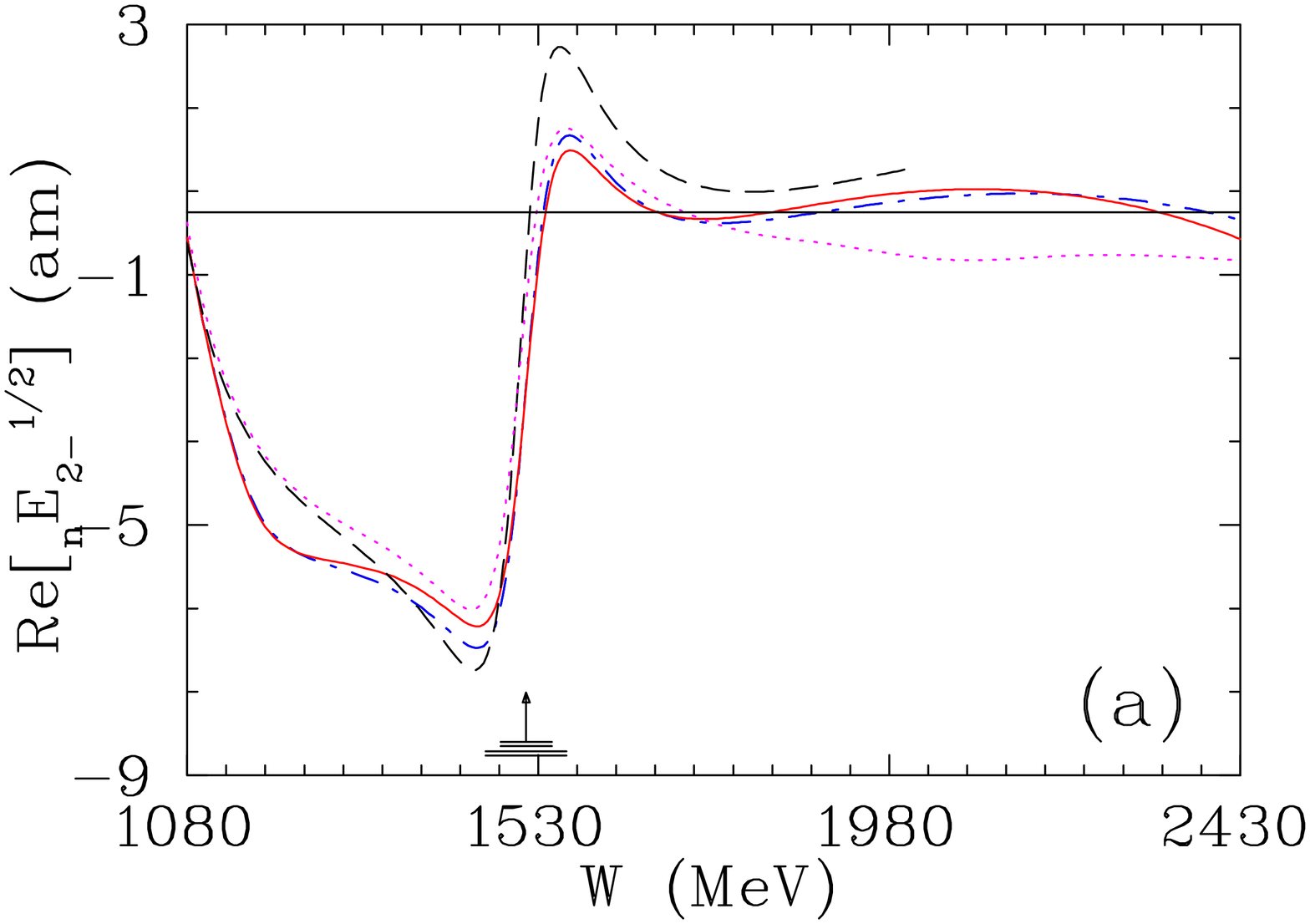}~~\includegraphics[height=0.4\textwidth,angle=90]{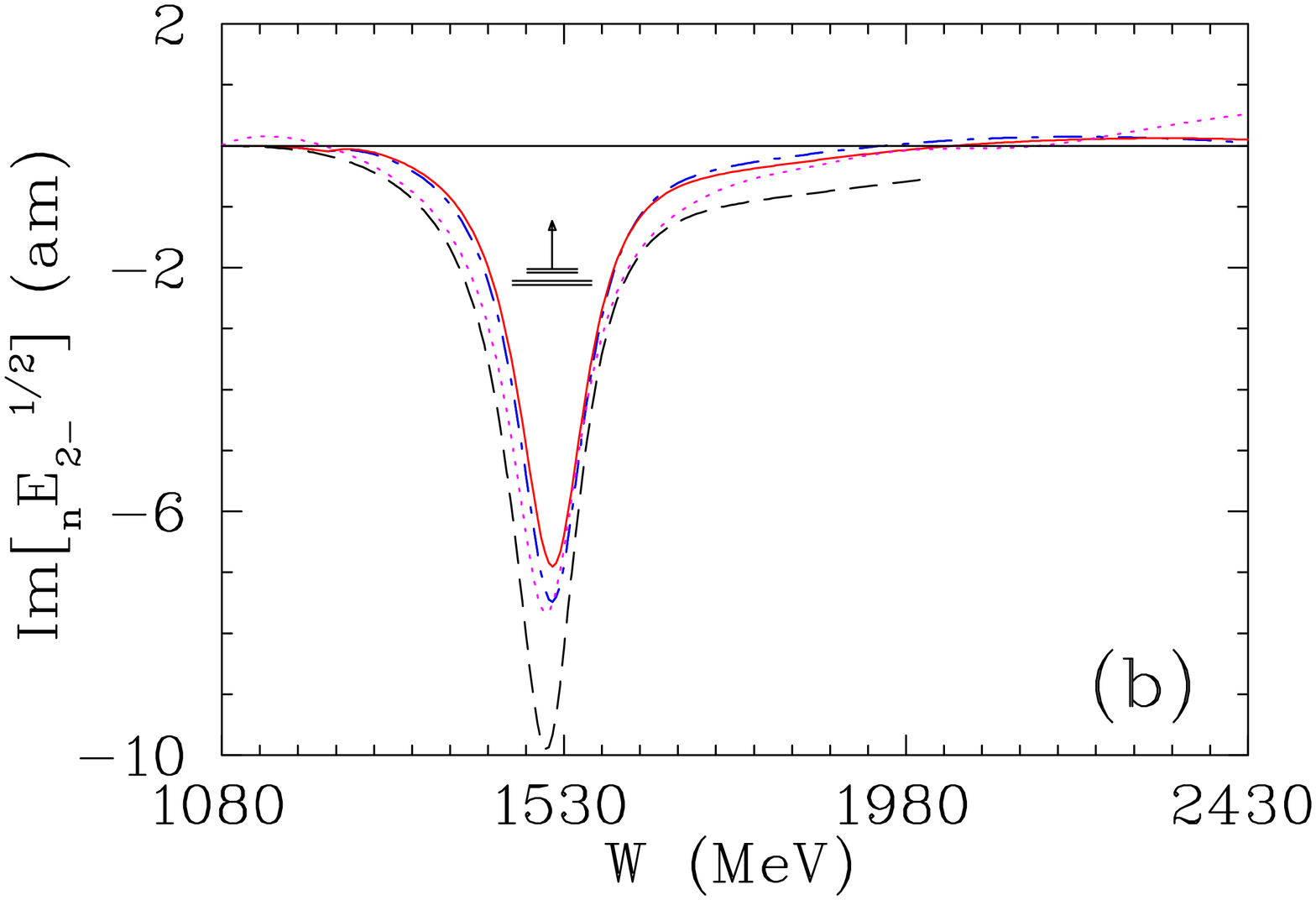}}
\centerline{\includegraphics[height=0.4\textwidth,angle=90]{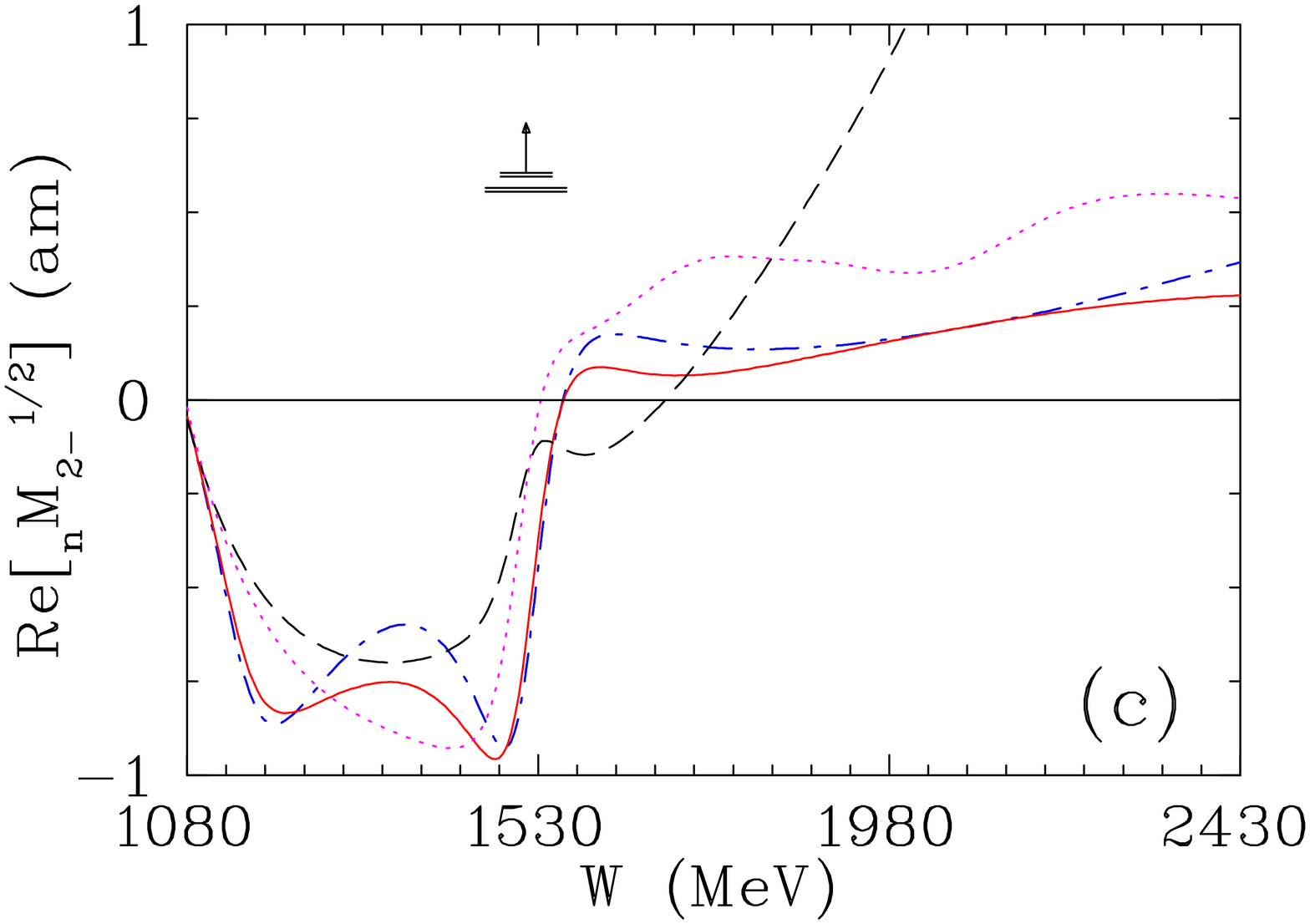}~~\includegraphics[height=0.4\textwidth,angle=90]{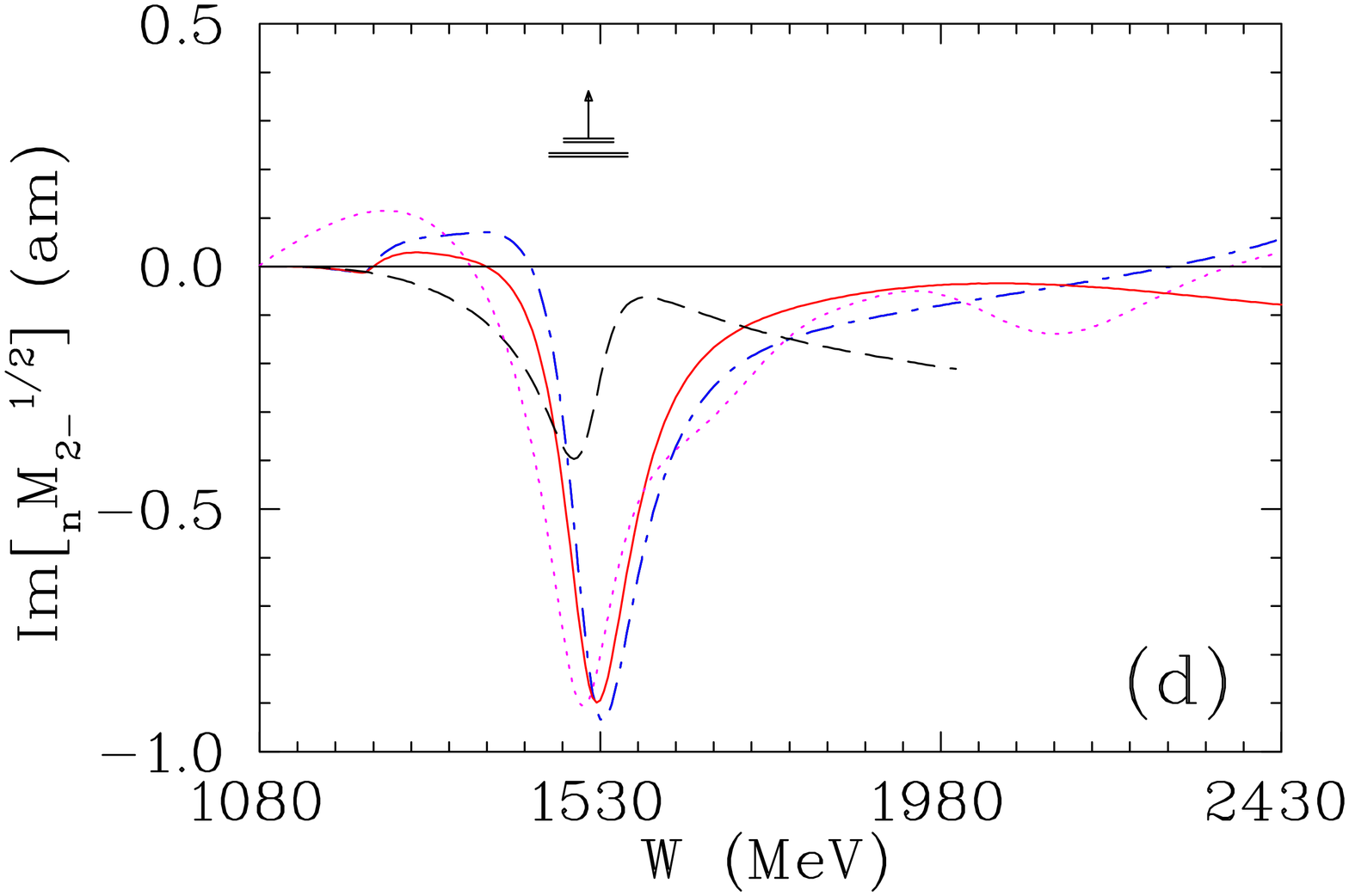}}
\centerline{\includegraphics[height=0.4\textwidth,angle=90]{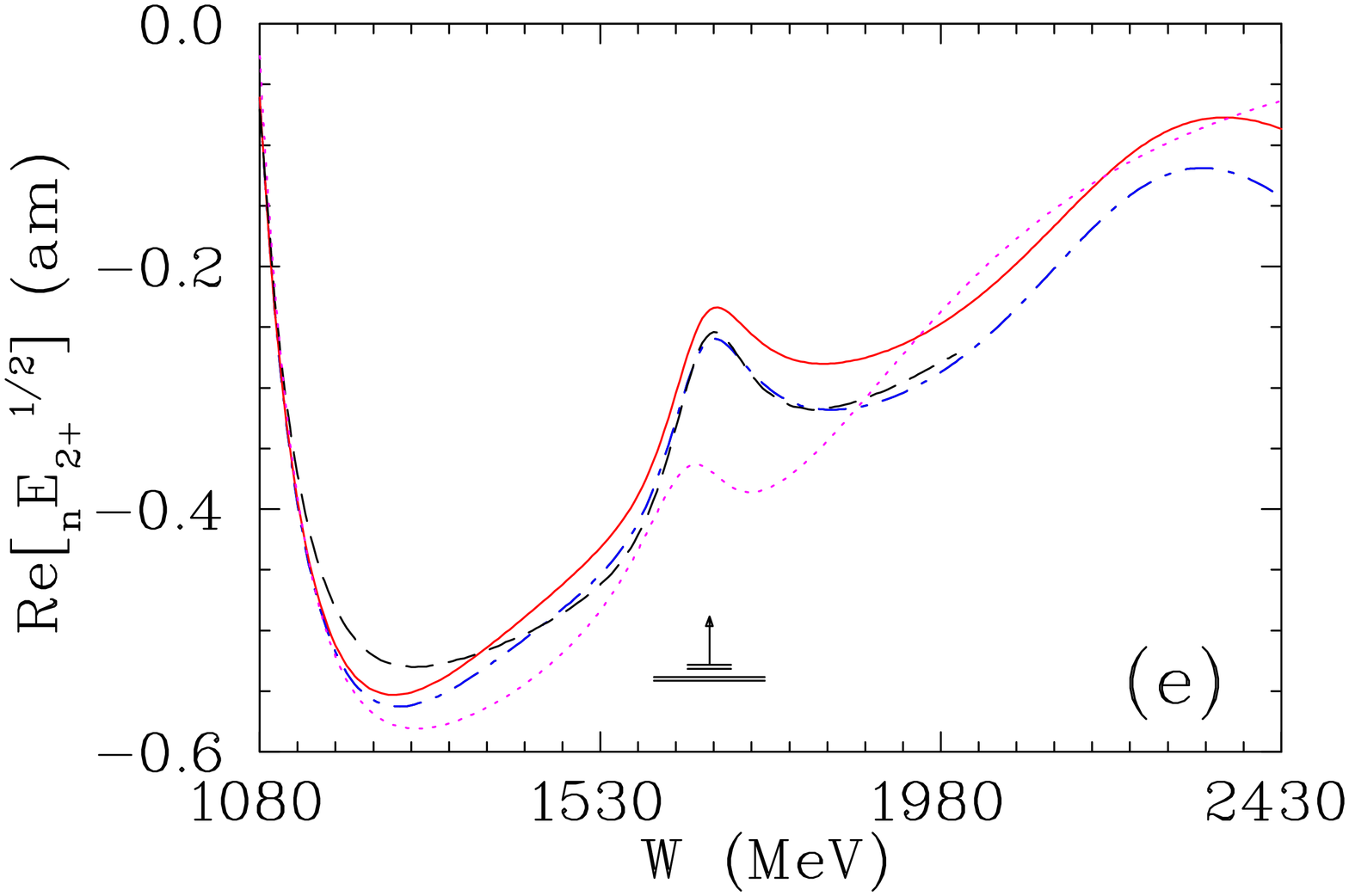}~~\includegraphics[height=0.4\textwidth,angle=90]{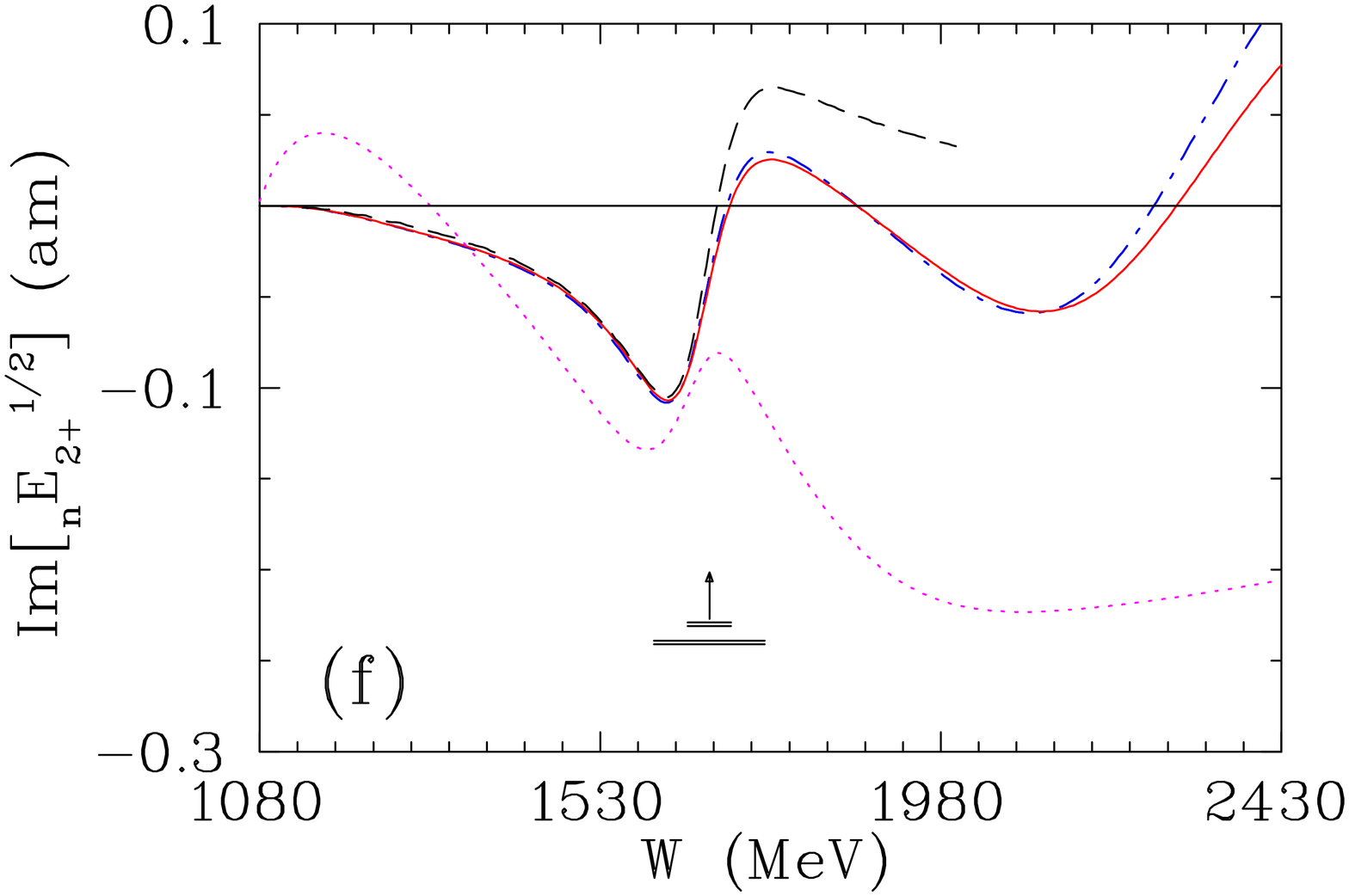}}
\centerline{\includegraphics[height=0.4\textwidth,angle=90]{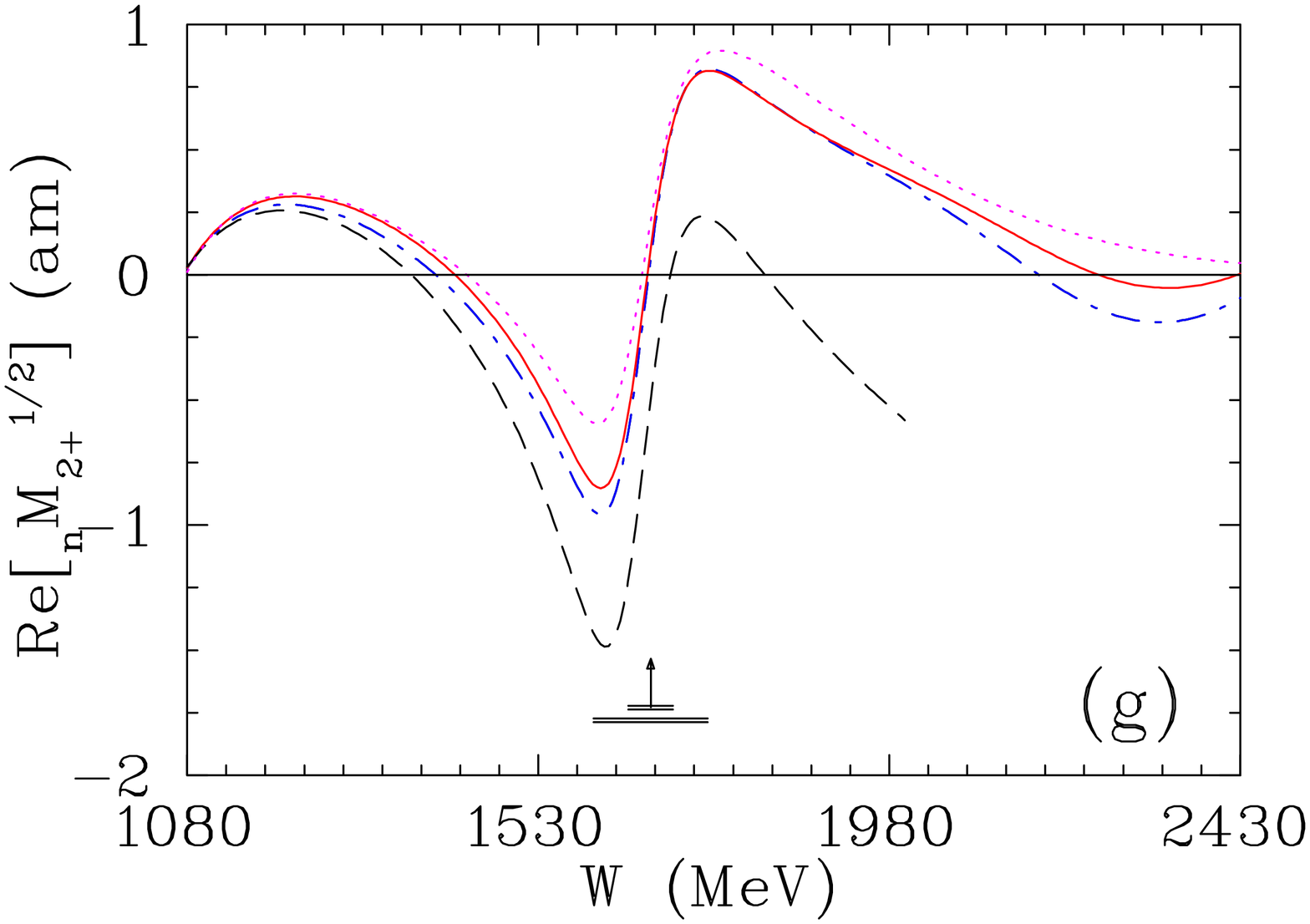}~~\includegraphics[height=0.4\textwidth,angle=90]{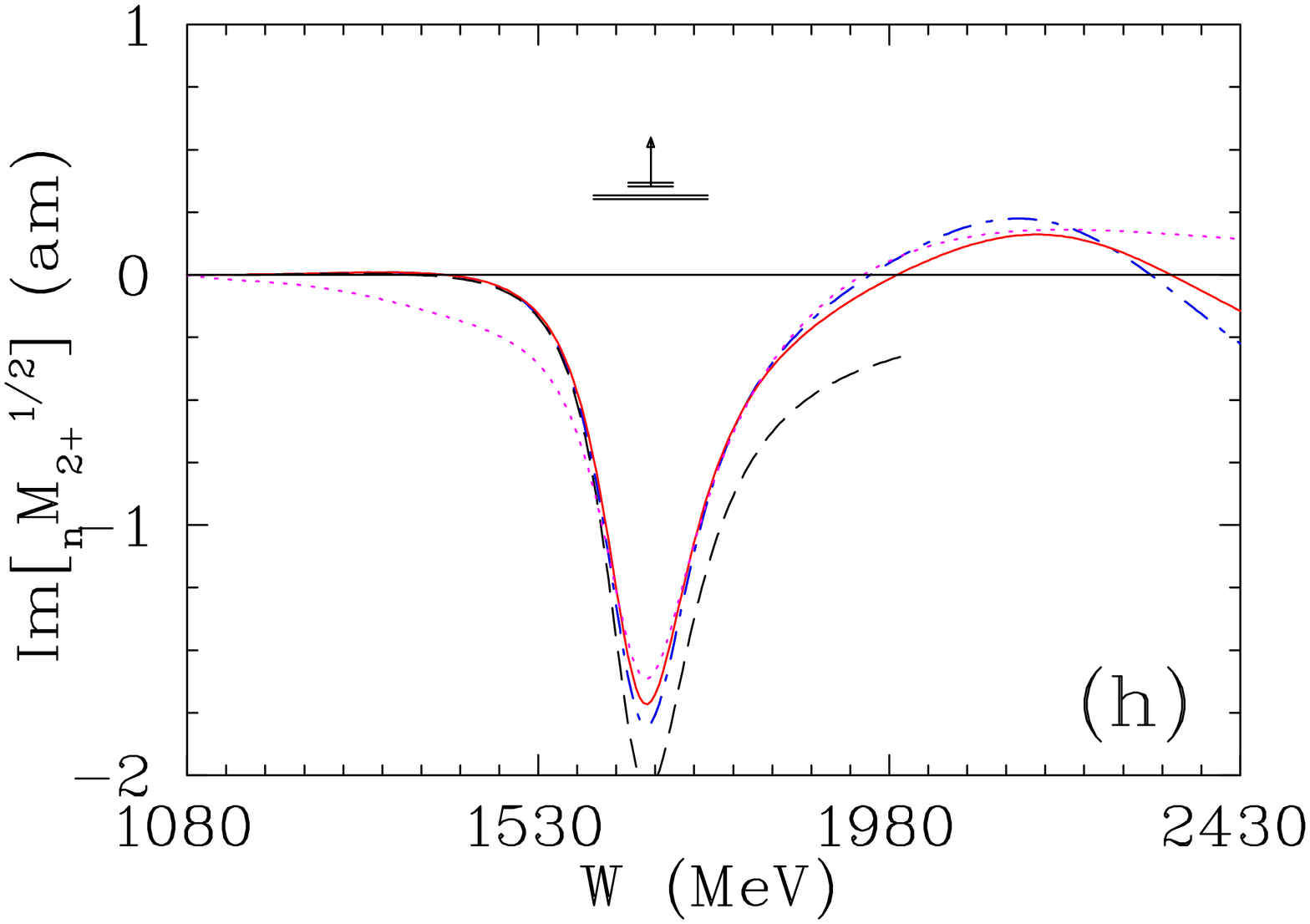}}

\protect\caption{(Color online) Neutron multipole $I$ = 1/2 amplitudes (in 
        attometer - am units) from threshold to $W$ = 2.43~GeV ($E_\gamma$ = 
        2.7~GeV). The notation of the multipoles is the same as in 
        Fig.~\protect\ref{fig:g5}. \label{fig:g6}}
\end{figure*}

\begin{figure*}[th]
\centerline{\includegraphics[height=0.4\textwidth,angle=90]{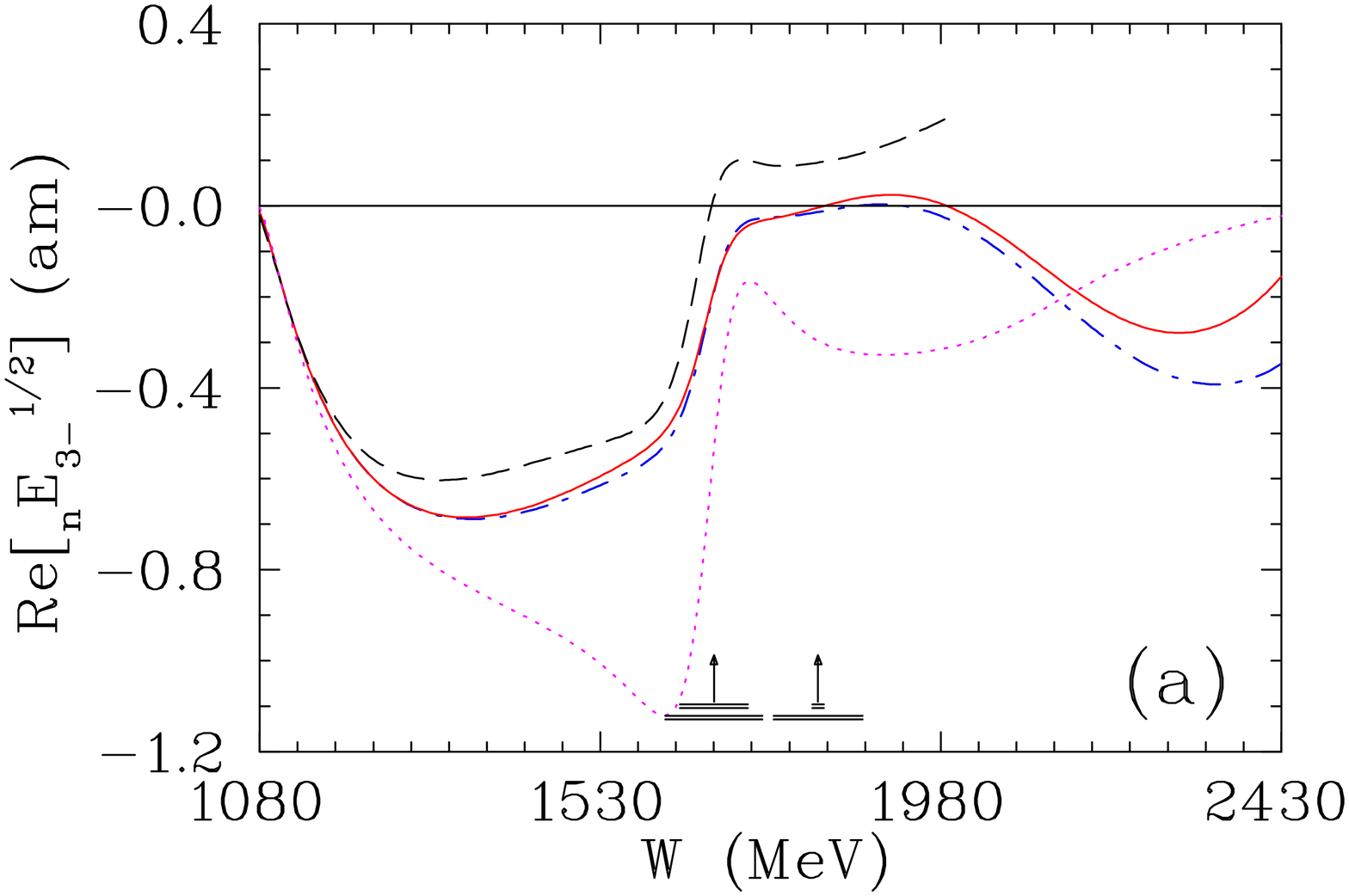}~~\includegraphics[height=0.4\textwidth,angle=90]{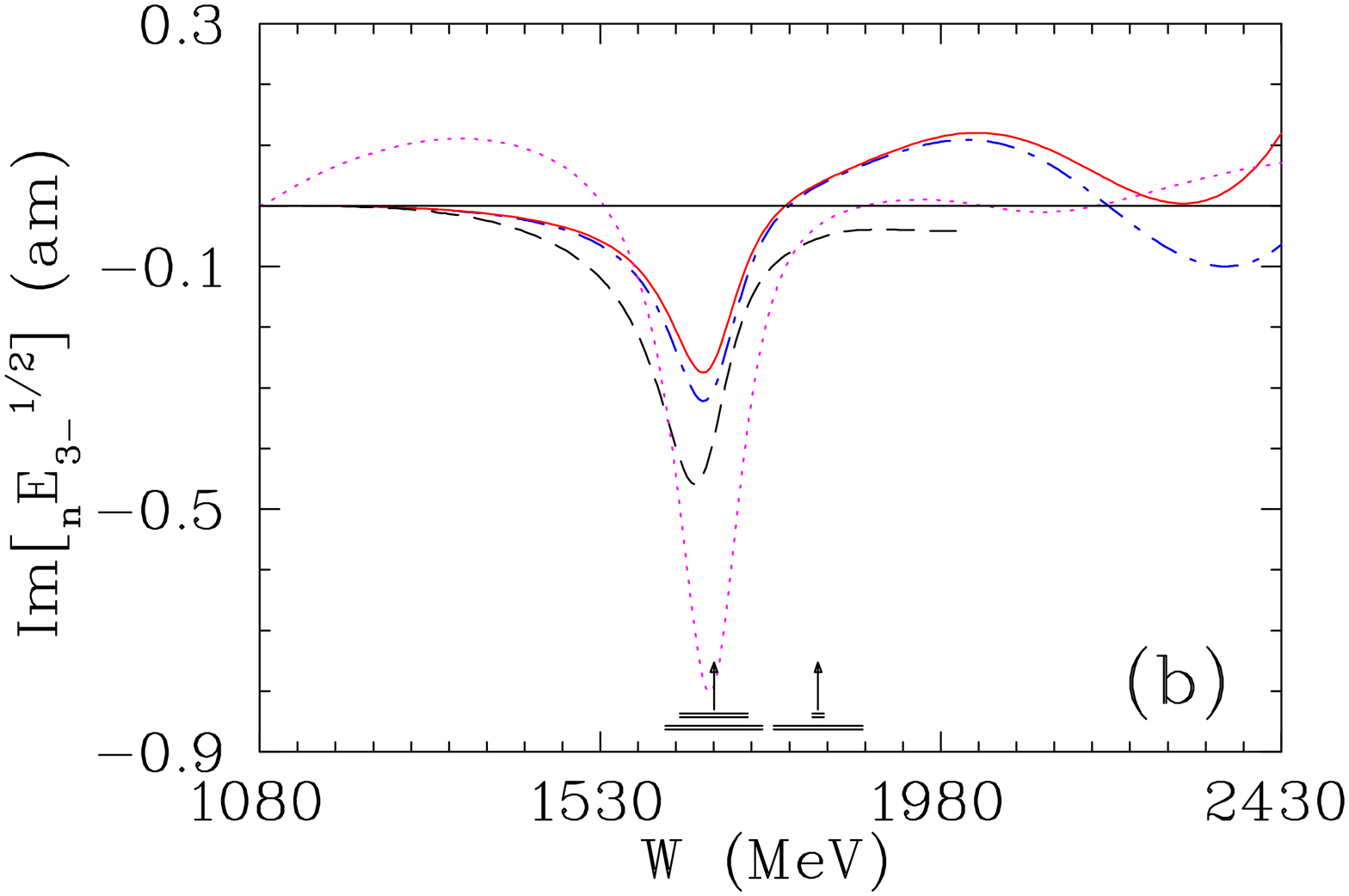}}
\centerline{\includegraphics[height=0.4\textwidth,angle=90]{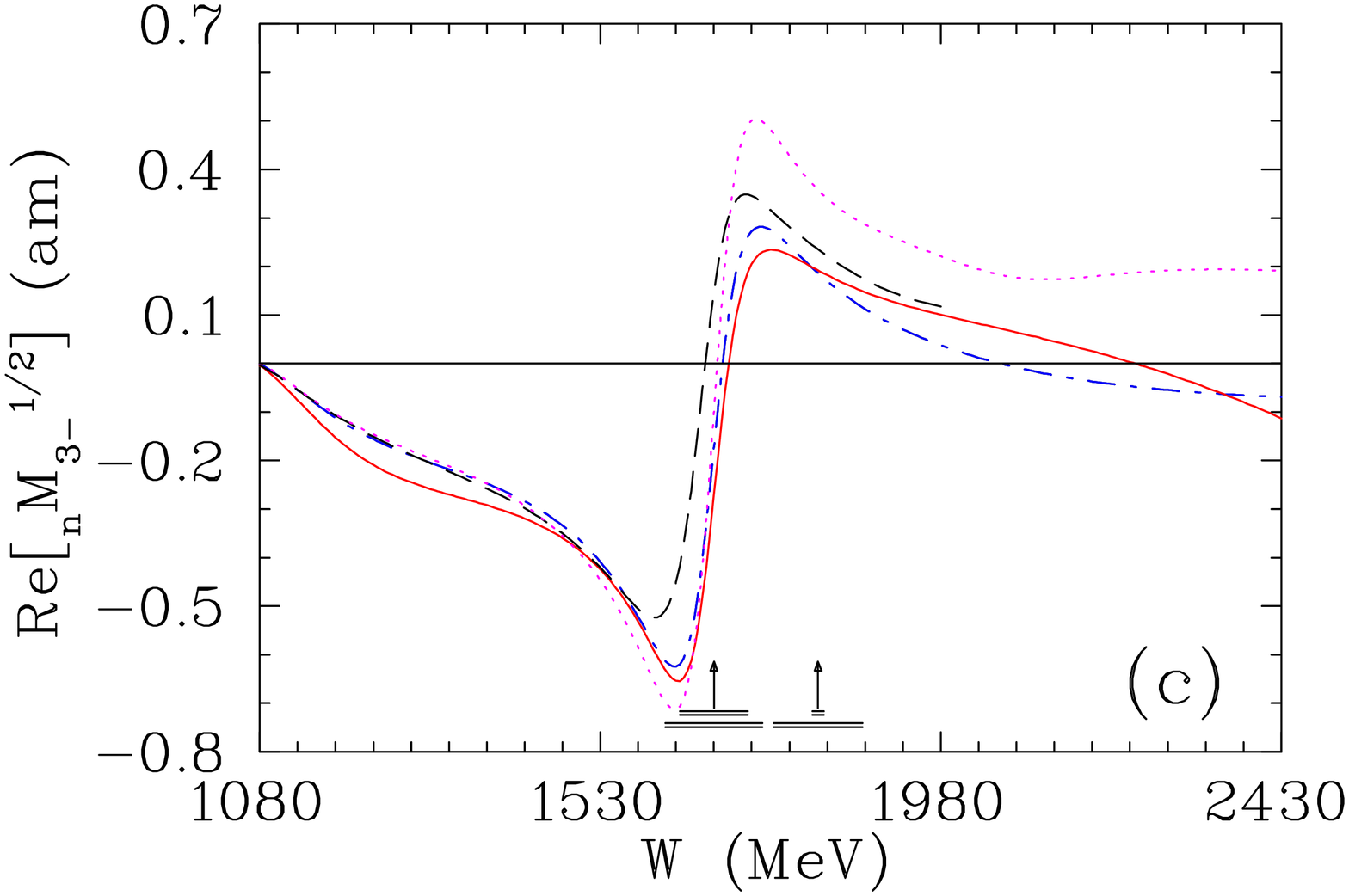}~~\includegraphics[height=0.4\textwidth,angle=90]{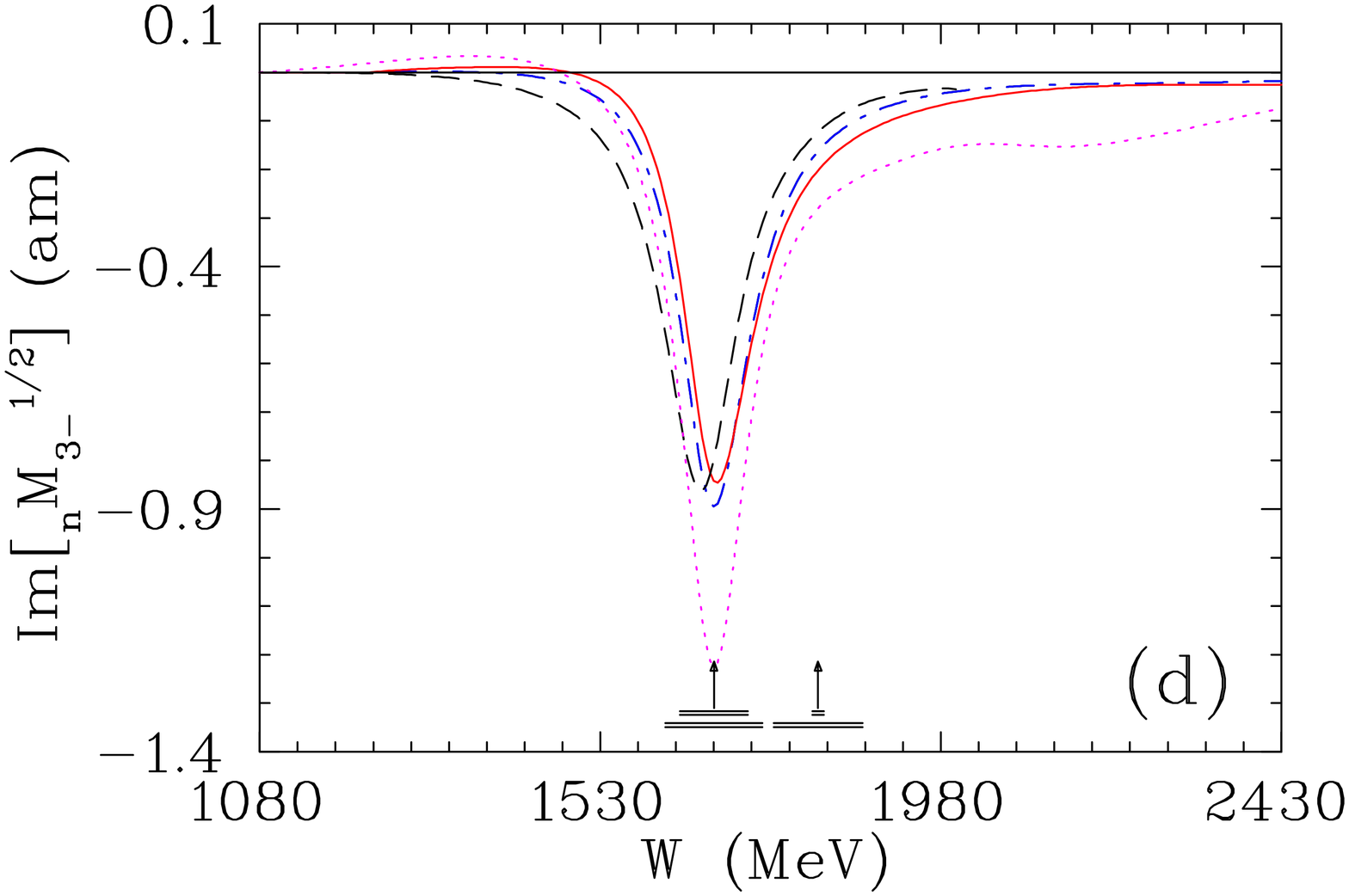}}

\protect\caption{(Color online) Neutron multipole $I$ = 1/2 amplitudes (in 
        attometer - am units) from threshold to $W$ = 2.43~GeV ($E_\gamma$ = 
        2.7~GeV). The notation of the multipoles is the same as in 
        Fig.~\protect\ref{fig:g5}. \label{fig:g7}}
\end{figure*}

In Figs.~\ref{fig:g5} to \ref{fig:g7}, $I=1/2$ multipole amplitudes from the 
present and previous SAID fits are compared to predictions from the MAID and 
Bonn-Gatchina groups. The Bonn-Gatchina analysis has been regularly 
updated, whereas the MAID fit was published in 2007 and therefore does not 
include any results from the past decade, including the recent CLAS g10 cross 
section measurements of Ref.~\cite{gb12}. The cross section requires $I=3/2$ 
multipoles as well, but these are highly constrained by proton-target 
measurements and have not changed significantly with the addition of 
neutron-target measurements (and therefore are not shown here). In the 
multipole plots, the subscript $n$ denotes a neutron target and $\ell \pm$ 
gives the value of $j=\ell \pm 1/2$, while the superscript gives the isospin 
index.

Changes in the multipole amplitudes can be seen in a comparison of the SAID 
curves in Figs.~\ref{fig:g5} to \ref{fig:g7}. Consistency among the analyses 
is visible in multipoles containing a dominant resonance, such as the $_nE_{2-}^{1/2}$
and $_nM_{2+}^{1/2}$ multipoles. However, the $_nE_{1+}^{1/2}$ and 
$_nM_{1+}^{1/2}$ multipoles differ even at the qualitative level. This discrepancy 
is evident in the proton-target multipoles as well. 

The full world database of $\gamma n\to \pi^- p$ experiments above $E_\gamma=1.2$~GeV 
contains mainly differential cross sections, apart from some $\Sigma$ beam asymmetry 
measurements from Yerevan~\cite{Yerevan}, GRAAL~\cite{GRAAL}, and CEA~\cite{CEA}. 
Ultimately, more measurements of the polarization observables are needed in the 
$\pi^-p$ and $\pi^0n$ channels in order to fully constrain the underlying reaction 
amplitudes. New $\gamma n$ measurements from the CLAS g14 dataset~\cite{sandorfiE} 
will significantly add to the available polarization observable measurements.

%------------------------------------------
%------------------------------------------
%------------------------------------------
%------------------------------------------
%------------------------------------------
%------------------------------------------

Looking for significant changes in the imaginary parts of the multipoles 
(Figs.~\ref{fig:g5} to \ref{fig:g7}) in the energy region below the older set 
of CLAS g10 cross sections~\cite{gb12}, several $N^*\to \gamma n$ photo-decay 
amplitudes have been extracted at their pole positions on the complex plane. 
This is the first determination of these amplitudes for the $N(1440)1/2^+$, 
$N(1535)1/2^-$, $N(1650)1/2^-$, and $N(1720)3/2^+$ states. A new approach has been 
applied to determine the pole positions and residues from the pion 
photoproduction multipoles~\cite{Svarc}. The method is based on a Laurent 
expansion of the multipoles, $M(W)$, with a Pietarinen series representing 
the regular (non-pole) part of the energy dependence as

\begin{equation}
\label{eq:Laurent-Pietarinen}
M(W) = \sum_{i=1}^k \frac{a_{-1}^{(i)}}{W-W_i}+ B^L(W).
\end{equation}

\noindent
Here $W$, $a_{-1}^{(i)}$, and $W_i$ are complex numbers representing the c.m. energy,
residues, and pole positions for the $i^{th}$ pole, respectively, and $B^L(W)$ is
a regular function in the whole complex plane. A general unknown analytic function 
$B(W)$ can be expanded into a power series of Pietarinen functions as:

\begin{eqnarray}
B^L(W)&=& \sum_{n=0}^M c_n\, X(W)^n + \sum_{n=0}^N d_n\, Y(W)^n \\
      &~& ~~~~~ + \sum_{n=0}^N e_n\, Z(W)^n + \cdots, \nonumber  \\
X(W)  &=& \frac{\alpha-\sqrt{x_P-W}}{\alpha+\sqrt{x_P - W }}, \nonumber \\  
Y(W)  &=& \frac{\beta-\sqrt{x_Q-W }}{\beta+\sqrt{x_Q-W }},    \nonumber \\
Z(W)  &=& \frac{\gamma-\sqrt{x_R-W }}{\gamma+\sqrt{x_R-W }}, \nonumber 
\end{eqnarray}

\noindent
where $c_n,d_n,e_n$ and $\alpha,\beta,\gamma$ are real numbers that represent tuning 
parameters and coefficients of the Pietarinen functions $X(W)$, $Y(W)$, and $Z(W)$, 
respectively. A variable number of series was used, depending on the structure of 
the non-pole part of each amplitude, and $x_P$, $x_Q$, and $x_R$ represent the branch
points for each Pietarinen function. Once the pole position and residue were 
determined, the photo-decay amplitude at the pole could be constructed, as described 
in Ref.~\cite{photo-pole}. The residue of the corresponding $\pi N$ elastic scattering 
amplitude, required in this construction, was taken from the SAID analysis of elastic 
scattering data~\cite{piN}.

The $A_{1/2}(n)$ and $A_{3/2}(n)$ neutron helicity amplitudes for $N(1440)1/2^+$, 
$N(1535)1/2^-$, $N(1650)1/2^-$, and $N(1720)3/2^+$ for the new SAID MA27 solution 
are compared in Table~\ref{tab:tbl2} to the recent SAID GB12~\cite{gb12} and 
BG2013~\cite{BnGa13} solutions that were based on fits to all available data at 
the time, including the CLAS g10 dataset~\cite{gb12}. From this table, the 
photo-decay amplitudes determined from the MA27 solution can be directly compared 
against the Breit-Wigner determinations. In addition, Table~\ref{tab:tbl2} 
includes a comparison to the older MAID2007~\cite{MAID07} solution, to the 
relativized quark model predictions of Ref.~\cite{Capstick}, and to the current 
PDG values~\cite{PDG16}. The uncertainties on the modulus and phase quoted in 
Table~\ref{tab:tbl2} for the new MA27 solution were derived by comparing the 
global energy-dependent and energy-independent, single-energy amplitudes (see 
Ref.~\cite{ed-se} for a discussion on the two approaches). The comparison gave 
residues with uncertainties. Extracting the photo-decay amplitudes and considering 
the spread of possible values gave the MA27 uncertainties. These comparisons showed 
that the parameters from the MA27 solution are reasonably well under control.

The pole-valued and Breit-Wigner amplitudes from the fits are generally consistent 
in terms of the moduli. The comparisons are reasonable for the $N(1440)1/2^+$ and 
the $N(1535)1/2^-$. For the $N(1650)1/2^-$, the change from GB12~\cite{gb12} is 
significant and the result is in reasonable agreement with BG2013~\cite{BnGa13}, 
which used the FSI corrected g10 $\gamma n \to \pi^- p$ cross sections that were 
used for GB12. The $N(1650)1/2^-$ state has been difficult to describe as it is 
so close to the $N(1535)1/2^-$ and the $\eta N$ cusp, however, this pole-valued 
determination is believed to be more model-independent than the Breit-Wigner 
amplitude~\cite{Svarc}, which is reflected in the quoted uncertainties shown in
Table~\ref{tab:tbl2}. For the $N(1720)3/2^+$, the differences with respect to 
the BG2013 solution~\cite{BnGa13} are significant and indicate that the CLAS g13 
data provide tighter constraints in the coupled-channel model fits.

Comparing the new SAID MA27 solution with the relativized quark model predictions 
of Ref.~\cite{Capstick}, there are significant differences in the helicity 
amplitudes for the $N(1440)1/2^+$ and $N(1650)1/2^-$, while the helicity 
amplitudes for the $N(1535)1/2^-$ and $N(1720)3/2^+$ are in good agreement. With
respect to the current PDG values~\cite{PDG16}, Table~\ref{tab:tbl2} shows
good correspondence with the MA27 solution for $N(1440)1/2^+$ and $N(1535)1/2^-$,
but sizable disagreements for the higher-lying states.

A direct comparison of the quoted uncertainties on the neutron 
helicity amplitudes from the different solutions presented in Table~\ref{tab:tbl2} 
must be made with some caution. For the MA27, GB12, and BG2013 listings, the 
uncertainties do not take into account the significant model dependence in fitting 
the sparse database. In fact, the variance of the extracted results from different 
solutions fitting the same database would provide a reasonable estimate for this 
model dependence. However, this direct comparison is not possible given the 
different data sets employed for the different solutions shown in Table~\ref{tab:tbl2}. 
Considering this issue, it is still meaningful that the overall quoted uncertainties 
for the helicity amplitudes from the MA27 solution are noticeably reduced relative to 
the BG2013 solution and to the GB12 solution (in particular for the $N(1650)1/2^-$) 
due to a combination of two factors. The first is the increased size of the database 
for MA27 that includes the new g13 $\gamma n$ cross sections and the second is the 
reduced model dependence of the pole fit approach employed for MA27.

%------------------------------------------
\begin{table*}[htb!]

\centering \protect\caption{Moduli (in $\rm (GeV)^{-1/2}$) and phases (in degrees) 
        of the neutron helicity amplitudes $A_{1/2}(n)$ and $A_{3/2}(n)$ from the 
        SAID MA27 solutions (third column). The Breit-Wigner neutron photo-decay 
        amplitudes are compared from SAID GB12~\protect\cite{gb12} (fourth column) 
        from BG2013~\protect\cite{BnGa13} (fifth column), and from MAID2007
        \protect\cite{MAID07} (sixth column). The relativized quark model 
        predictions from Ref.~\protect\cite{Capstick} (seventh column) are 
        included along with the PDG values (eighth column)~\cite{PDG16}.}
\vspace{2mm}
{\begin{tabular}{|c|c|c|c|c|c|c|c|} \hline
Resonance      & Coupling     & MA27               & GB12             & BG2013           & MAID2007 & Capstick & PDG 2016         \tabularnewline
               &              & modulus, phase     &                  &                  &          &          &                  \tabularnewline\hline
$N(1440)1/2^+$ & $A_{1/2}(n)$ & 0.065$\pm$0.005,   5$^\circ$$\pm$3$^\circ$  & 0.048$\pm$ 0.004 &  0.043$\pm$0.012 &  0.054   & -0.006   &  0.040$\pm$0.010 \tabularnewline
$N(1535)1/2^-$ & $A_{1/2}(n)$ &-0.055$\pm$0.005,   5$^\circ$$\pm$2$^\circ$  &-0.058$\pm$ 0.006 & -0.093$\pm$0.011 & -0.051   & -0.063   & -0.075$\pm$0.020 \tabularnewline
$N(1650)1/2^-$ & $A_{1/2}(n)$ & 0.014$\pm$0.002, -30$^\circ$$\pm$10$^\circ$ &-0.040$\pm$ 0.010 &  0.025$\pm$0.020 &  0.009   & -0.035   & -0.050$\pm$0.020 \tabularnewline
$N(1720)3/2^+$ & $A_{1/2}(n)$ &-0.016$\pm$0.006,  10$^\circ$$\pm$5$^\circ$  &                  & -0.080$\pm$0.050 & -0.003   &  0.004   & -0.080$\pm$0.050 \tabularnewline
$N(1720)3/2^+$ & $A_{3/2}(n)$ & 0.017$\pm$0.005,  90$^\circ$$\pm$10$^\circ$ &                  & -0.140$\pm$0.065 & -0.031   &  0.011   & -0.140$\pm$0.065 \tabularnewline
\hline
\end{tabular}} \label{tab:tbl2}
\end{table*}
%------------------------------------------

%------------------------------------------
\section{Summary and Conclusions}
\label{sec:conc}

A comprehensive set of $\gamma n\to\pi^-p$ differential cross sections at 157 
photon energies has been determined with CLAS using a tagged-photon beam at 
incident photon energies from 0.445~GeV to 2.510~GeV. These data provide a factor 
of nearly three increase to the world's data for this channel at these energies. 
To extract the $\gamma n$ cross section from the $\gamma d$ data, FSI 
corrections were included using a diagrammatic technique that takes into account 
a kinematic cut with momenta below (above) 200~MeV/c to select slow (fast) outgoing 
protons. In this analysis, the FSI correction factor depended on the photon energy 
and meson production angle, and was averaged over the rest of the variables in the 
region of the quasi-free process on the neutron.

The data collected in this CLAS g13 dataset spans a broad energy range, from just 
above the $\Delta$ isobar through the second, third, and fourth resonance regions.
These data extend far into the poorly studied high-mass region above
$W \sim 1.8$~GeV where many resonances are expected to exist but have not been
firmly established. The precision of the data can be seen not only in the presented 
differential cross sections, but also through the uncertainties on the extracted 
Legendre coefficients. This approach of fitting the excitation functions with a 
Legendre series presents the data in a more compact and visual manner. These results 
will be useful for performing detailed phase shift analyses to better understand the
resonant amplitudes.

On the experimental side, further improvements in the partial wave analyses await 
more precision data, specifically in the region above $E_\gamma=0.5$~GeV involving 
polarized photons and/or polarized targets. The data that are presently available 
are provided in Ref.~\cite{SAID}. Due to the closing of hadron facilities, new 
$\pi^-p \to \gamma n$ experiments are not planned, and only $\gamma n \to \pi^-p$ 
measurements are possible at electromagnetic facilities using deuterium targets. 
The agreement of these new $\gamma n \to \pi^- p$ cross section data with existing 
inverse $\pi^-$ photoproduction measurements indicates that these g13 measurements 
are reliable despite the use of deuterium as an effective neutron target.

As part of this new dataset for $\gamma n \to \pi^- p$, a new SAID multipole 
analysis called MA27 has been completed. This energy-dependent solution, which 
includes the CLAS g13 data, provides an improved understanding of the $N^*$ 
resonance parameters for several states, compared to the previous GB12 SAID 
solution that does not include the g13 CLAS data. In the MA27 solution, several 
photo-decay amplitudes $N^* \to \gamma n$ have been extracted at their pole 
positions on the complex plane with very small uncertainties. This is the 
first-ever determination of the excited neutron multipoles for the $N(1440)1/2^+$, 
$N(1535)1/2^-$, $N(1650)1/2^-$, and $N(1720)3/2^+$ resonances, contributing a 
crucial complement to the excited proton spectra. In addition, these new precision 
$\gamma n \to \pi^- p$ data will provide important and necessary constraints to 
advance coupled-channel analysis fits that are sorely lacking $\gamma n$ data over 
nearly the full nucleon resonance region.

%------------------------------------------
\begin{acknowledgments}
The authors acknowledge the outstanding efforts of the staff of the Accelerator 
and the Physics Divisions at Jefferson Lab that made this experiment possible. 
This work was supported by the US Department of Energy, the National Science 
Foundation, the Scottish Universities Physics Alliance (SUPA), the United 
Kingdom's Science and Technology Facilities Council, the National Research 
Foundation of Korea, the Italian Instituto Nazionale di Fisica Nucleare, the 
French Centre National de la Recherche Scientifique, the French Commissariat \`{a} 
l'Energie Atomique, and the Deutsche Forschungsgemeinschaft (SFB 1044). This 
material is based upon work supported by the U.~S.~Department of Energy, Office 
of Science, Office of Nuclear Physics under Contract No. DE-AC05-06OR23177. The 
authors A.E.K. and V.E.T. also acknowledge the support of the grant RFBR 
16-02-00767.
\end{acknowledgments}

%------------------------------------------

\end{document}